\documentclass[12pt,a4paper]{article}
\usepackage{amssymb,amsmath}
\usepackage[dvips]{lscape,graphicx}

\newcommand{\bc}{\begin{center}}
\newcommand{\ec}{\end{center}}
\newcommand{\bd}{\begin{displaymath}}
\newcommand{\ed}{\end{displaymath}}
\newcommand{\be}{\begin{equation}}
\newcommand{\ee}{\end{equation}}
\newcommand{\ba}{\begin{array}}
\newcommand{\ea}{\end{array}}
\newcommand{\bt}{\begin{tabular}}
\newcommand{\et}{\end{tabular}}

\newcommand{\ds}{\displaystyle}

\begin{document}

\begin{titlepage}

\vspace*{-15mm}
\begin{flushright}
{CERN-PH-TH/2008-058}\\
\end{flushright}
\vspace*{5mm}

\begin{center}
{
\LARGE
On the origin of
approximate custodial symmetry in the Two-Higgs
Doublet
Model.}\\[8mm]
{\large C.~D.~Froggatt$^{a,b}$, R.~Nevzorov$^a$\,${}^{1}$,
H.~B.~Nielsen$^b$, D.~Thompson$^a$\\[3mm]
\itshape{$^a$ Department of Physics and Astronomy,}\\[2mm]
\itshape{Glasgow University, Glasgow, Scotland}\\[2mm]
\itshape{$^b$ The Niels Bohr Institute, Copenhagen, Denmark}
}\\[1mm]
\end{center}
\vspace*{0.75cm}

\begin{abstract}{
\noindent We argue that the consistent implementation of the
multiple point principle (MPP) in the general non-supersymmetric
two Higgs doublet model (2HDM) can lead to a set of approximate
global custodial symmetries that ensure CP conservation in the
Higgs sector and the absence of flavour changing neutral currents
(FCNC) in the considered model. In particular the existence of a
large set of degenerate vacua at some high energy scale $\Lambda$
caused by the MPP can result in approximate $U(1)$ and $Z_2$ 
symmetries that suppress FCNC and CP--violating interactions 
in the 2HDM. We explore the renormalisation group (RG) flow of 
the Yukawa and Higgs couplings within the MPP inspired 2HDM 
with approximate custodial symmetries and show that the solutions 
of the RG equations are focused near quasi--fixed points at low 
energies if the MPP scale scale $\Lambda$ is relatively high.
We study the Higgs spectrum and couplings near the quasi--fixed
point at moderate values of $\tan\beta$ and compute a theoretical
upper bound on the lightest Higgs boson mass. If $\Lambda\gtrsim
10^{10}\,\mbox{GeV}$ the lightest CP--even Higgs boson is always
lighter than $125\,\mbox{GeV}$. When the MPP scale is low, the
mass of the lightest Higgs particle can reach
$180-220\,\mbox{GeV}$ while its coupling to the top quark can be
significantly larger than in the SM, resulting in the enhanced
production of Higgs bosons at the LHC. Other possible scenarios
that appear as a result of the implementation of the MPP in the
2HDM are also discussed.}
\end{abstract}

\vspace{1cm} \footnoterule{\noindent${}^{1}$
On leave of absence from the Theory Department, ITEP, Moscow, Russia}

\end{titlepage}

\newpage
\section{Introduction}

The understanding of the origin of the strong suppression of
flavour changing neutral current (FCNC) transitions observed in
Nature together with the origin of CP violation, are among the
major outstanding problems in particle physics. In the standard 
model (SM) CP violation arises from the phase of the CKM matrix 
\cite{ckm10}-\cite{ckm11} and from the ``$\theta$-term" in the QCD Lagrangian.
Within the SM the particle content, gauge invariance and
renormalizability imply the absence of FCNC transitions at the
tree level. At one--loop, they are further suppressed by light
quark masses (when compared to $M_W$), i.e. through the GIM
mechanism \cite{Glashow:1970gm}, and by small mixing between the
third and the first two generations. 

However because of the possible presence of new physics the SM 
should be regarded as an effective ``low energy" theory which, up 
to some scale $\Lambda$, is a good approximation to the more 
fundamental underlying theory. Therefore the renormalizable 
interactions of the SM are in general supplemented by higher 
dimensional interaction terms suppressed by some powers of the scale 
$\Lambda$. These new interactions introduce new sources of CP
violation. In the considered case $SU(3)_C\times SU(2)_W\times U(1)_Y$ 
invariance is not sufficient any more to protect the observed 
strong suppression of the FCNC processes. Under these circumstances 
we may expect that either the scale $\Lambda$ is 
huge\footnote{The strongest bounds are obtained from
$K^0-\overline{K^0}$ mixing and CP violation in $K$ meson decay
measurements that forbid any $\Lambda$ below $10^4\,\mbox{TeV}$.
The measurements of CP violation in $B$ meson decay as well as in
$D^0-\overline{D^0}$ and $B^0-\overline{B^0}$ mixings imply that
$\Lambda\gtrsim 10^3\,\mbox{TeV}$ \cite{Grossman:2003qi}.} or
dangerous new interactions are absent because of symmetries of the
underlying theory. If the only suppression of FCNC processes is
due to the scale $\Lambda$, then there is a tension between the new
physics scale which is required in order to solve the hierarchy
problem and the one which is needed in order to satisfy the
experimental bounds from flavour physics.
This is the so--called new physics flavour 
problem \cite{Grossman:2003qi}.

In this article we consider the multiple point principle (MPP)
\cite{mpp01}-\cite{mpp03} as a possible mechanism for the suppression of the
flavour changing neutral current and CP--violation effects within
the general non-supersymmetric two Higgs doublet extension of the
SM \cite{2hdm}--\cite{Sher:1988mj}. The violation of CP invariance
and the existence of tree--level flavour--changing neutral
currents are generic features of $SU(2)_W\times U(1)_Y$ theories
with two and more Higgs doublets. Potentially large FCNC
interactions appear in these models, because the diagonalization
of the quark mass matrix does not automatically lead to the
diagonalization of the two or even more Yukawa coupling matrices,
which describe the interactions of Higgs bosons with fermionic
matter. Moreover the Higgs potential of the two--Higgs doublet
model (2HDM) contains a lot of new couplings. Some of them may be
complex, resulting in CP violation in the Higgs sector
\cite{HiggsCPV01}-\cite{HiggsCPV06}. Although
one can eliminate the violation of CP invariance in the Higgs
sector and tree--level FCNC transitions by imposing a discrete
$Z_2$ symmetry, such a symmetry leads to the formation of domain
walls in the early Universe \cite{Zeldovich:1974uw} which create
unacceptably large anisotropies in the cosmic microwave background
radiation \cite{Vilenkin:1984ib}. Therefore in practice it is
necessary to impose only an approximate symmetry, typically broken
by soft mass terms.

The MPP postulates the existence of many phases
with the same energy density which are allowed by a given theory
\cite{mpp01}-\cite{mpp03}. When applied to the SM, the multiple point principle
implies that the Higgs effective potential possesses two
degenerate minima taken to be at the electroweak and Planck scales
respectively. The degeneracy of vacua at the electroweak and 
Planck scales can be achieved only if 
(see \cite{Froggatt:1995rt})
\be
M_t=173\pm 4\,\mbox{GeV}\, ,\qquad M_H=135\pm 9\, \mbox{GeV}\, .
\label{2hdm2} 
\ee
This MPP prediction for the Higgs mass lies on the SM
vacuum stability curve \cite{Sher:1988mj}, \cite{101}--\cite{208}
corresponding to the cut-off $\Lambda = M_{Pl}$
\footnote{The requirement of the validity of perturbation
theory up to the Planck scale leads to an upper bound on $M_H$ in
the SM, which is about $180-190\,\mbox{GeV}$ \cite{101}-\cite{105}.}.
The hierarchy between the electroweak and Planck scales might 
also be explained by MPP within the pure SM, if there exists 
a third degenerate vacuum \cite{mpp-sm01}-\cite{mpp-sm03}.

If we require the vacuum we live in to be just metastable
w.r.t.~decay into the second vacuum, rather than being
exactly degenerate with it,
and otherwise make similar assumptions to those in
\cite{Froggatt:1995rt}, the energy density in the second vacuum
falls below that of the vacuum in which we live. Consequently
the Higgs mass is then predicted to be a bit smaller. With the
value used in this article for the top quark mass \cite{Brubaker:2006xn},
$M_t =171.4 \pm 2.1$ GeV,
the value predicted for the Higgs mass from borderline
metastability of our vacuum, which we call
meta-MPP \cite{Yasutaka},
becomes $M_H = 118.4 \pm 5$ GeV. This is
remarkably close to the two-standard deviation hint of a
Higgs signal seen in LEP \cite{lep115} at 115 GeV.

In previous papers \cite{sugra01}-\cite{sugra03} the MPP assumption has been
adapted to models based on $(N=1)$ local supersymmetry --
supergravity, in order to provide an explanation for the small
deviation of the cosmological constant from zero. Recently we also
considered the application of the MPP to the SUSY inspired two
Higgs doublet model of type II \cite{Froggatt:2006zc}. We
established MPP conditions in this model and discussed the
restrictions on the mass of the SM--like Higgs boson caused by the
MPP. Here we are going to extend this analysis to the general
2HDM.

In the next section we specify the model. In section 3 we
present the derivation of the MPP conditions that result in
approximate custodial
$U(1)$ and $Z_2$ symmetries. The renormalisation group (RG) flow
of Yukawa couplings within these MPP inspired two Higgs doublet
models is considered in section 4. In particular, we establish the
positions of quasi--fixed points and argue that the quasi--fixed
point scenarios with large $\tan\beta$ lead to unacceptably large
values of the top quark mass. In section 5 we study the evolution
of Higgs self--couplings and analyse the spectrum of Higgs bosons
and their couplings near the quasi--fixed point at moderate
$\tan\beta$. We examine the phenomenological viability of other
possible MPP solutions in section 6. Our results are summarised in
section 7. In Appendix A the $\beta$--functions of Higgs
self--couplings in the general two Higgs doublet extension of the
SM are presented. The derivation of the other MPP conditions that do
not give rise to an approximate
custodial $U(1)$ symmetry is discussed in Appendix B.

\section{Two Higgs doublet extension of the SM}
\label{2HDM}

The most general renormalizable $SU(2)_W\times U(1)_Y$ gauge
invariant potential of the model involving two Higgs doublets is
given by \be
\begin{array}{c}
V_{eff}(H_1, H_2) = m_1^2(\Phi)H_1^{\dagger}H_1 + m_2^2(\Phi)H_2^{\dagger}H_2 -
\biggl[m_3^2(\Phi) H_1^{\dagger}H_2+h.c.\biggr]+\\[3mm]
\ds\frac{\lambda_1(\Phi)}{2}(H_1^{\dagger}H_1)^2 + \frac{\lambda_2(\Phi)}{2}(H_2^{\dagger}H_2)^2 +
\lambda_3(\Phi)(H_1^{\dagger}H_1)(H_2^{\dagger}H_2) + \lambda_4(\Phi)|H_1^{\dagger}H_2|^2\\[3mm]
\ds + \biggl[\frac{\lambda_5(\Phi)}{2}(H_1^{\dagger}H_2)^2 + \lambda_6(\Phi)(H_1^{\dagger}H_1)(H_1^{\dagger}H_2)+
\lambda_7(\Phi)(H_2^{\dagger}H_2)(H_1^{\dagger}H_2)+h.c. \biggr]
\end{array}
\label{2hdm3}
\ee
where
$$
H_n=\left(
\ba{c}
\chi^+_n\\[2mm]
(H_n^0+iA_n^0)/\sqrt{2}
\ea
\right) \qquad n=1,2\,.
$$
It is easy to see that the number of couplings in the two Higgs
doublet model potential compared with the SM grows from two to
ten. Furthermore, four of them $m_3^2$, $\lambda_5$, $\lambda_6$
and $\lambda_7$ can be complex, inducing CP--violation in the
Higgs sector. In what follows we suppose that the mass parameters
$m_i^2$ and Higgs self--couplings $\lambda_i$ of the effective
potential (\ref{2hdm3}) only depend on the overall sum of the
squared norms of the Higgs doublets, i.e.
$$
\Phi^2=\Phi_1^2+\Phi_2^2\,,\qquad \Phi_n^2=H_n^{\dagger}H_n =
\frac{1}{2}\biggl[(H_n^0)^2+(A_n^0)^2\biggr]+|\chi_n^+|^2\,.
$$
The dependence of $m_i^2$ and $\lambda_i$ on $\Phi$ is described
by the renormalization group equations, where the renormalization
scale is replaced by $\Phi$.

At the physical minimum of the scalar potential (\ref{2hdm3}) the
Higgs fields develop vacuum expectation values 
\be
<H_1^0>=v_1\,,\qquad\qquad <H_2^0>=v_2
\label{2hdm4} 
\ee 
breaking the $SU(2)_W\times U(1)_Y$ gauge
symmetry to $U(1)_{em}$ associated with electromagnetism and
generating the masses of all bosons and fermions. 
The overall Higgs norm
$<\Phi>=\sqrt{\dfrac{|v_1|^2+|v_2|^2}{2}}=\dfrac{v}{\sqrt{2}}=174\,\mbox{GeV}$ is fixed by the
Fermi scale. At the same time the ratio of the Higgs vacuum
expectation values remains arbitrary. Hence it is convenient to
introduce $\tan\beta=|v_2|/|v_1|$.

As has been already mentioned in the Introduction, the Yukawa
interactions of the Higgs fields $H_1$ and $H_2$ with quarks and
leptons generate phenomenologically unwanted FCNC transitions. In
particular these interactions contribute to the amplitude of
$K^0-\overline{K}^0$ oscillations and give rise to new channels of
muon decay like $\mu\to e^{-}e^{+}e^{-}$. The common way to
suppress flavour changing processes is to impose a certain
protecting custodial $Z_2$ symmetry that forbids potentially
dangerous couplings of the Higgs fields to quarks and leptons
\cite{Glashow:1976nt}. Such a custodial symmetry requires the
vanishing of the Higgs couplings $\lambda_6$ and $\lambda_7$. It
also requires the down-type quarks to couple to just one Higgs
doublet, $H_1$ say, while the up-type quarks couple either to the
same Higgs doublet $H_1$ (Model I) or to the second Higgs doublet
$H_2$ (Model II) but not both\footnote{Similarly the leptons are
required to only couple to one Higgs doublet, usually chosen to be
the same as the down-type quarks. However there are variations of
Models I and II, in which the leptons couple to $H_2$ rather than
to $H_1$.}. In fact, as we shall use in subsection 3.3,
it is possible to generalise the idea of such a $Z_2$ symmetry
so that each fermion couples to just one Higgs field ($H_1$ or
$H_2$) but in a generation dependent way.
The custodial $Z_2$ symmetry forbids the mixing term
$m_3^2(\Phi) (H_1^{\dagger}H_2)$ in the Higgs effective potential
(\ref{2hdm3}). But usually a soft violation of the $Z_2$ symmetry by
dimension--two terms is allowed, since it does not induce
Higgs--mediated tree--level flavor changing neutral currents
(FCNC).

The set of RG equations that determines the running of Yukawa and
Higgs couplings in the two Higgs doublet model with exact and
softly broken $Z_2$ symmetry can be found in
\cite{Hill:1985tg}--\cite{rg02}. The constraints on the Higgs masses
in the 2HDM with an unbroken $Z_2$ symmetry have been examined in
a number of publications \cite{rg01}--\cite{Nie:1998yn}. The
analysis of \cite{Nie:1998yn} was performed assuming vacuum
stability and the applicability of perturbation theory up to a
high energy scale (of order the grand unification scale),
revealing that then all Higgs boson masses lie below
$200\,\mbox{GeV}$. A very stringent restriction on the masses of
the charged and pseudoscalar states was found. They do not exceed
$150\,\mbox{GeV}$. However such a light charged Higgs boson is
ruled out by the direct searches for the rare B--meson decays
($B\to X_s\gamma$) in the Model II of the 2HDM, which cannot
therefore be valid with an unbroken $Z_2$ symmetry up to the
unification scale. The theoretical restrictions on the mass of the
SM--like Higgs boson within the 2HDM with a softly broken $Z_2$
symmetry were studied in \cite{Kanemura:1999xf}.

We emphasize that, in this article, we do not impose any custodial
symmetry but rather consider the general Higgs potential
(\ref{2hdm3}). Instead we require that at some high energy scale
($M_Z<<\Lambda\lesssim M_{Pl}$), which we shall refer to as the
MPP scale $\Lambda$, a large set of degenerate vacua allowed by
the 2HDM is realized. In compliance with the MPP, these vacua and
the physical one must have the same energy density. Thus the MPP
implies that the couplings $\lambda_i(\Lambda)$ should be adjusted
so that an appropriate cancellation among the quartic terms in the
effective potential (\ref{2hdm3}) takes place.

Here and further we impose a certain hierarchical structure on the
Yukawa couplings. To explain the observed mass hierarchy in the
quark and lepton sectors, we assume that the Yukawa couplings of
the quarks and leptons of the third generation are considerably
larger than the quark and lepton Yukawa couplings of the first two
generation. In this approximation the part of the 2HDM Lagrangian
describing the interactions of quarks and leptons with the Higgs
doublets $H_1$ and $H_2$ reduces to 
\be 
\ba{l}
\mathcal{L}_{Yuk}\simeq h_t(H_2\varepsilon
Q)\bar{t}_R+g_b(H_2^{\dag}Q)\bar{b}_R
+g_{\tau}(H_2^{\dag}L)\bar{\tau}_R+\qquad\qquad\qquad\qquad\\[2mm]
\qquad\qquad\qquad\qquad+g_t(H_1\varepsilon
Q)\bar{t}_R+h_b(H_1^{\dag}Q)\bar{b}_R
+h_{\tau}(H_1^{\dag}L)\bar{\tau}_R+h.c.\,, 
\ea 
\label{2hdm5} 
\ee
where $Q$ and $L$ are left--handed doublets of quarks and leptons
of the third generation, while $\tau_R$, $t_R$ and $b_R$ are
right--handed $SU(2)_W$ singlet components of $\tau$--lepton,
$t$-- and $b$--quarks. The running of the Yukawa couplings of the 
third generation obey the following set of differential equations:
\be
\ba{l}
\ds\frac{d g_t}{dt}=\frac{1}{16\pi^2}\biggl[g_t\biggl(\frac{9}{2}|g_t|^2+\frac{9}{2}|h_t|^2+
\frac{3}{2}|h_b|^2+\frac{1}{2}|g_b|^2+|h_{\tau}|^2
\biggl)+h_t\biggl(g_b h^{*}_b+g_{\tau}h_{\tau}^{*}\biggr)-\\[2mm]
\qquad\ds-g_t\left(8g_3^2+\frac{9}{4}g_2^2+\frac{17}{12}g_1^2\right)\biggr]\,,\\[2mm]
\ds\frac{d h_t}{dt}=\frac{1}{16\pi^2}\biggl[h_t\biggl(\frac{9}{2}|g_t|^2+\frac{9}{2}|h_t|^2+
\frac{1}{2}|h_b|^2+\frac{3}{2}|g_b|^2+|g_{\tau}|^2
\biggl)+g_t\biggl(h_b g^{*}_b+h_{\tau}g_{\tau}^{*}\biggr)-\\[2mm]
\qquad\ds-h_t\left(8g_3^2+\frac{9}{4}g_2^2+\frac{17}{12}g_1^2\right)\biggr]\,,\\[2mm]
\ds\frac{d h_b}{dt}=\frac{1}{16\pi^2}\biggl[h_b\biggl(\frac{3}{2}|g_t|^2+\frac{1}{2}|h_t|^2+
\frac{9}{2}|h_b|^2+\frac{9}{2}|g_b|^2+|h_{\tau}|^2\biggl)+g_b\biggl(h_t g^{*}_t+h_{\tau}g_{\tau}^{*}\biggr)-\\[2mm]
\qquad\ds-h_b\left(8g_3^2+\frac{9}{4}g_2^2+\frac{5}{12}g_1^2\right)\biggr]\,,\\[2mm]
\ds\frac{d g_b}{dt}=\frac{1}{16\pi^2}\biggl[g_b\biggl(\frac{1}{2}|g_t|^2+\frac{3}{2}|h_t|^2+
\frac{9}{2}|h_b|^2+\frac{9}{2}|g_b|^2+|g_{\tau}|^2\biggl)+h_b\biggl(g_t h^{*}_t+g_{\tau}h_{\tau}^{*}\biggr)-\\[2mm]
\qquad\ds-g_b\left(8g_3^2+\frac{9}{4}g_2^2+\frac{5}{12}g_1^2\right)\biggr]\,,
\ea
\label{2hdm6}
\ee
$$
\ba{l}
\ds\frac{dh_{\tau}}{dt}=\frac{1}{16\pi^2}\biggl[h_{\tau}\biggl(3|g_t|^2+3|h_b|^2+
\frac{5}{2}|h_{\tau}|^2+\frac{5}{2}|g_{\tau}|^2\biggl)+
3g_{\tau}\biggl(h_b g^{*}_b+h_t g_t^{*}\biggr)-\qquad\qquad\\[2mm]
\qquad\ds-h_{\tau}\left(\frac{9}{4}g_2^2+\frac{15}{4}g_1^2\right)\biggr]\,,\\[2mm]
\ds\frac{dg_{\tau}}{dt}=\frac{1}{16\pi^2}\biggl[g_{\tau}\biggl(3|h_t|^2+3|g_b|^2+
\frac{5}{2}|h_{\tau}|^2+\frac{5}{2}|g_{\tau}|^2\biggl)+
3h_{\tau}\biggl(g_b h^{*}_b+g_t h_t^{*}\biggr)-\\[2mm]
\qquad\ds-g_{\tau}\left(\frac{9}{4}g_2^2+\frac{15}{4}g_1^2\right)\biggr]\,,
\ea
$$
where $t=\ln\, \mu$ and $\mu$ is the renormalization scale. Also
the $g_i(\mu)$ are here the gauge couplings for the $U(1)_Y$,
$SU(2)_W$ and $SU(3)_C$ interactions.

\section{MPP conditions as an origin of the approximate 
custodial symmetries in the 2HDM}

\subsection{Philosophy of using MPP in 2HDM}
\label{philosophy}

It is our philosophy that, unless the parameters - couplings and 
masses - are restricted by some symmetry or other principle, we 
expect them to be essentially random. A priori the 2HDM has only 
the gauge symmetries and the general Poincar\'e symmetries.
Just imposing the approximate symmetry needed to rescue the 2HDM
from immediate disagreement with experimental facts, such as the
absence of FCNC, looks like an ad hoc invention to cure the model. 

In this paper we argue that such an approximate custodial symmetry
can originate from the multiple point principle (MPP).
Indeed we know that the MPP requirement of many degenerate vacua 
immediately gets fulfilled in models which possess extra 
global symmetries. One of the most famous examples of a symmetry 
that leads to a set of degenerate vacua is supersymmetry (SUSY). 
In exact SUSY models there are typically many minima (often even 
flat directions) in the scalar potential with just zero vacuum 
energy density. However, in phenomenologically acceptable models
based on softly broken supersymmetry, MPP conditions are realised 
automatically but only up to soft terms in the Lagrangian. Therefore 
here we shall similarly postulate ``hard MPP", in which we impose 
the degeneracy of vacua with only a limited accuracy set by 
the size of the soft mass terms\footnote{It should be noted here 
that in practice we also neglected the soft Higgs mass term in 
the vacuum degeneracy condition used in the application of MPP 
to the SM \cite{Froggatt:1995rt}.} in the Lagrangian. In contrast 
with an exact MPP, ``hard MPP" gives rise to approximate global
symmetries.

As a concrete realisation of such ``hard MPP", we ignore the mass 
terms in the potential (\ref{2hdm3}) and establish a relation 
between the MPP and global custodial $U(1)$ symmetries within 
the 2HDM which we shall refer to as a generalised (i.e.~generation 
dependent) Peccei-Quinn symmetry. The MPP requirement of a set of 
degenerate vacua at some high energy scale $\Lambda$ leads to 
a spontaneously broken global $U(1)$ custodial symmetry. Then we 
take into account the contribution of mass terms in the potential 
(\ref{2hdm3}) and allow vacua to be approximately degenerate at 
the MPP scale. This gives rise to a set of custodial symmetry 
violating couplings. These couplings allows to avoid any problems 
related with the presence of Nambu-Goldstone bosons in the particle 
spectrum that are usually unacceptable phenomenologically.

\subsection{Symmetry derivations from MPP in the leading approximation}
\label{treelevel}
Now, we aim to specify the largest possible set of global minima
of the 2HDM scalar potential with almost vanishing energy density,
which may exist at the MPP high energy scale $\Lambda$ where the
mass terms in the potential (\ref{2hdm3}) can be neglected. The
most general vacuum configuration takes the form:
\be
<H_1>=\Phi_1\left( \ba{c}
0\\[2mm]
1
\ea
\right)\,,\qquad
<H_2>=\Phi_2\left(
\ba{c}
\sin\theta\\[2mm]
\cos\theta\, e^{i\omega}
\ea
\right)\,,\\[2mm]
\label{2hdm7} 
\ee 
where $\Phi_1^2+\Phi_2^2=\Lambda^2$. Here, the gauge is fixed so that only 
the real part of the lower component of $H_1$ gets a vacuum expectation value.

Let us assume that the 2HDM scalar potential (\ref{2hdm3})
possesses a set of vacua in which the energy density goes to zero
for all possible values of the phase $\omega$. The degeneracy of
the vacuum configuration (\ref{2hdm7}) with respect to $\omega$
implies that $\cos\theta$, $\Phi_1$ and $\Phi_2$ gain non--zero
values at the corresponding minima. It also requires that the 2HDM
scalar potential and all its partial derivatives are independent 
of $\omega$ at the MPP scale, i.e.
\be
V_{\omega}=\frac{\lambda_5(\Phi)}{2}\Phi_1^2\Phi_2^2\cos^2\theta\,
e^{2i\omega}+
\biggl[\lambda_6(\Phi)\Phi_1^3\Phi_2+\lambda_7(\Phi)\Phi_1\Phi_2^3
\biggr]\cos\theta\, e^{i\omega}+h.c.=0 
\label{2hdm8} 
\ee 
\begin{eqnarray}
\lefteqn{
\frac{\partial V_{\omega}}{\partial \Phi_1} =
\left[ \lambda_5(\Phi) \Phi_1 \Phi_2^2  +
\beta_{\lambda_5}(\Phi) \frac{\Phi_1^3 \Phi_2^2}{2\Phi^2}
\right] \cos^2\theta\, e^{2i\omega} +}  \nonumber \\
&\left[3\lambda_6(\Phi) \Phi_1^2 \Phi_2   +
\beta_{\lambda_6}(\Phi) \ds\frac{\Phi_1^4 \Phi_2}{\Phi^2}  +
\lambda_7(\Phi) \Phi_2^3  +\beta_{\lambda_7}(\Phi)\ds\frac{\Phi_1^2 \Phi_2^3}{\Phi^2}\right]
\cos\theta\, e^{i\omega} +h.c\Biggl|_{\Phi=\Lambda}=0
\label{2hdm9}
\end{eqnarray}
\begin{eqnarray}
\lefteqn{\frac{\partial V_{\omega}}{\partial \Phi_2} =
\left[ \lambda_5(\Phi) \Phi_1^2 \Phi_2  +
\beta_{\lambda_5}(\Phi) \frac{\Phi_1^2 \Phi_2^3}{2\Phi^2}
\right]\cos^2\theta\, e^{2i\omega}  + } \nonumber \\
&\left[\lambda_6(\Phi) \Phi_1^3   +
\beta_{\lambda_6}(\Phi) \ds \frac{\Phi_1^3 \Phi_2^2}{\Phi^2}  +
3\lambda_7(\Phi) \Phi_2^2 \Phi_1  +
\beta_{\lambda_7}(\Phi) \ds \frac{\Phi_1 \Phi_2^4}{\Phi^2}
\right]\cos\theta \,e^{i\omega} +h.c.\Biggl|_{\Phi=\Lambda}=0\,.
\label{2hdm10}
\end{eqnarray}
Here $\beta_{\lambda_i}(\Phi) =\ds \frac{d \lambda_i(\Phi)}{d \ln
\Phi}$ is the renormalisation group beta function for the Higgs
self-coupling $\lambda_i(\Phi)$. It is readily verified that the
vanishing of the coefficients of $e^{i\omega}$ and $e^{2i\omega}$
in Eqs.~(\ref{2hdm8}) - (\ref{2hdm10}) leads to the conditions:
\be 
\lambda_5(\Lambda)=\lambda_6(\Lambda)=\lambda_7(\Lambda)=0,\qquad
\beta_{\lambda_5}(\Lambda) =
\beta_{\lambda_6}\Phi_1^2 + \beta_{\lambda_7}\Phi_2^2 = 0.
\label{2hdm11} 
\ee 

Taking into account the derived MPP conditions (\ref{2hdm11}) and
substituting the vacuum configuration (\ref{2hdm7}) into the
quartic part of the 2HDM potential, one finds for any
$\Phi\simeq \Lambda$: 
\be 
\ba{rcl}
V(H_1,H_2)&\approx&\ds\frac{1}{2}\biggl(\sqrt{\lambda_1(\Phi)}\Phi_1^2-
\sqrt{\lambda_2(\Phi)}\Phi_2^2\biggr)^2+\\[3mm]
&&+\left(\sqrt{\lambda_1(\Phi)\lambda_2(\Phi)}+\lambda_3(\Phi)+
\lambda_4(\Phi)\cos^2\theta\right) \Phi_1^2\Phi_2^2\, . 
\ea
\label{2hdm13} 
\ee 
The Higgs scalar potential (\ref{2hdm13})
attains its minimal value for $\cos\theta=0$ if
$\lambda_4(\Lambda)>0$ or $\cos\theta=\pm 1$ when
$\lambda_4(\Lambda)<0$. Since the degeneracy of the vacuum
configuration (\ref{2hdm7}) with respect to $\omega$ may be
realised only if $\cos\theta$ has a non--zero value, the
self--consistent implementation of the MPP requires
$\lambda_4(\Lambda)$ to be negative. Then around the minimum 
the scalar potential can be written as 
\be
V(H_1,H_2)\approx\ds\frac{1}{2}\biggl(\sqrt{\lambda_1(\Phi)}\Phi_1^2-
\sqrt{\lambda_2(\Phi)}\Phi_2^2\biggr)^2+\tilde{\lambda}(\Phi)\Phi_1^2\Phi_2^2\,,
\label{2hdm14} 
\ee 
where $\tilde{\lambda}=\sqrt{\lambda_1\lambda_2}+\lambda_3+\lambda_4\,.$
If near the MPP scale the combination of the Higgs self--couplings
$\tilde{\lambda}(\Phi)$ is less than zero, then there exists a
minimum with huge and negative energy density that hampers the MPP
implementation. Otherwise when $\tilde{\lambda}(\Phi)>0$ the Higgs
potential (\ref{2hdm14}) is always positive definite, which spoils
the consistent implementation of the MPP as well. Thus, in order
to get a set of degenerate vacua in which the energy density tends
to zero for all possible values of the phase $\omega$ at the MPP
scale, one has to assume that $\tilde{\lambda}(\Lambda)=0$\,.
Then $V(H_1,H_2)$ reaches a minimal value at 
\be
\Phi_1=\Lambda\cos\gamma\,,\qquad
\Phi_2=\Lambda\sin\gamma\,,\qquad
\tan\gamma=\Biggl(\ds\frac{\lambda_1}{\lambda_2}\Biggr)^{1/4}\,.
\label{2hdm16} 
\ee

Next we should also require the vanishing of partial derivatives
of the scalar potential (\ref{2hdm14}) with respect to $\Phi_1$
and $\Phi_2$ \footnote{The partial derivative $\partial
V/\partial\theta$ goes to zero when $\cos\theta\to \pm 1$.}. This
results in another MPP condition: 
\be
\beta_{\tilde{\lambda}}(\Lambda)=\frac{1}{2}\beta_{\lambda_1}(\Lambda)
\sqrt{\frac{\lambda_2(\Lambda)}{\lambda_1(\Lambda)}}+
\frac{1}{2}\beta_{\lambda_2}(\Lambda)\sqrt{\frac{\lambda_1(\Lambda)}{\lambda_2(\Lambda)}}
+\beta_{\lambda_3}(\Lambda)+\beta_{\lambda_4}(\Lambda)=0\,.
\label{2hdm17} 
\ee 
The MPP conditions mentioned above give rise to the following set 
of MPP scale vacua
\be <H_1>=\left(
\begin{array}{c}
0\\ \Phi_1
\end{array}
\right)\,,
\qquad <H_2>=\left(
\begin{array}{c}
0\\ \Phi_2\, e^{i\omega}
\end{array}
\right)\,, 
\label{2hdm18} 
\ee 
which have zero energy density for any $\omega$. The Higgs field norms $\Phi_1$ 
and $\Phi_2$ in Eq.~(\ref{2hdm18}) are defined by the equations for the extrema
of the 2HDM potential whose solution is given by Eq.~(\ref{2hdm16}). 

Combining Eqs.~(\ref{2hdm11}) and (\ref{2hdm16})
and using the explicit form of the $\beta$--functions for the
Higgs self--couplings given in Apendix A, one obtains two
conditions that quark and lepton Yukawa couplings should obey if
MPP is realised in Nature: 
\be
3 h^2_b(\Lambda) g^{*2}_b(\Lambda)+h^2_{\tau}(\Lambda) g^{*2}_{\tau}(\Lambda)=0\,,\\
\label{2hdm19}
\ee
\begin{eqnarray}
\lefteqn{3 h_b(\Lambda) g^{*}_b(\Lambda) \biggl[\sqrt{\lambda_2(\Lambda)}|h_b(\Lambda)|^2
+\sqrt{\lambda_1(\Lambda)}|g_b(\Lambda)|^2\biggr]+} \nonumber \\
&& + h_{\tau}(\Lambda) g^{*}_{\tau}(\Lambda) \biggl[\sqrt{\lambda_2(\Lambda)}|h_{\tau}(\Lambda)|^2+
\sqrt{\lambda_1(\Lambda)}|g_{\tau}(\Lambda)|^2\biggr]=0\,.
\label{2hdm20}
\end{eqnarray}
To simplify calculations we use here the basis in the field space
where only one Higgs doublet $H_2$ interacts with the top--quark at
the scale $\Lambda$, i.e. $g_t(\Lambda)=0$. Conditions
(\ref{2hdm19})--(\ref{2hdm20}) are fulfilled simultaneously only
if 
\be 
\ba{clcl}
(I)\,& h_b(\Lambda)=h_{\tau}(\Lambda)=0\,;&\qquad\qquad (II)\,& g_b(\Lambda)=g_{\tau}(\Lambda)=0\,;\\
(III)\,& h_b(\Lambda)=g_{\tau}(\Lambda)=0\,;&\qquad\qquad (IV)\,& g_b(\Lambda)=h_{\tau}(\Lambda)=0\,.
\ea
\label{2hdm21}
\ee
The solutions $(I)-(IV)$ correspond to the 2HDM Model I and Model II Yukawa couplings
and their leptonic variations. In these models the MPP conditions reduce to
\be
\left\{
\ba{l}
\lambda_5(\Lambda)=\lambda_6(\Lambda)=\lambda_7(\Lambda)=0\,,\\[2mm]
\tilde{\lambda}(\Lambda)=\beta_{\tilde{\lambda}}(\Lambda)=0\,. 
\ea
\right. 
\label{2hdm22} 
\ee 
The MPP conditions were formulated in exactly this form in \cite{Froggatt:2006zc}, 
where the multiple point principle was applied to the Model II of the two Higgs
doublet extension of the SM. It is worth noting that the relations
corresponding to Eq.(\ref{2hdm22}) are satisfied identically in
the minimal SUSY model (MSSM) at any scale lying higher than the
masses of the superparticles.

Usually the existence of a large set of degenerate vacua is
associated with an enlarged global symmetry of the Lagrangian of
the considered model. The 2HDM is not an exception. In all the
models $(I-IV)$, the quartic part of the Higgs effective potential
(\ref{2hdm3}) and the Lagrangian describing the interactions of
quarks and leptons with the Higgs fields (\ref{2hdm5}) are
invariant under $Z_2$ symmetry transformations at the MPP scale,
which prevent the appearance of flavour changing neutral currents
at the tree level. Moreover when $m_3^2$, $\lambda_5$, $\lambda_6$
and $\lambda_7$ vanish, the full Lagrangian of the 2HDM is
invariant under the transformations of an $SU(2)\times [U(1)]^2$
global symmetry. The mixing term $m_3^2(H_1^{\dagger}H_2)$ in the 
Higgs effective potential (\ref{2hdm3}), which we have neglected at 
the MPP scale, softly breaks the $Z_2$ and extra $U(1)$ (Peccei--Quinn) 
symmetries but it does not create new sources of CP--violation or 
FCNC transitions. Indeed, the renormalization group flow preserves the 
invariance of the quartic part of the Higgs effective potential 
(\ref{2hdm3}), as well as the invariance of the Lagrangian of the 
interactions of fermions with the Higgs fields, under the transformations 
of the $Z_2$ and Peccei--Quinn symmetries. This means that, if 
the Peccei--Quinn symmetry violating Yukawa or Higgs couplings are set 
to zero (or small) at some scale $\Lambda$, they will remain zero (or 
small) at any scale below $\Lambda$. In the Higgs sector of the 
general 2HDM, only the imaginary parts of $m_3^2$, $\lambda_5$,
$\lambda_6$ and $\lambda_7$ cause CP--non-conservation. Since MPP
suppresses the Higgs self--couplings which are responsible for the
violation of the CP--invariance and the complex phase of $m_3^2$
can be easily absorbed by the appropriate redefinition of the
Higgs fields, MPP protects the CP--invariance within the two Higgs
doublet extension of the SM. The tree--level FCNC transitions also
do not emerge after the soft breakdown of the $Z_2$ and
Peccei--Quinn symmetries, simply because the structure of the
interactions of the quarks and leptons with the Higgs doublets
remains intact.

Of course, one can argue that we only derive the custodial $Z_2$
symmetry for the interactions of the quarks and leptons of the
third generation, while the most stringent restrictions on the
Peccei--Quinn symmetry violating Yukawa couplings come from the
FCNC processes involving quarks and leptons of the first two
generations. Indeed, the MPP conditions for the Yukawa couplings
of the third generation obtained above (see Eq.(\ref{2hdm21}))
cannot be generalised to the three generation case in a
straightforward way at leading order.

\subsection{MPP symmetry derivation to one loop accuracy}
\label{oneloop}
In this subsection we shall discuss symmetry derivation from MPP
to one loop accuracy and shall indeed achieve the derivation of
a Peccei-Quinn-like symmetry even for the lower mass generations.
In the case when three generations of quarks and leptons have
non--negligible couplings to the Higgs doublets, all the SM bosons
and fermions contribute to the Higgs effective potential. It is
then convenient to present the potential in the following form
\be
V_{eff}(H_1, H_2)=\sum_{n=0}^{\infty}V_n(H_1, H_2),
\label{2hdm221}
\ee
where $V_0$ corresponds to the tree level
Higgs boson potential, while $V_n$ represents the $n$--loop
contribution to $V_{eff}$. In the one--loop approximation we have
\be
V_1=\ds\frac{1}{64\pi^2}Str\,|M|^4\biggl[\log\frac{|M|^2}{\mu^2}-C\biggr],
\label{2hdm222}
\ee
where $M$ is the mass matrix for the bosons and fermions in the model.
Here the supertrace operator counts positively
(negatively) the number of degrees of freedom for the different
bosonic (fermionic) fields, $C$ is a diagonal matrix which depends
on the renormalisation scheme, while $\mu$ is the renormalisation
scale. Previously we restricted our consideration to the leading
log approximation, i.e. we replaced $\log\ds\frac{|M|^2}{\mu^2}$
by $\log\ds\frac{\Phi^2}{\mu^2}$ in Eq.~(\ref{2hdm222}) and summed
all the leading logs using the renormalisation group equations. A
more accurate analysis, which we shall perform in the next
paragraph, requires us to take into account all terms
in Eq.~(\ref{2hdm222}).

The independence of the Higgs effective potential on $\omega$ at
the MPP scale implies that any order partial derivatives of
$V_{eff}$ with respect to $\omega$ vanish at this scale. From
Eq.~(\ref{2hdm222}) it becomes clear that this can be achieved
only when the masses of all the particles are independent of
$\omega$ near the MPP scale vacua\footnote{This is expected
intuitively and can be proved formally by considering $V_1$ as an
analytic function of $z=e^{i\omega}$, which is required to be
constant for $|z|=1$}. Near the vacuum configuration parametrized
by Eqs.~(\ref{2hdm16}) and (\ref{2hdm18}), the mass terms of the
quarks and leptons take the form
\be
\mathcal{L}_{mass}=\sum_{f=u,d,l}\biggl(\bar{f}_R\,\bar{f}_L\biggr)\left(
\begin{array}{cc}
0             & M_f\\
M^{\dagger}_f & 0
\end{array}
\right)
\left(
\begin{array}{c}
f_R \\
f_L
\end{array}
\right),
\label{2hdm223}
\ee
\be
\begin{array}{c}
M_u=H_u \Phi_2 e^{i\omega}+G_u \Phi_1,\qquad
M_d=G_d \Phi_2 e^{i\omega}+H_d \Phi_1,\\[2mm]
M_l=G_l \Phi_2 e^{i\omega}+H_l \Phi_1.
\end{array}
\label{2hdm224}
\ee
Here $H_f$ and $G_f$ are $3\times 3$ matrices
that replace $h_f$ and $g_f$ in the Lagrangian (\ref{2hdm5}).
Instead of the mass matrices of quarks and leptons
($\mathcal{M}_f$) one can consider
\be
\mathcal{M}_{f}\mathcal{M}^{\dagger}_{f}= \left(
\begin{array}{cc}
M_f M^{\dagger}_f &  0\\
      0           & M^{\dagger}_f M_f
\end{array}
\right), \label{2hdm225}
\ee
whose eigenvalues are positive
definite and equal to the absolute values of the fermion masses
squared. The eigenvalues of
$\mathcal{M}_{f}\mathcal{M}^{\dagger}_{f}$ will not depend on
$\omega$ only if \be G^{\dagger}_f H_f=H^{\dagger}_f G_f=G_f
H^{\dagger}_f=H_f G^{\dagger}_f=0. \label{2hdm226} \ee The
conditions (\ref{2hdm226}) can be satisfied only when either
$\mbox{det}\, G_f=0$ or $\mbox{det}\, H_f=0$ or both determinants
vanish. By means of unitary transformations of the right--handed
and left--handed states, one can easily diagonalise the Yukawa
matrices $H_f$. Then it can be readily shown that the solutions of
Eq.~(\ref{2hdm226}) can be written as follows:
\be H_f=\left(
\begin{array}{ccc}
h_{f_1} & 0       & 0\\
0       & h_{f_2} & 0\\
0       & 0       & h_{f_3}
\end{array}
\right),\qquad
G_f=\left(
\begin{array}{ccc}
g_{f_1} & 0       & 0\\
0       & g_{f_2} & 0\\
0       & 0       & g_{f_3}
\end{array}
\right),\qquad h_{f_i}\cdot g_{f_i}=0\,.
\label{2hdm227}
\ee

The solutions obtained above guarantee the suppression of FCNC
processes at the MPP scale vacua. Since according to the solutions
(\ref{2hdm227}) either $h_{f_i}$ or $g_{f_i}$ equals zero, each
fermion eigenstate couples to only one Higgs doublet (either $H_1$
or $H_2$) so that the conditions of the Glashow-Weinberg theorem
\cite{Glashow:1976nt} are satisfied and non--diagonal flavour
transitions are forbidden at the tree level. Moreover the MPP
solution for the Yukawa couplings derived above implies that the
Lagrangian for the Higgs--fermion interactions is invariant under
the symmetry transformations:
\be
\ba{c} 
H_1\to e^{i\alpha}\,H_1,
\qquad\quad u'_{R_i}\to e^{i\alpha}\,u'_{R_i}, \qquad\quad
d'_{R_i}\to e^{-i\alpha}\,d'_{R_i},\qquad\quad e'_{R_i}\to e^{-i\alpha}\,e'_{R_i},\\
H_2\to e^{i\beta}\,H_2, \qquad\quad u''_{R_i}\to
e^{i\beta}\,u''_{R_i}, \qquad\quad d''_{R_i}\to
e^{-i\beta}\,d''_{R_i},\qquad\quad e''_{R_i}\to
e^{-i\beta}\,e''_{R_i}. 
\ea 
\label{2hdm229}
\ee
Here $u'_{R_i}$,
$d'_{R_i}$, $e'_{R_i}$ are right--handed quarks and leptons which
couple to $H_1$ while $u''_{R_i}$, $d''_{R_i}$, $e''_{R_i}$ are
right--handed fermions that interact with $H_2$. These two global
$U(1)$ symmetries (\ref{2hdm229}) are responsible for the
suppression of FCNC effects in the considered MPP scenario\footnote{
In principle we should have included a kinetic mixing term
$\varkappa \biggl[(D_{\mu} H_1)^{\dagger} (D_{\mu} H_2)+h.c.\biggr]$
in the Lagrangian. However one can show that the existence of a set 
of degenerate vacua with respect to $\omega$ implies that 
$\varkappa=0$ at the scale $\Lambda$. Nevertheless small custodial
symmetry violating couplings would induce a small non--zero 
value of $\varkappa$}.
Really one linear combination of these two symmetries -- namely the
one for which $\alpha = \beta$ -- is just a symmetry inherited from
the Standard Model. It just corresponds to a combination of the
well-known accidental symmetries of baryon number and lepton number
conservation together with the weak hypercharge gauge symmetry. The
other $U(1)$ symmetry, corresponding to $\alpha = - \beta$, is a
generalisation of the Peccei-Quinn chiral $U(1)$ symmetry. It is, of
course, only the latter that is truly derived from MPP. It should
be remarked that the $Z_2$ subgroup of this generalised Peccei-Quinn
symmetry, obtained by setting $\alpha = \pi$ and $\beta = 0$, acts as
a custodial symmetry to prevent FCNC.

The renormalisation group flow of Yukawa couplings does not spoil the
invariance of the Lagrangian describing the interactions of quarks
and leptons with the Higgs bosons under the custodial symmetry
transformations (\ref{2hdm229}). As a consequence the same
symmetries forbid non--diagonal flavour transitions near the
electroweak scale vacuum at the tree level. Thus MPP provides a
reliable mechanism for the suppression of FCNC processes. Some of
the MPP solutions given by Eq.~(\ref{2hdm227}) are very well
known. For example, when either $G_f$ or $H_f$ vanishes the
suppression of non--diagonal flavour processes is caused by the
usual Peccei--Quinn symmetry. In this sense the MPP solutions derived
above may be considered as generalisations of the well-known
Peccei--Quinn symmetric solution of the FCNC problem.

There is an important feature that may allow us to distinguish the
softly broken Peccei--Quinn symmetric solution of the FCNC problem from
the MPP inspired two Higgs doublet models.
The point is that, in the MPP inspired two Higgs doublet extension
of the SM, the Peccei--Quinn--like symmetry violating Yukawa couplings
can have non--zero values. Actually, one may notice that we did
not require exact degeneracy of vacua at the electroweak and MPP
scale. Since we ignore all mass terms in the 2HDM potential
(\ref{2hdm3}) during the derivation of the MPP conditions, the
energy density of the vacua at the scale $\Lambda$ is expected to
be of the order of $v^2\Lambda^2$ while the total vacuum energy
density at the physical vacuum is set by $v^4$. Thus MPP here
postulates the degeneracy of all vacua with the accuracy
$v^2\Lambda^2$. It means that the Higgs self--couplings
$\lambda_{5,\,6,\,7}(\Lambda)$ which break the Peccei--Quinn--like
symmetry may take on small but non--zero values, i.e.
\be
\left|\lambda_5(\Lambda)\right|,\,
\left|\lambda_6(\Lambda)\right|,\, \left|\lambda_7(\Lambda)\right|
\lesssim \dfrac{v^2}{\Lambda^2}. \label{2hdm228}
\ee
Because of
this the $\beta$--functions $\beta_{\lambda_5}(\Lambda)$ and
$\beta_{\lambda_6}(\Lambda)\Phi_1^2
+\beta_{\lambda_7}(\Lambda)\Phi_2^2$ appearing in
Eqs.~(\ref{2hdm9})--(\ref{2hdm10}) do not vanish exactly as well.
This permits us to establish constraints on the values of the
Peccei--Quinn--like symmetry violating Yukawa couplings
\be
\ba{c}
\beta_{\lambda_5}(\Lambda) \simeq  \ds\frac{h^2(\Lambda) g^2(\Lambda)}{(4\pi)^2}\lesssim v^2/\Lambda^2\,,\\[3mm]
\beta_{\lambda_6}(\Lambda)\cos^2\gamma +
\beta_{\lambda_7}(\Lambda)\sin^2\gamma \simeq
\ds\frac{h^3(\Lambda) g(\Lambda)}{(4\pi)^2}\lesssim
v^2/\Lambda^2\,, \label{2hdm23} \ea
\ee
where $h$ should be
associated with the Yukawa couplings which preserve the Peccei--Quinn--like
symmetry, while $g$ is a typical value of the Peccei--Quinn--like
symmetry violating Yukawa couplings. Here it is worth emphasizing
that any Yukawa coupling which breaks the Peccei--Quinn--like symmetry will
contribute to the left--hand side of Eqs.~(\ref{2hdm23}).
Therefore the inequalities (\ref{2hdm23}) constrain all
Peccei--Quinn--like symmetry violating Yukawa couplings, including the
ones which induce FCNC transitions of quarks and leptons of the
first two generations.

If $\Lambda$ is quite close to the Planck scale then all couplings
that break Peccei--Quinn--like or custodial $Z_2$ symmetries are really
tiny: $\lambda_{5,\,6,\,7}\lesssim 10^{-34}$ and $g\lesssim
10^{-32}/h^3$. When $h$ varies from $1$ to $10^{-5}$, the limits
on the Peccei--Quinn--like symmetry violating Yukawa couplings change
from $10^{-32}$ to $10^{-17}$. For such tiny values of
$\lambda_{5,\,6,\,7}$ and $g$, all CP--violation and FCNC effects
are extremely strongly suppressed and do not lead to any phenomena
which could be observed in the near future.

However, in order to get suitable suppression of the FCNC
transitions observed experimentally, the MPP scale does not
necessarily have to be as large as the Planck scale. For instance,
at very large values of $\tan\beta$ in model II of the 2HDM, when
$h_b\sim h_{\tau}\sim 1$, the appropriate suppression of the
Peccei--Quinn--like symmetry violating Yukawa couplings in $K$ meson
physics may be obtained even for $\Lambda\simeq
100-1000\,\mbox{TeV}$, if we assume that all the Peccei--Quinn--like
symmetry violating couplings are of the same order of magnitude or
that their pattern exhibits a hierarchical structure similar to
the one suggested by Cheng and Sher \cite{Cheng:1987rs}.

If the $b$--quark and $\tau$ lepton Yukawa couplings are as small as 
in the SM, i.e. $h\sim 10^{-2}$, then to ensure the absence of 
large FCNC transitions the MPP scale should be pushed up to
$10^5-10^6\,\mbox{TeV}$. The contribution of non--renormalisable terms,
arising from new physics at such high scales $\Lambda$, to FCNC
processes is also negligibly small. In this case
$|\lambda_{5,\,6,\,7}|\lesssim 10^{-12}-10^{-14}$.

\section{The running of the Yukawa couplings and the quasi--fixed point solutions}

Now we consider the RG flow of the Yukawa couplings within the MPP
inspired 2HDM with approximate generalised Peccei--Quinn and $Z_2$ symmetries.
When the MPP scale is relatively high the Peccei--Quinn--like symmetry
violating Yukawa and Higgs couplings are extremely small, which
allows us to suppress non--diagonal flavour transitions and
CP--violating interactions. Meanwhile, if the interval between the
MPP and electroweak scales is large enough, the solutions of the
RG equations are concentrated in the vicinity of the quasi--fixed points. 
The quasi--fixed point scenario in the 2HDM was analysed in
\cite{Hill:1985tg},\cite{Froggatt:1990wa01}--\cite{Froggatt:2007qp03}.

\subsection{Quasi--fixed point scenario at moderate $\tan\beta$ }

Because the Yukawa couplings of quarks and leptons of the first
two generation, as well as the Peccei--Quinn--like symmetry violating
Yukawa and Higgs couplings, are negligibly small they are
irrelevant for our analysis of RG flow. So we return back to the
four MPP solutions (\ref{2hdm21}) derived in the previous section.
Moreover at moderate values of $\tan\beta$ ($\tan\beta\lesssim
10$), the Yukawa couplings of the $b$--quark and $\tau$--lepton
are also very small and can be safely ignored. As a result the
renormalisation group equations (\ref{2hdm6}) are simplified
drastically and an exact analytic solution for the top quark
Yukawa coupling may be obtained. It can be written as follows:
\begin{equation}
\begin{gathered}
Y_t(\mu)=\frac{\dfrac{2 E(l)}{9 F(l)}}{1+\dfrac{2}{9 Y_t(\Lambda) F(l)}},\qquad
\tilde{\alpha}_i(\mu)=\frac{\tilde{\alpha}_i(\Lambda)}{1+b_i\tilde{\alpha}_i(\Lambda)\,l},\\
E(l)=\left[\frac{\tilde{\alpha}_3(\mu)}{\tilde{\alpha}_3(\Lambda)}\right]^{8/7}
\left[\frac{\tilde{\alpha}_2(\mu)}{\tilde{\alpha}_2(\Lambda)}\right]^{3/4}
\left[\frac{\tilde{\alpha}_1(\mu)}{\tilde{\alpha}_1(\Lambda)}\right]^{-17/84}
,\quad F(l)=\int\limits_0^l E(l')dl',
\end{gathered}
\label{2hdm31}
\end{equation}
where the index $i$ varies from $1$ to $3$, $b_1=7$, $b_2=-3$,
$b_3=-7$, $l=\ln(\Lambda^2/\mu^2)$,
$\tilde{\alpha}_i(\mu)=\left(\dfrac{g_i(\mu)}{4\pi}\right)^2$,
$Y_t(\mu)=\left(\dfrac{h_t(\mu)}{4\pi}\right)^2$. If the MPP scale
is very high and $h_t^2(\Lambda)\gtrsim 1$, the second term in the
denominator of the expression describing the evolution of
$Y_t(\mu)$ is much smaller than unity at the electroweak scale.
For example, when $\Lambda$ is of the order of the Planck scale
and $l=l_0=\ln(\Lambda^2/M_t^2)$, the term $\dfrac{2}{9
Y_t(\Lambda) F(l_0)}$ is approximately equal to $\dfrac{1}{7
h_t^2(\Lambda)}$. Due to the small numerical coefficient in front
of $1/h_t^2(\Lambda)$, the dependence of $h_t^2(\mu)$ on its
initial value $h_t^2(\Lambda)$ disappears when
$h^2_t(\Lambda)\gtrsim 1$ and all solutions of the RG equation for
the top quark Yukawa coupling are concentrated in a narrow
interval near the quasi--fixed point \cite{Hill:1985tg},
\cite{Hill:1980sq}:
\begin{equation}
Y_\text{QFP}(M_t)=\dfrac{2\, E(l_0)}{9\,F(l_0)}\,.
\label{2hdm32}
\end{equation}
Formally a solution of this type can be obtained in the limit when
$Y_t(\Lambda)$ is infinitely large. But in reality the convergence
of RG solutions to the quasi--fixed point (\ref{2hdm32}) does not
require extremely large values of the top quark Yukawa coupling at
the MPP scale, if $\Lambda$ is high enough.

In Fig.~1a we examine the deviations of the solutions of the RG
equations from the quasi--fixed point (\ref{2hdm32}) at the
electroweak scale as a function of the MPP scale. The dash--dotted, 
solid and dashed curves represent the quasifixed point solution 
($h_t(\Lambda)>>1$) and the solutions to the RG equations that 
correspond to $h^2_t(\Lambda)=10$ and $h^2_t(\Lambda)=2.25$. 
The dash--dotted, solid and dashed lines are rather close to each 
other at large values of $\Lambda$ ($\Lambda\gtrsim 10^{13}\,\mbox{GeV}$). 
This demonstrates that the solutions of the RG equation for $h_t(\mu)$ 
are attracted towards the quasi--fixed point relatively strongly. 
At low values of the MPP scale, $\Lambda\simeq 10^4-10^7\,\mbox{GeV}$, 
the convergence of $h_t(\mu)$ to the quasi--fixed point is quite weak, 
so that it is rather difficult to get a reasonable prediction for the 
top quark Yukawa coupling at the electroweak scale. Generally the
quasi--fixed point solution provides an upper bound on $h_t(M_t)$.

The convergence of the RG solutions to the quasi--fixed point
allows us to predict the value of the top quark Yukawa coupling at
the electroweak scale for each fixed value of the MPP scale. Then,
using the relation between the running mass and Yukawa coupling of
the $t$--quark
\begin{equation}
m_t(M_t)=\dfrac{h_t(M_t)}{\sqrt{2}}v\sin\beta, 
\label{2hdm34}
\end{equation}
one can find the value of $\tan\beta$ that corresponds to the
quasi--fixed point (\ref{2hdm32}). Here we use the world average
mass of the top quark $M_t=171.4\pm 2.1$ GeV (see
\cite{Brubaker:2006xn}) and the relationship between the
$t$--quark pole ($M_t$) and running ($m_t(\mu)$) masses
\cite{mtMS01}-\cite{mtMS05},
\begin{equation}
m_t(M_t)=M_t\biggl[1-1.333\ds\,\frac{\alpha_s(M_t)}{\pi}-
9.125\left(\ds\frac{\alpha_s(M_t)}{\pi}\right)^2\biggr]\,.
\label{2hdm35}
\end{equation}
We find that, in the two--loop
approximation, $m_t(M_t)\simeq 161.6\pm 2\,\mbox{GeV}$.

The results of our calculations are summarised in Fig.~1b, where
we set $m_t(M_t)\simeq 161.6\,\mbox{GeV}$. In Fig.~1b we plot the 
values of $\tan\beta$ that correspond to $h^2_t(\Lambda)>>1$, 
$h^2_t(\Lambda)=10$ and $h^2_t(\Lambda)=2.25$ (dash--dotted, solid 
and dashed lines respectively) as a function of the scale $\Lambda$. 
From this figure it becomes clear that, at large values of 
$h_t(\Lambda)\gtrsim 1.5$, the RG solutions for the top quark Yukawa 
coupling are gathered in the vicinity of $\tan\beta=1$ at the 
electroweak scale. The dash--dotted curve in Fig.~1(b) corresponds 
to the maximal possible value of $h_t(M_t)$ and, as a consequence, 
represents the lower bound on $\tan\beta$.

\subsection{Quasi--fixed point solutions at large $\tan\beta$ }

When the values of $\tan\beta$ are large the solutions of the RG
equations are also focused near quasi--fixed points, if the
appropriate Yukawa couplings at the MPP scale are relatively
large. But in this case the position of the quasi--fixed point is
model dependent. For example, in the model $(I)$ (see
Eq.(\ref{2hdm21})) $g_b(\Lambda)$ and $g_{\tau}(\Lambda)$ cannot
be large, because this results in large masses for the $b$--quark
and $\tau$--lepton $\sim 100\,\mbox{GeV}$. Therefore, in model
$(I)$ of the MPP inspired 2HDM, there is only one
phenomenologically acceptable quasi--fixed point solution which is
given by Eq.~(\ref{2hdm32}).

In the case of the MPP solution $(IV)$ both $h_t(\Lambda)$ and
$h_{b}(\Lambda)$ are allowed to be large, because the masses of
the top and bottom quarks are generated by two different Higgs
doublets whose vacuum expectation values can be very different and
be used to induce a large hierarchy between $m_t(M_t)$ and
$m_b(M_t)$. To ensure that $m_t(M_t)>>m_{\tau}(M_t)$ in the
considered model, the $\tau$--lepton Yukawa coupling has to be
always much smaller than the top quark one. Therefore we can
neglect $g_{\tau}$ in our analysis of the RG flow. Then the two
remaining RG equations, which describe the evolution of $h_t(\mu)$
and $h_b(\mu)$, are invariant under the interchange
$h_t(\mu)\leftrightarrow h_b(\mu)$, if we ignore the $U(1)_Y$
gauge coupling which does not much affect the running of the
Yukawa couplings. In the limit when $g_1\to 0$ and
$Y_t(\Lambda)=Y_b(\Lambda)=Y_0$, the solutions of the RG equations
take the form
\begin{equation}
\begin{gathered}
Y_t(\mu)\simeq Y_b(\mu)\simeq \frac{\dfrac{E_1(l)}{5 F_1(l)}}{1+\dfrac{1}{5 Y_0 F_1(l)}},\\
E_1(l)=\left[\frac{\tilde{\alpha}_3(\mu)}{\tilde{\alpha}_3(\Lambda)}\right]^{8/7}
\left[\frac{\tilde{\alpha}_2(\mu)}{\tilde{\alpha}_2(\Lambda)}\right]^{3/4},\qquad\qquad
F_1(l)=\int\limits_0^l E_1(l')dl',
\end{gathered}
\label{2hdm36}
\end{equation}
where $Y_b(\mu)=\left(\dfrac{h_b(\mu)}{4\pi}\right)^2$.
Eq.~(\ref{2hdm36}) demonstrates that, at large values of $Y_0$,
the solutions of the RG equations for $Y_t(\mu)$ and $Y_b(\mu)$
approach the quasi--fixed point at the electroweak scale:
\begin{equation}
Y_t(M_t)\simeq Y_b(M_t)\simeq \dfrac{E_1(l_0)}{5\, F_1(l_0)}\,.
\label{2hdm37}
\end{equation}

Substituting the obtained prediction for the $b$--quark Yukawa
coupling into the equation
\begin{equation}
m_b(M_t)=\dfrac{h_b(M_t)}{\sqrt{2}}v\cos\beta, \label{2hdm38}
\end{equation}
which relates the running $b$--quark mass $m_b(M_t)$ with its
Yukawa coupling at the electroweak scale, one can determine the
value of $\tan\beta$ that corresponds to the quasi--fixed point
solution (\ref{2hdm37}). Because $h_t(M_t)\simeq h_b(M_t)$, one
finds that $\tan\beta\simeq m_t(M_t)/m_b(M_t)=55-60$. For such
large values of $\tan\beta$ the running mass of the $t$--quark
does not depend on $\tan\beta$, i.e. $m_t(M_t)\approx
\ds\frac{h_t(M_t)}{\sqrt{2}}v$, and can also be predicted. The
results of our numerical computations are presented in Table 1.
From Table 1 it is obvious that the quasi--fixed point solution
(\ref{2hdm37}) for the MPP solution (IV) results in an
unacceptably large value for the running top quark mass.

Another quasi--fixed point arises in the framework of the MPP
solution $(III)$ (see Eq.~(\ref{2hdm21})). In this case the
non--zero $b$--quark Yukawa coupling $g_b$ has to be negligibly
small to guarantee that $m_b(M_t)<< m_t(M_t)$. Nevertheless the
$\tau$--lepton Yukawa coupling may be comparable with the top
quark one at the MPP scale, since the masses of the $t$--quark and
$\tau$--lepton are induced by different Higgs doublets. In the
limit when the $b$--quark Yukawa coupling goes to zero, the RG
flow of $h_t(\mu)$ and $h_{\tau}(\mu)$ is described by two
independent first order differential equations which can be solved
analytically. The corresponding analytic solution for the top
quark Yukawa coupling is given by Eq.~(\ref{2hdm31}), while the
solution for the $\tau$--lepton one can be written in the
following form:
\begin{equation}
\begin{gathered}
Y_{\tau}(\mu)=\frac{\dfrac{2 E_2(l)}{5 F_2(l)}}{1+\dfrac{2}{5 Y_{\tau}(\Lambda) F_2(l)}},\\
E_2(l)=\left[\frac{\tilde{\alpha}_2(\mu)}{\tilde{\alpha}_2(\Lambda)}\right]^{3/4}
\left[\frac{\tilde{\alpha}_1(\mu)}{\tilde{\alpha}_1(\Lambda)}\right]^{-15/28},\qquad\qquad
F_2(l)=\int\limits_0^l E_2(l')dl',
\end{gathered}
\label{2hdm39}
\end{equation}
where $Y_{\tau}(\mu)=\left(\dfrac{h_{\tau}(\mu)}{4\pi}\right)^2$.

In the $(\rho_t,\,\rho_{\tau})$ plane, where
$\rho_t(\mu)=Y_t(\mu)/\tilde{\alpha}_3(\mu) =
(h_t(\mu)/g_3(\mu))^2$ and
$\rho_{\tau}(\mu)=Y_{\tau}(\mu)/\tilde{\alpha}_3(\mu) =
(h_{\tau}(\mu)/g_3(\mu))^2$, the allowed part of the parameter
space at the electroweak scale is limited by two perpendicular
lines
\begin{equation}
\rho_t=\dfrac{2\, E(l_0)}{9\, \tilde{\alpha}_3(M_t)
F(l_0)},\qquad\qquad \rho_{\tau}=\dfrac{2\, E_2(l_0)}{5\,
\tilde{\alpha}_3(M_t) F_2(l_0)}, \label{2hdm40}
\end{equation}
where $l_0=\ln(\Lambda^2/M_t^2)$. The two lines (\ref{2hdm40})
together form a quasi--fixed (or Hill type effective) line. The
solutions of the RG equations (\ref{2hdm31}) and (\ref{2hdm39})
are gathered near this line, when the Yukawa couplings at the MPP
scale increase. At the same time if $l/(4\pi)\gtrsim 1$,
$\rho_t(\mu)$ and $\rho_{\tau}(\mu)$ are attracted towards the
invariant line, which can be parametrised as:
\begin{equation}
\left\{
\begin{array}{rcl}
\rho_t(\mu) & = & \dfrac{2\, E(l)}{9\, \tilde{\alpha}_3(\mu) F(l)}\\[3mm]
\rho_{\tau}(\mu) & = & \dfrac{2\, E_2(l)}{5\, \tilde{\alpha}_3(\mu) F_2(l)}.
\end{array}
\right.
\label{2hdm41}
\end{equation}
Infrared fixed lines and surfaces, as well as their properties,
were studied in detail in \cite{infr-line01}-\cite{infr-line03}. 
When $l=\ln{(\Lambda^2/\mu^2)}$ goes to zero,
the invariant line (\ref{2hdm41}) approaches its asymptotic limit
where $\rho_t, \rho_{\tau}>>1$ and $\rho_{\tau}\to 1.8\, \rho_t$,
which is a fixed point of the RG equations for the Yukawa
couplings in the gaugeless limit ($g_1=g_2=g_3=0$). The invariant
line connects this fixed point with the infrared stable fixed
point $\left(\dfrac{2}{9},\,0\right)$ to which all solutions
converge when either $l\to\infty$ or $g_3(\mu)$ approaches a
Landau pole. At the electroweak scale $\rho_t(\mu)$ and
$\rho_{\tau}(\mu)$ are concentrated near the quasi--fixed point
$\left(\dfrac{2\, E(l_0)}{9\, \tilde{\alpha}_3(M_t) F(l_0)},
\dfrac{2\, E_2(l_0)}{5\, \tilde{\alpha}_3(M_t) F_2(l_0)}\right)$,
which coincides with the intersection point of the invariant and Hill type
effective lines \cite{qfp01}-\cite{qfp03}.

The value of $\tan\beta$ at which this quasi--fixed point solution
is realised can be found from the relation between the running
mass and the Yukawa coupling of the $\tau$--lepton:
\begin{equation}
m_{\tau}(M_t)=\dfrac{h_{\tau}(M_t)}{\sqrt{2}}v\cos\beta.
\label{2hdm42}
\end{equation}
Eq.~(\ref{2hdm42}) results in extremely large values of
$\tan\beta=90-100$. At these values of $\tan\beta$, the running
mass of the $t$--quark is set by $h_t(M_t)$ alone. This permits us
to evaluate $m_t(M_t)$ in the vicinity of the considered
quasi--fixed point. However the prediction obtained for $m_t(M_t)$
is considerably higher than the experimental running mass of the
$t$--quark calculated by means of Eq.~(\ref{2hdm35}) (see Table
1).

The method of computation of the quasi--fixed point coordinates
discussed above can be applied to the determination of the
position of the quasi--fixed point in the MPP scenario $(IV)$ as
well. Once again the Hill type effective line restricts the
allowed range of the parameter space in the $(\rho_t, \rho_{b})$
plane at the electroweak scale. Here
$\rho_b(\mu)=Y_b(\mu)/\tilde{\alpha}_3(\mu) =
(h_b(\mu)/g_3(\mu))^2$. Outside this range the solutions of the
renormalisation group equations for $h_t(t)$ and $h_b(t)$ develop
a Landau pole below the scale $\Lambda$. The quasi--fixed point in
this model appears as a result of the intersection of the Hill
type effective line and the invariant line, which connects a fixed
point of the RG equations in the gaugeless limit $(\rho_t=\rho_b)$
with the infrared stable fixed point $(\rho_t=\rho_b=1/5)$
\cite{qfp01}-\cite{qfp03}.

The most difficult case for the analysis of RG flow is the MPP
solution $(II)$ where $h_t(\Lambda)$, $h_b(\Lambda)$ and
$h_{\tau}(\Lambda)$ can be large simultaneously, while the mass
hierarchy within the third generation of fermions is caused by
large values of $\tan\beta$. In the model $(II)$, for each allowed
set of top quark and $b$--quark Yukawa couplings at the
electroweak scale, the interval of variation of $h_{\tau}(M_t)$ is
limited from above. This theoretical restriction comes from the
requirement of the validity of perturbation theory up to the MPP
scale. A change of $h_t(M_t)$ and $h_b(M_t)$ leads either to a
growth or a reduction in the upper limit on $h_{\tau}(M_t)$. As a
result, at the electroweak scale, the allowed range of Yukawa
couplings in the $(\rho_t, \rho_b, \rho_{\tau})$ space is limited
by a Hill type effective surface. With increasing $h_t(\Lambda)$,
$h_b(\Lambda)$ and $h_{\tau}(\Lambda)$, the solutions of the RG
equations are gathered near this surface. At the same time if the
interval of evolution is relatively large, i.e.
$l_0=\ln\dfrac{\Lambda^2}{M_t^2}\gtrsim 4\,\pi$, the solutions of
the RG equations for $h_t(\mu)$, $h_b(\mu)$ and $h_{\tau}(\mu)$
are also attracted to the invariant line, which joins together a
fixed point of the RG equations in the gaugeless limit
$\left(\rho_b=\dfrac{11}{15}\rho_t,\, \rho_{\tau}=\dfrac{16}{15}\rho_t\right)$ 
and an infrared stable fixed point $(\rho_t=\rho_b=1/5, \rho_{\tau}=0)$ 
\cite{infr-line01}-\cite{infr-line03}.
Thus, at the electroweak scale, the solutions of the RG equations
for the Yukawa couplings are concentrated near the quasi--fixed
point, which is located at the intersection of the invariant line
with the Hill type effective surface \cite{qfp01}-\cite{qfp03}.

The values of $\rho_t(M_t)$, $\rho_b(M_t)$ and $\rho_{\tau}(M_t)$
that correspond to this quasi--fixed point are given in Table 1. As
before, using the relation between the running $\tau$--lepton mass
and $h_{\tau}(M_t)$ (\ref{2hdm42}), one can find $\tan\beta$.
Substituting the obtained value of $\tan\beta$ into
Eqs.~(\ref{2hdm34}) and (\ref{2hdm38}), the running quark masses
$m_t(M_t)$ and $m_b(M_t)$ can be predicted anew. From Table 1 it
becomes clear that the running top quark mass is still
unacceptably large, while the prediction for $m_b(M_t)$ is too
small compared to the experimental value \cite{bmass} of
$m_b(M_t) = 2.75 \pm 0.09$ GeV.
As a consequence we conclude that, in the MPP scenarios
considered above, it is rather difficult to get a self--consistent
solution if two or three Yukawa couplings are greater than unity
at the scale $\Lambda$, because in the dominant part of parameter
space $m_t(M_t)$ tends to be significantly higher than
$160-170\,\mbox{GeV}$.

\section{Higgs phenomenology}

\subsection{The RG flow of the Higgs self--couplings near the quasi--fixed point}

Nevertheless a self--consistent solution can be obtained in the
case when only $h_t(\Lambda)\gtrsim 1$, while all other Yukawa
couplings are small. In this limit only the top quark Yukawa coupling 
is relevant and the solutions of the RG equations for $h_t(\mu)$
are attracted to the quasi--fixed point (\ref{2hdm32}). With
increasing $h_t(\Lambda)$, the solutions of the RG equations for the Higgs
self--couplings are also gathered near the quasi--fixed points. To
establish the positions of the quasi--fixed points for
$\lambda_i(\mu)$, we apply the method of determination developed
in Section 4. For the purposes of our RG studies, it is convenient
to introduce
\begin{equation}
\rho_i(\mu)=\dfrac{\lambda_i(\mu)}{g_3^2(\mu)},\qquad\qquad
R_i(\mu)=\dfrac{\rho_i(\mu)}{\rho_t(\mu)}=\dfrac{\lambda_i(\mu)}{h_t^2(\mu)},
\label{2hdm43}
\end{equation}
where the index $i$ runs from $1$ to $4$. When $\lambda_1(\Lambda)$,
$\lambda_2(\Lambda)$, $\lambda_3(\Lambda)$, $\lambda_4(\Lambda)$
and the top quark Yukawa coupling at the MPP scale grow, the
corresponding solutions of the RG equations are focused near the
intersection point of the invariant line and the Hill type
effective surface that sets an upper limit on the values of
$\rho_1, \rho_2, \rho_3$ and $\rho_4$ at the electroweak scale
(see \cite{qfp01}-\cite{qfp03}).

As was revealed in the previous subsection, the invariant line
connects a stable fixed point of the RG equations in the gaugeless
limit with an infrared fixed point. When the strong gauge coupling
approaches a Landau pole, all solutions of the RG equations are
concentrated near the infrared fixed point
\begin{equation}
\rho_t=\dfrac{2}{9},\qquad \rho_1=0,\qquad \rho_2=\dfrac{\sqrt{689}-25}{36}\simeq 0.0347,
\qquad \rho_3=0,\qquad \rho_4=0.
\label{2hdm44}
\end{equation}
This is the only stable fixed point in the infrared region. In the
gaugeless limit fixed points obey the following system of
nonlinear algebraic equations:
\begin{equation}
\left\{
\ba{l}
12 R_1^2+ 4 R_3^2+ 4 R_3 R_4+ 2 R_4^2- 9 R_1 = 0\\
12 R_2^2+ 4 R_3^2+ 4 R_3 R_4+ 2 R_4^2+ 3 R_2-12 =  0\\
2 (R_1+R_2)(3R_3+R_4)+ 4 R_3^2+ 2 R_4^2- 3 R_3 = 0\\
2 R_4 (R_1+R_2+4R_3+2R_4)-3R_4 = 0.
\ea
\right.
\label{2hdm45}
\end{equation}
The equations (\ref{2hdm45}) come from the requirement that the
beta--functions of the $R_i$ vanish in the limit $g_i\to 0$.
The numerical solutions $R_i^0$ of Eqs.~(\ref{2hdm45}) are given
in Table 2. The solutions with negative or zero values of $R_1^0$
and $R_2^0$ do not satisfy the vacuum stability constraints: \be
\lambda_1(\Phi)>0,\qquad \lambda_2(\Phi)>0,\qquad
\tilde{\lambda}(\Phi)>0. \label{2hdm46} \ee The conditions
(\ref{2hdm46}) must be fulfilled everywhere from the electroweak
scale to the MPP scale. Otherwise another minimum of the Higgs
effective potential (\ref{2hdm3}) arises at some intermediate
scale, destabilising the physical and MPP scale vacua.

Near the fixed points the RG equations for $R_i(t)$ can be
linearised, i.e $R_i(t)\simeq R_i^0+r_i(t)$.
The linearised system of RG equations for $r_i(t)$ can be written
in the following form
\begin{equation}
\dfrac{d r_i}{dt}=\sum_{j=1}^4 \dfrac{\partial \beta_{R_i}}{\partial R_j}\biggl|_{R_i=R_i^0} r_j,\qquad\qquad
\dfrac{\partial \beta_{R_i}}{\partial R_j}\biggl|_{R_i=R_i^0}=\dfrac{h_t^2}{16\pi^2}\,a_{ij}
\label{2hdm461}
\end{equation}
where
$$
a_{ij}=
\left(
\begin{array}{cccc}
24 R_1^0-9      & 0               & 8 R_3^0+4 R_4^0          & 4(R_3^0+R_4^0) \\[0mm]
0               & 24 R_2^0+3      & 8 R_3^0+4 R_4^0          & 4(R_3^0+R_4^0) \\[0mm]
6 R_3^0+2 R_4^0 & 6 R_3^0+2 R_4^0 & 6(R_1^0+R_2^0)+8 R_3^0-3 & 2(R_1^0+R_2^0+2R_4^0)\\[0mm]
2 R_4^0         & 2 R_4^0         & 8 R_4^0                  & 2(R_1^0+R_2^0+4 R_3^0+4 R_4^0)-3
\end{array}
\right)
$$
The fixed point is stable when all the eigenvalues of the matrix
$a_{ij}$ are positive. Only in this case do all the $r_i(t)$ tend
to zero in the infrared region.

The analysis of the convergence of the solutions of the linearised
system of RG equations (\ref{2hdm461}), in the vicinity of the
fixed points listed in Table 2, reveals that there is only one
stable fixed point solution which corresponds to \be
R_1=\dfrac{3}{4},\qquad R_2=\dfrac{\sqrt{65}-1}{8}\simeq
0.883,\qquad R_3=0,\qquad R_4=0. \label{2hdm47} \ee Choosing a
large value of the top quark Yukawa coupling at the scale
$\Lambda$ (say $h_t^2(\Lambda)=10$) and using the fixed point
solution (\ref{2hdm47}) as a boundary condition for the RG
equations, one establishes the position of the quasi--fixed point
at the electroweak scale. If the MPP scale is close to $M_{Pl}$ we
get
\begin{equation}
\ba{c}
\rho_t(M_t)\simeq 1.174,\qquad \rho_1(M_t)\simeq 0.341,\qquad \rho_2(M_t)\simeq 0.694,\\
\rho_3(M_t)\simeq -0.011,\qquad \rho_4(M_t)\simeq -0.013.
\ea
\label{2hdm48}
\end{equation}

It turns out that, for large values of the top quark Yukawa
coupling at the scale $\Lambda$, the allowed range of the Higgs
self--couplings is rather strongly constrained. Stringent
constraints on $\lambda_i(\Lambda)$ come from the MPP conditions
(\ref{2hdm22}). Using the equations $\tilde{\lambda}(\Lambda)=0$
and $\beta_{\tilde{\lambda}}(\Lambda)=0$, one can express
$\lambda_3(\Lambda)$ and $\lambda_4(\Lambda)$ in terms of the
other gauge, Yukawa and Higgs self--couplings, i.e.
\be
\lambda_3(\Lambda)=-\sqrt{\lambda_1(\Lambda)\lambda_2(\Lambda)}-\lambda_4(\Lambda)\,,
\label{2hdm49} \ee \be \ba{c}
\lambda_4^2(\Lambda)=\ds\frac{6h_t^4(\Lambda)\lambda_1(\Lambda)}
{(\sqrt{\lambda_1(\Lambda)}+\sqrt{\lambda_2(\Lambda)})^2}
-2\lambda_1(\Lambda)\lambda_2(\Lambda)\\[4mm]
\ds-\frac{3}{8}\biggl(3g_2^4(\Lambda)+2g_2^2(\Lambda)g_1^2(\Lambda)+g_1^4(\Lambda)\biggr)\,,
\ea \label{2hdm50}
\ee
where $\lambda_4(\Lambda)<0$. Thus the RG
flow of the Higgs self--couplings, in the MPP inspired 2HDM with
an approximate generalised Peccei--Quinn symmetry, is determined by
$h_t(\Lambda)$, $\lambda_1(\Lambda)$ and $\lambda_2(\Lambda)$.
Varying $\lambda_1(\Lambda)$ and $\lambda_2(\Lambda)$, one can
obtain the restrictions on their values. Because
$\lambda_4(\Lambda)$ is a real quantity, Eq.~(\ref{2hdm50}) limits
the allowed range of $\lambda_1(\Lambda)$ and $\lambda_2(\Lambda)$
from above. For instance, when
$\lambda_1(\Lambda)=\lambda_2(\Lambda)=\lambda_0$ the quantity
$\lambda_4^2(\Lambda)$ remains positive only if
$\lambda_0<\ds\frac{\sqrt{3}}{2}h_t^2(\Lambda)$. The lower bound
on the Higgs self--couplings originates from the vacuum stability
conditions (\ref{2hdm46}). Indeed, if
$\lambda_1(\Lambda)=\lambda_2(\Lambda)=\lambda_0$ is sufficiently
small then $\tilde{\lambda}(\mu)$ tends to be negative at some
intermediate scale, destabilising the physical and MPP scale
vacua. Our numerical studies show that, for $\Lambda=M_{Pl}$ and
$R_1(M_{Pl})=R_2(M_{Pl})=R_0$, the value of $R_0$ can vary only
within a very narrow interval from $0.79$ to $0.87$ if
$h_t(\Lambda)\gtrsim 1.5$. Moreover the allowed range of $R_0$
shrinks further when $h_t(\Lambda)$ increases. For
$h_t(\Lambda)\gtrsim 2.5$ the value of $R_0$ can vary only between
$0.83$ and $0.87$.

In Figs.~2a and 2b we present the restrictions on the Higgs
self--couplings $\lambda_1(\Lambda)$ and $\lambda_2(\Lambda)$ for
$h_{t}(\Lambda)=3$ and two different values of the MPP scale
$\Lambda=M_{Pl}$ and $\Lambda=10\,\mbox{TeV}$. In these plots the
allowed region of the parameter space in the
$R_1(\Lambda)-R_2(\Lambda)$ plane is limited by the dotted and
solid lines. The dotted line represents the vacuum stability
constraints (\ref{2hdm46}). For any point in the
$R_1(\Lambda)-R_2(\Lambda)$ plane below the dotted curve, the
vacuum stability conditions are violated at some intermediate
scale between $\Lambda$ and $M_t$ preventing the consistent
implementation of the MPP. The solid line constrains the allowed
range of the Higgs self--couplings from above. For any point above
this line $\lambda_4^2(\Lambda)$ is negative. From Fig.~2a one can
see that vacuum stability and MPP conditions set stringent
constraints on the Higgs self--couplings for $R_1(\Lambda)>0.3$,
if the MPP scale is relatively high. In the considered case only a
very narrow strip in the $R_1(\Lambda)-R_2(\Lambda)$ plane is not
ruled out. When the MPP scale decreases, the allowed range of
$\lambda_1(\Lambda)$ and $\lambda_2(\Lambda)$ enlarges. From
Fig.~2b it follows that the position of the solid line does not
change significantly with decreasing scale $\Lambda$, whereas the
restrictions on the Higgs self--couplings caused by the vacuum
stability constraints (\ref{2hdm46}) become less stringent. It is
worth noticing here that, independently of the MPP scale, the
stable fixed point (\ref{2hdm47}), which is shown as an open circle
in Fig.~2, always lies in the allowed region of parameter space.

In Fig.~3 we plot the RG flow of the Higgs self--couplings from
$\Lambda=M_{Pl}$ to the electroweak scale. As boundary conditions
we use a set of points $R_1(\Lambda)$ and $R_{2}(\Lambda)$ from
the allowed part of the parameter space shown in Fig.~2a. The
Higgs self--couplings $\lambda_3(\Lambda)$ and
$\lambda_4(\Lambda)$ are chosen so that the MPP conditions
(\ref{2hdm49})--(\ref{2hdm50}) are fulfilled. Figs.~3a and 3b
demonstrate that the trajectories, which represent different
solutions of the RG equations for $\lambda_1(\mu)$,
$\lambda_2(\mu)$, and $\lambda_3(\mu)$, are focused in a narrow
region near the quasi--fixed points at low energies. At the same
time the trajectories in the $R_4(\mu)$--$R_1(\mu)$ plane are
rather spread out in the infrared region (see Fig.~3c). This is an
indication that the solutions of the RG equations for
$\lambda_4(\mu)$ are attracted very weakly to the corresponding
quasi--fixed point.

In Fig.~4 we examine the convergence of the solutions of the RG
equations to the quasi--fixed points, as a function of the MPP
scale. In Figs.~4a--4d we set $R_1(\Lambda)=0.75$ and
$R_2(\Lambda)=0.883$ and choose $R_3(\Lambda)$ and $R_4(\Lambda)$
so that the MPP conditions (\ref{2hdm49})--(\ref{2hdm50}) are
satisfied. The solid and dashed lines in Fig.~4 represent the
dependence of the $R_i(M_t)$ on the scale $\Lambda$ for
$h_t^2(\Lambda)=10$ and $h_t^2(\Lambda)=2.25$ respectively. From
Fig.~4a and 4b one can see that $\lambda_1(M_t)$ and
$\lambda_2(M_t)$ do not change substantially, when
$h_t^2(\Lambda)$ varies from $10$ to $2.25$. This demonstrates the
good convergence rate of the solutions of the RG equations for
$\lambda_1(\mu)$ and $\lambda_2(\mu)$ to the corresponding
quasi--fixed points. The values of $\lambda_3(M_t)$ are quite
sensitive to the choice of the MPP scale and $h_t(\Lambda)$ (see
Fig.~4c). If $\Lambda$ is relatively high ($\Lambda\gtrsim
10^{13}\,\mbox{GeV}$), the solutions of the RG equations for
$\lambda_3(\mu)$ are gathered near zero at the electroweak scale.
But for relatively low $\Lambda$ ($\Lambda\lesssim
10^{3}\,\mbox{TeV}$), the value of $\lambda_3(M_t)$ changes
considerably when $h_t^2(\Lambda)$ is reduced from $10$ to $2.25$.
In general the convergence of the solutions of the RG equations
for $\lambda_1(\mu)$, $\lambda_2(\mu)$ and $\lambda_3(\mu)$
becomes worse when the MPP scale decreases. Finally Fig.~4d
indicates a strong dependence of $\lambda_4(M_t)$ on the scale
$\Lambda$ and $h_t^2(\Lambda)$, which makes it rather difficult to
get any reasonable prediction for the value of this Higgs
self--coupling at the electroweak scale.

\subsection{Higgs masses and couplings}

Relying on the results of the analysis of the RG flow in the MPP
inspired 2HDM, we can explore the Higgs spectrum at the
electroweak scale. The Higgs sector of the 2HDM involves
two charged and three neutral scalar states.
Since our MPP solutions conserve CP, one of the neutral Higgs bosons 
is purely CP--odd. The charged and pseudoscalar Higgs states 
gain masses: 
\be
m^2_{\chi^{\pm}}=m_A^2-\ds\frac{\lambda_4}{2}v^2\,,\qquad\qquad
m_A^2=\frac{2m_3^2}{\sin 2\beta}\,. 
\label{2hdm51} 
\ee 
In the case of the MPP solution (II), the direct searches for 
the rare B--meson decays ($B\to X_s\gamma$) place a
lower limit on the charged Higgs boson mass \cite{mch01}-\cite{mch02}: 
$m_{\chi^{\pm}}> 350\,\mbox{GeV}$\,. 

The CP--even states are mixed and form a $2\times 2$ mass matrix.
It is convenient to introduce a new
field space basis $(h,\,H)$ rotated by the angle $\beta$
with respect to the initial one:
\be \ba{c}
H_1^0=(h \cos\beta- H \sin\beta+v_1)\,, \\[0mm]
H_2^0=(h \sin\beta+ H \cos\beta+v_2)\,.
\ea
\label{2hdm53}
\ee
Then the field $h$ is the analogue of the SM Higgs field with
vacuum expectation value $<h> = v$ and is solely responsible
for the symmetry breaking, while the field $H$ has zero vacuum
expectation value and is irrelevant for symmetry breaking. In
this new basis the mass matrix of the Higgs scalars takes the form
(see also \cite{higgs-basis01}-\cite{higgs-basis03})
\be
M^2=\left(
\ba{ll}
M^2_{11} & M^2_{12}\\[1mm]
M^2_{21} & M^2_{22}
\ea
\right)=
\left(
\ba{ll}
\frac{\ds\partial^2V}{\ds\partial \upsilon^2} \qquad~~~
\frac{\ds 1}{\ds \upsilon}\frac{\ds\partial^2V}{\ds\partial \upsilon\partial\beta}\\[1mm]
\frac{\ds 1}{\ds \upsilon}\frac{\ds\partial^2V}{\ds \partial \upsilon\partial\beta} \qquad
\frac{\ds 1}{\ds \upsilon^2}\frac{\ds \partial^2V}{\ds \partial\beta^2}
\ea
\right)\,,
\label{2hdm54}
\ee
$$
\ba{rcl}
M_{11}^2&=&\biggl(\lambda_1\cos^4\beta+\lambda_2\sin^4\beta+\ds\frac{\lambda}{2}\sin^22\beta\biggr)v^2\,,\\[0mm]
M_{12}^2&=&M_{21}^2=\ds\frac{v^2}{2}\biggl(-\lambda_1\cos^2\beta+\lambda_2\sin^2\beta+\lambda\cos 2\beta
\biggr)\sin 2\beta\,,\\[0mm]
M_{22}^2&=&m_A^2+\ds\frac{v^2}{4}\biggl(\lambda_1+\lambda_2-2\lambda\biggr)\sin^22\beta\,,
\ea
$$
where $\lambda=\lambda_3+\lambda_4$. 
The masses of the two CP--even eigenstates obtained by
diagonalizing the matrix (\ref{2hdm54}) are given by
\be
m_{h_1,\,h_2}^2=\frac{1}{2}\left(M^2_{11}+
M^2_{22}\mp\sqrt{(M_{22}^2-M_{11}^2)^2+4M^4_{12}}\right)\,.
\label{2hdm55}
\ee
The qualitative pattern of the Higgs spectrum
depends very strongly on the mass $m_A$ of the pseudoscalar Higgs
boson. With increasing $m_A$ the masses of all the Higgs particles
grow. At very large values of $m_A$ ($m_A^2>>v^2$), the lightest
Higgs boson mass approaches its theoretical upper limit
$\sqrt{M_{11}^2}$. 

In the rotated field basis $(h, H)$ the trilinear part of the 
Lagrangian, which determines the interactions of the neutral Higgs 
states with the $Z$--boson, is simplified 
\cite{coupl-basis01}-\cite{coupl-basis03}:
\be
L_{AZH}=\ds\frac{\bar{g}}{2}
M_{Z}Z_{\mu}Z_{\mu}h+\frac{\bar{g}}{2}Z_{\mu}
\biggl[H(\partial_{\mu}A)-(\partial_{\mu}H)A\biggr]~.
\label{2hdm56}
\ee
where $\bar{g} = \sqrt{g_2^2 + g_1^2}$.
Following the traditional notations we 
define normalised $R$--couplings of the neutral Higgs states to 
vector bosons as follows:
$g_{VVh_i}=R_{VVh_i}\times$ SM coupling
$\left(\mbox{i.e.}\,\,\dfrac{\bar{g}}{2}M_V\right)$;
$g_{ZAh_i}=\ds\frac{\bar{g}}{2}R_{ZAh_i}$, where $V$ is a
$W^{\pm}$ or a $Z$ boson. The relative couplings $R_{ZZh_i}$ and
$R_{ZAh_i}$ are given in terms of the angles $\alpha$ and $\beta$
\cite{Carena:2002es}: 
\be 
\ba{c}
R_{ZZh_1}=R_{WWh_1}=-R_{ZAh_2}=\sin(\beta-\alpha)\,,\\
R_{ZZh_2}=R_{WWh_2}=R_{ZAh_1}=\cos(\beta-\alpha)\,, 
\ea
\label{2hdm57} 
\ee 
where the angle $\alpha$ is defined as follows:
\be 
\ba{rcl}
h_1&=&-(H_1^0-v_1)\sin\alpha+(H_2^0-v_2)\cos\alpha\,,\\[0mm]
h_2&=&(H_1^0-v_1)\cos\alpha+(H_2^0-v_2)\sin\alpha\,,
\ea
\label{2hdm58}
\ee
$$
\tan\alpha=\dfrac{(\lambda v^2-m_A^2)\sin\beta\cos\beta}{m_A^2\sin^2\beta+\lambda_1 v^2\cos^2\beta-m_{h_1}^2}\,.
$$
The absolute values of the $R$--couplings $R_{VVh_i}$ and
$R_{ZAh_i}$ vary from zero to unity.

The couplings of the Higgs eigenstates to the top quark
$g_{t\bar{t}h_i}$ can also be presented as a product of the
corresponding SM coupling and the $R$--coupling $R_{t\bar{t}h_i}$:
\be
R_{t\bar{t}h_1}=\dfrac{\cos\alpha}{\sin\beta}\,,\qquad\qquad\qquad
R_{t\bar{t}h_2}=\dfrac{\sin\alpha}{\sin\beta}\,. 
\label{2hdm61}
\ee 
Since the $R_{t\bar{t}h_i}$ are inversely proportional to
$\sin\beta$ and near the quasi--fixed point $\tan\beta\lesssim 1$, 
the values of $R_{t\bar{t}h_i}$ can be substantially larger than unity.

As follows from Eqs.~(\ref{2hdm51})--(\ref{2hdm61}), the spectrum
and couplings of Higgs bosons in the MPP inspired 2HDM, with softly
broken Peccei--Quinn and $Z_2$ symmetry, is parametrized in terms
of $m_A$, $\tan\beta$ and four Higgs self--couplings
$\lambda_1(M_t),\,\lambda_2(M_t),\,\lambda_3(M_t)$ and
$\lambda_4(M_t)$. In our study of the phenomenology of the Higgs
sector, we concentrate on the quasi--fixed point scenario. In
particular, at the MPP scale we set $R_1(\Lambda)=0.75$,\,
$R_2(\Lambda)\simeq 0.883$ and $h_t^2(\Lambda)=10$, which
correspond to the quasi--fixed point solution. At the same time we
do not keep $R_3(\Lambda)=R_4(\Lambda)=0$. Instead, we find
appropriate values of $\lambda_3(\Lambda)$ and
$\lambda_4(\Lambda)$ that obey the MPP conditions
(\ref{2hdm49})--(\ref{2hdm50}). Then we evolve the top quark
Yukawa and Higgs couplings down to the electroweak scale. The results
of our calculations have already been discussed in the previous
sections. According to our analysis near the quasi--fixed points,
the Higgs self--couplings, the top quark Yukawa coupling and
$\tan\beta$ depend only on the MPP scale (see Figs.~1 and 4). As a
result, in the considered scenario, all Higgs masses and couplings
are functions of the scale $\Lambda$ and the pseudoscalar mass
$m_A$. Therefore, at the next stage, we examine the dependence of
the Higgs masses and couplings on the pseudoscalar mass for each
fixed value of the MPP scale.

The results of our investigations are summarised in Fig.~5--7. In
Fig.~5 we plot the masses and couplings of the CP--even Higgs
eigenstates for the MPP scale $\Lambda=M_{Pl}$. From Fig. 5a it is
clear that the masses of the Higgs particles change considerably
when $m_A$ varies. In particular, the masses of the heaviest
CP--even and charged Higgs states rise with increasing
pseudoscalar mass. At large values of $m_A\gtrsim
300\,\mbox{GeV}$, the corresponding Higgs states are almost
degenerate around $m_A$. The mass of the lightest CP--even Higgs
boson is not so sensitive to the variations of the pseudoscalar
mass. It varies from $80\,\mbox{GeV}$ to $120\,\mbox{GeV}$. The
lightest Higgs scalar $h_1$ can be identified as being
predominantly the SM-like superposition $h$ of the neutral components of
Higgs doublets, because its relative coupling to a $Z$ pair is
always close to unity (see Fig.~5b). The contribution of the
orthogonal combination of neutral components of Higgs doublets $H$
to $h_1$ is considerably smaller. As a result the coupling of the
lightest CP--even Higgs state to the Higgs pseudoscalar and $Z$ is
suppressed. But, at low values of $m_A\lesssim 100\,\mbox{GeV}$,
the R--coupling $R_{ZAh_1}$ is still large enough that the
lightest CP--even and pseudoscalar Higgs states could have been
produced in $e^{+}e^{-}$ collisions at LEP. Because $R_{ZZh_1}$ is
rather close to unity, the associated production of the lightest
Higgs scalar and the $Z$ boson would also have been possible.
Consequently the non-observation of the SM--like Higgs particle at
LEP rules out most of the parameter space near the quasi--fixed
point solution if the scale $\Lambda$ is relatively high, i.e.
$\Lambda\gtrsim 10^{15}$. This is a consequence of the stringent
bound on the mass of the SM--like Higgs caused by the RG flow of
Higgs self--couplings from $\Lambda$ to the electroweak scale. In
Fig.~6 the theoretical upper bound on $m_{h_1}$ as a function of
the MPP scale $\Lambda$ is presented. If $\Lambda\gtrsim
10^{10}\,\mbox{GeV}$ the lightest CP--even Higgs boson is lighter
than $125\,\mbox{GeV}$. The upper bound on $m_{h_1}$ grow from
$125\,\mbox{GeV}$ to $140\,\mbox{GeV}$, when the MPP scale is
lowered from $10^{10}\,\mbox{GeV}$ to $10^{7}\,\mbox{GeV}$ (see
Fig.~6)

When $\Lambda$ is near the Planck scale the $H_1^0$ component
of the lightest CP--even Higgs scalar is larger than the $H_2^0$
component for $m_A < 400$ GeV. This is essentially because,
at large values of the MPP scale, $\lambda_1(M_t)$ is less than
$\lambda_2(M_t)$ while $v_1\simeq v_2$ in the vicinity of the
quasi--fixed point.
Since $H_1^0$ is the larger component, the coupling of the
lightest CP--even Higgs eigenstate $h_1$ to the top quark
is smaller than the coupling of the heaviest one for $m_A < 400$ GeV. The
dependence of the relative couplings of the CP--even Higgs bosons
to the top quark, $R_{t\bar{t}h_i}$, on $m_A$ is examined in
Fig.~5c. From this figure one can see that $R_{t\bar{t}h_2}$ is more
than twice as big as $R_{t\bar{t}h_1}$ at low values of
$m_A\lesssim 100\,\mbox{GeV}$. However such small values of the pseudoscalar
mass are excluded by the unsuccessful Higgs searches at LEP. With
increasing $m_A$ the heaviest CP--even, CP--odd and charged Higgs
states decouple. As a consequence, the couplings of the lightest
Higgs boson to a $Z$ pair and to the top quark approach the SM
ones (see Fig.~5c), i.e. $h_1\simeq h$. Our numerical analysis
reveals that for $\Lambda\gtrsim 10^{8}\,\mbox{GeV}$ the relative
coupling $R_{t\bar{t}h_1}\lesssim 1$.

The situation changes significantly when the MPP scale is
relatively low. In Fig.~7 we study the dependence of the masses and
couplings of the Higgs bosons on $m_A$ for the MPP scale
$\Lambda=100\,\mbox{TeV}$. As before the masses of the heaviest
CP--even, CP--odd and charged Higgs states are set by $m_A$. When
$m_A$ grows, all Higgs masses increase and at large values of the
pseudoscalar Higgs mass $m_{\chi^{\pm}}\simeq m_{h_2}\simeq m_A$
(see Fig.~7a). However the upper
bound on the lightest CP--even Higgs scalar mass $m_{h_1}$ increases
significantly. If $\Lambda\simeq 100\,\mbox{TeV}$ the
upper bound on the mass of the lightest CP--even Higgs state
changes from $140\,\mbox{GeV}$ to $180\,\mbox{GeV}$. Once again
the main contribution to the wave function of the lightest Higgs
scalar corresponds to the SM-like superposition of neutral components of Higgs
doublets $h$ so that $R_{ZZh_1}\approx 1$ (see Fig.~7b). However
the values of $m_A \gtrsim M_Z$ are not excluded by LEP data,
because the associated lightest Higgs scalar production with
either a $Z$ boson or a Higgs pseudoscalar is kinematically
forbidden. At low values of the pseudoscalar mass ($m_A < 250$ GeV)
the $H_2^0$ component of the lightest CP--even Higgs state is now
larger than the $H_1^0$ component. Despite
$\lambda_2(M_t)$ still being larger than $\lambda_1(M_t)$, the
vacuum expectation value $v_2$ becomes considerably smaller than
$v_1$ ($\tan\beta\sim 0.5$) resulting in $\lambda_1 v_1^2>
\lambda_2 v_2^2$. This gives rise to a realignment in the Higgs
spectrum. As can be seen from Fig.~7c, the change in content of $h_1$
leads to a substantial increase in the coupling of the lightest
Higgs scalar to the top quark. Our numerical studies demonstrate
that, for values of the MPP scale $\Lambda$ below
$1000\,\mbox{TeV}$, there is some range of $m_A$ in the
quasi--fixed point scenario where $R_{t\bar{t}h_1}\gtrsim
R_{t\bar{t}h_2}\gtrsim 1$. Due to the significant growth of the
coupling of the lightest CP--even Higgs state to the top quark,
the production cross section of the SM--like Higgs in the 2HDM can
be $1.5-2$ times larger than 
in the SM \cite{Froggatt:2007qp01}-\cite{Froggatt:2007qp03}. The
enhanced production of the SM--like Higgs boson allows us to
distinguish the quasi--fixed point scenario in the MPP inspired
2HDM with low MPP scale from the SM and its supersymmetric
extensions, even if the extra Higgs states are heavy ($m_A\gtrsim
400-500\,\mbox{GeV}$).

\section{Other MPP solutions}

The MPP solution that corresponds to the set of degenerate vacua,
in which the energy density vanishes near the scale $\Lambda$ for
any $\omega$, might not be a unique one in the two Higgs doublet
extension of the SM. Indeed, in Appendix B we present the
derivation of other MPP conditions, which correspond to the set of
vacua which have zero vacuum energy density at the MPP scale for
any choice of $\theta$ or $\gamma$. These scenarios were not
discussed in our previous article \cite{Froggatt:2006zc}, where we
considered the implementation of the multiple point principle in
the two Higgs doublet model of type II. 

It turns out that it is quite difficult to achieve a
self--consistent realization of these other MPP solutions, if we
restrict our consideration to the simplest two Higgs doublet
extension of the SM with a minimal matter content. Indeed, in the
Higgs field basis where only $H_2$ couples to the $t$--quark at
the MPP scale, the observed mass hierarchy within the third
generation of fermions implies that $h_t(\Lambda)>>g_b(\Lambda),
g_{\tau}(\Lambda)$. At the same time, for small values of
these Peccei--Quinn symmetry violating Yukawa couplings, some of
the MPP conditions derived in the Appendix B cannot be satisfied.

For instance, let us consider the MPP conditions (\ref{B7}), that
result in the vacuum configuration in which the energy density
goes to zero for arbitrary values of the ratio of the Higgs vacuum
expectation values $\Phi_2/\Phi_1$ when
$\lambda_4(\Lambda)<|\lambda_5(\Lambda)|$. Substituting the
explicit expressions for $\beta_{\lambda_3}$, $\beta_{\lambda_4}$
and $\beta_{\lambda_5}$ in the last MPP condition of
Eqs.~(\ref{B7}), we find
\be
\ba{c}
\beta_{\lambda_3}(\Lambda)+\beta_{\lambda_4}(\Lambda)+\mbox{Re}\,\beta_{\lambda_5}(\Lambda)=
\ds\frac{1}{16\pi^2}\biggl[
2\lambda_3^2(\Lambda)+4\lambda_5^2(\Lambda)+\ds\frac{9}{4}g_2^4(\Lambda)+\\[2mm]
+\ds\frac{3}{2}g_2^2(\Lambda)g_1^2(\Lambda)+\ds\frac{3}{4}g_1^4(\Lambda)-24|h_b(\Lambda)|^2|g_b(\Lambda)|^2-
8|h_{\tau}(\Lambda)|^2|g_{\tau}(\Lambda)|^2-\\[2mm]
-\biggl(6h_b^2(\Lambda)g^{*2}_b(\Lambda)+2h^2_{\tau}(\Lambda)g^{*2}_{\tau}(\Lambda)+h.c.\biggr)
\biggr]=0\,.
\ea
\label{2hdm24}
\ee
Here we have redefined the Higgs fields, so that $\lambda_5(\Lambda)$ is real and negative. 
From Eq.~(\ref{2hdm24}) it becomes clear that the positive contribution of the Higgs and 
gauge couplings to the corresponding combination of $\beta$--functions cannot be compensated
by the negative contribution coming from the Yukawa interactions if
$g_b(\Lambda)\sim g_{\tau}(\Lambda)\sim 10^{-2}$, unless $|h_{b}(\Lambda)|^2$ or
$|h_{\tau}(\Lambda)|^2\gtrsim 10$. However such large values of $|h_{b}(\Lambda)|$
and $|h_{\tau}(\Lambda)|$ would spoil the validity of perturbation theory.

Due to similar reasons, it is not possible to achieve the
degeneracy of vacua with respect to $\theta$. The MPP conditions
that ensure the existence of such a set of degenerate minima of
the Higgs effective potential at the MPP scale are given by
Eqs.~(\ref{B12}).  After the substitution of explicit expressions
for $\beta_{\lambda_4}$ and $\beta_{\lambda_5}$, one of the MPP
conditions, $\beta_{\lambda_4}(\Lambda)+
\mbox{Re}\,\beta_{\lambda_5}(\Lambda)=0$, reduces to
\be
\ba{c}
12|h_t(\Lambda)|^2|h_b(\Lambda)|^2 +3
g_2^2(\Lambda)g_1^2(\Lambda)-12|h_b(\Lambda)|^2|g_b(\Lambda)|^2-
4|h_{\tau}(\Lambda)|^2|g_{\tau}(\Lambda)|^2-\\[2mm]
-\biggl(6h_b^2(\Lambda)g^{*2}_b(\Lambda)+2h^2_{\tau}(\Lambda)g^{*2}_{\tau}(\Lambda)+h.c.\biggr)=0\,.
\ea
\label{2hdm25}
\ee
To satisfy the MPP condition
(\ref{2hdm25}), either $g_b(\Lambda)$ or
$g_{\tau}(\Lambda)$ should be large.
This makes the generation of the observed mass
hierarchy rather problematic.

Nevertheless there is one new set of the MPP conditions whose
realisation does not require large Peccei--Quinn symmetry
violating Yukawa couplings. Indeed the MPP conditions (\ref{B5}),
which lead to the presence of vacua in which the energy density
tends to zero for any ratio of the Higgs vacuum expectation values
$\Phi_2/\Phi_1$ at the MPP scale when
$\lambda_4(\Lambda)>|\lambda_5(\Lambda)|$, can be fulfilled even
if $g_b(\Lambda)$ and $g_{\tau}(\Lambda)$ are negligibly small.
The corresponding set of degenerate minima of the Higgs effective
potential
\be
<H_1>=\left(
\begin{array}{c}
0\\ \Phi_1
\end{array}
\right)\,, \qquad <H_2>=\left(
\begin{array}{c}
\Phi_2\\ 0
\end{array}
\right) \,, \qquad \Phi_1^2+\Phi_2^2=\Lambda^2,
\label{2hdm26}
\ee
arises if the Higgs self--couplings $\lambda_{1}(\Lambda)$,
$\lambda_{2}(\Lambda)$, $\lambda_{3}(\Lambda)$ and their
$\beta$--functions  $\beta_{\lambda_1}(\Lambda)$,
$\beta_{\lambda_2}(\Lambda)$, $\beta_{\lambda_3}(\Lambda)$ vanish.
The vanishing of the three $\beta$--functions (see
Eq.~(\ref{A2})) for the Higgs
self--couplings $\lambda_1$, $\lambda_2$ and $\lambda_3$ at the
MPP scale can be achieved only if the Yukawa couplings of the
third generation obey two relationships:
\be
\ba{c}
|h_t(\Lambda)|^4=|h_b(\Lambda)|^4+\ds\frac{1}{3}|h_{\tau}(\Lambda)|^4\,,\\
|h_t(\Lambda)|^4=|h_t(\Lambda)|^2|h_b(\Lambda)|^2+\ds\frac{1}{4}g_2^2(\Lambda)g_1^2(\Lambda)\,.
\ea 
\label{2hdm27}
\ee
In Eq.~(\ref{2hdm27}) we neglect $g_b(\Lambda)$ and $g_{\tau}(\Lambda)$.

The relations (\ref{2hdm27}) allows us to express the $b$--quark
and $\tau$--lepton Yukawa couplings at the scale $\Lambda$ in
terms of $h_t(\Lambda)$. Thus, for each fixed value of the top
quark Yukawa coupling at the MPP scale, one can calculate the RG
flow of $h_t(\mu)$, $h_b(\mu)$ and $h_{\tau}(\mu)$ from the scale
$\Lambda$ to $\mu=M_t$. Then Eq.~(\ref{2hdm42}) can be used for the 
determination of $\tan\beta$. Since at the electroweak scale 
$h_t(M_t)\sim h_b(M_t)\sim h_{\tau}(M_t)$, the relation (\ref{2hdm42}) 
results in large values of $\tan\beta\sim m_t/m_b$. In the considered 
part of the parameter space $m_t(M_t)$ is almost independent of 
$\tan\beta$, i.e. $m_t(M_t)\simeq h_t(M_t)v/\sqrt{2}$.

On the other hand $m_t(M_t)$ can be determined rather precisely
from experiment, using the relationship between the top quark pole
and running masses (\ref{2hdm35}). In the MPP scenario discussed
here, the top quark Yukawa coupling at the scale $\Lambda$ can be
adjusted so that the observed value of $M_t$ is reproduced. This
permits us to evaluate all Yukawa couplings at the electroweak
scale and to predict the values of $\tan\beta$ and $m_b(M_t)$
using Eqs.~(\ref{2hdm38}),\,(\ref{2hdm42}). The results of our
numerical studies are summarised in Table 3, where we explore the
dependence of $\tan\beta$ and $m_b(M_t)$ on the scale $\Lambda$.
One can see that the value of $\tan\beta$ is always rather close
to $50$, while $m_b(M_t)$ changes from $3.2$ GeV to $2.6$ GeV when
the MPP scale grows from $10\,\mbox{TeV}$ to $M_{Pl}$.

The prediction for the running $b$--quark mass at the electroweak
scale can be easily improved, if we include the Peccei--Quinn
symmetry violating Yukawa couplings $g_b$ and $g_{\tau}$. Small
values of these couplings affect neither the relations between the
Yukawa couplings (\ref{2hdm27}) nor the running of $h_t(\mu)$,
$h_b(\mu)$ and $h_{\tau}(\mu)$. However even very small values of
the corresponding couplings ($\sim 10^{-2}$) change the
predictions for $m_b(M_t)$ and $\tan\beta$ significantly.
As a result one can easily reproduce the experimental value of 
the running $b$--quark mass, $m_b(M_t) = 2.75 \pm 0.09$ GeV. 
But even zero values of the Peccei--Quinn symmetry violating 
Yukawa couplings lead to a reasonable prediction for $m_b(M_t)$ 
for large values of $\Lambda$.

A stringent restriction on the MPP scale in the considered
scenario comes from the non--observation of the Higgs particle at
LEP. In Tables 3 and 4 we examine the upper bound on the mass of
the SM--like Higgs particle as a function of the scale $\Lambda$
and the Higgs self--couplings $\lambda_4(\Lambda)$ and
$\lambda_5(\Lambda)$. In order to ensure that
the vacua (\ref{2hdm26}) are stable at the MPP scale,
$\lambda_4(\Lambda)$ has to be positive and the absolute value of
$\lambda_5(\Lambda)$ should be less than $\lambda_4(\Lambda)$. To
guarantee that the Higgs effective potential is positive definite
everywhere between the MPP and electroweak scales, which makes the
consistent implementation of the MPP possible, the following
conditions must be fulfilled:
\begin{equation}
\lambda_1(\Phi)>0,\qquad \lambda_2(\Phi)>0,
\label{2hdm351}
\end{equation}
\begin{equation}
\widehat{\lambda}(\Phi)=\sqrt{\lambda_1(\Phi)\lambda_2(\Phi)}+\lambda_3(\Phi)+
\mbox{min}\{0,\,\lambda_4(\Phi)-|\lambda_5(\Phi)|\}>0\,.
\label{2hdm352}
\end{equation}
For $\lambda_5=0$, the inequalities
(\ref{2hdm351})--(\ref{2hdm352}) coincide with the vacuum
stability constraints derived in our previous work
\cite{Froggatt:2006zc}. It turns out that
$\widehat{\lambda}(\Phi)$ is only positive for any value of $\Phi$
between $\Lambda$ and $M_t$ when
$|\lambda_5(\Lambda)|<0.83\cdot\lambda_4(\Lambda)$. In Fig.~8a and
8b we plot the running of $\lambda_1(\mu)$, $\lambda_2(\mu)$ and
$\widehat{\lambda}(\mu)$ for $\lambda_5=0$
and two different values of the MPP scale: $\Lambda=M_{Pl}$ and
$\Lambda=10\,\mbox{TeV}$.

The upper bound $m_h$ on the mass of the SM--like Higgs boson, which is
given by
\begin{equation}
\ba{c}
m_{h_1}^2\lesssim m_h^2 = v^2\biggl(\lambda_1(M_t)\cos^4\beta+\lambda_2(M_t)\sin^4\beta+
\ds\frac{\lambda(M_t)}{2}\sin^2 2\beta\biggr),\\[2mm]
\lambda(M_t)=\lambda_3(M_t)+\lambda_4(M_t)+\lambda_5(M_t),
\ea
\label{2hdm353}
\end{equation}
does not vary substantially when $\lambda_4(\Lambda)$ and
$\lambda_5(\Lambda)$ change (see Table 4). This weak dependence of the
theoretical restriction on the SM--like Higgs mass on
$\lambda_4(\Lambda)$ and $\lambda_5(\Lambda)$ is a result of the
suppression of their contribution to $m_{h_1}^2$ at large values of
$\tan\beta$. At the same time the upper bound on $m_{h_1}^2$ decreases
significantly when the MPP scale $\Lambda$ varies from $M_{Pl}$ to
$10\,\mbox{TeV}$. In the case when $\Lambda\simeq 10\,\mbox{TeV}$
the SM--like Higgs mass does not exceed $75\,\mbox{GeV}$. Such
small values of $m_{h_1}$ have been already ruled out by LEP. To
satisfy LEP constraints on the mass of the Higgs boson, the MPP
scale should be larger than $10^8\,\mbox{GeV}$. In the considered
MPP scenario the upper bound on the SM--like Higgs mass attains
its maximum value of $140\,\mbox{GeV}$ for $\Lambda\simeq M_{Pl}$.

Although the MPP scenario discussed here is not excluded from the
phenomenological point of view, it seems to be rather problematic
to achieve the degeneracy of vacua at the MPP scale with the
accuracy $v^2\Lambda^2$ in this case. Indeed, in order to
guarantee that the vacua at the MPP scale are really degenerate
with respect to either $\gamma$ or $\theta$, we have to require,
as in the case of degeneracy of vacua with respect to $\omega$,
that the masses of all the fermions and bosons should not change
when $\gamma$ or $\theta$ varies. Otherwise quantum corrections to
the Higgs boson potential (\ref{2hdm222}) spoil the degeneracy of
the considered vacua. In general, when the Higgs fields acquire
vacuum expectation values (\ref{2hdm7}), the $SU(2)\times U(1)$
gauge bosons gain the following masses
\be
\ba{c}
M_{1,2}^2=\ds\frac{g_2^2}{2}\Phi^2,\quad
M_{3,4}^2=\biggl[\ds\frac{g_2^2+g_1^2}{2}\Phi^2\pm
\sqrt{\left(\ds\frac{g_2^2+g_1^2}{2}\Phi^2\right)^2-4\,g_2^2\,
g_1^2\,\Phi_2^2\, \Phi_1^2\, \sin^2\theta}\biggr]. 
\ea
\label{2hdm28}
\ee
In the vicinity of the MPP scale
$\Phi^2=\Lambda^2$, $\Phi_1=\Lambda\cos\gamma$ and
$\Phi_2=\Lambda\sin\gamma$. Eq.~(\ref{2hdm28}) reveals that the
degeneracy of the MPP scale vacua with respect to $\gamma$ can be
achieved only for $\sin\theta=0$ \footnote{One can show that in
the limit $\sin\theta\to 0$ the set of degenerate vacua with
respect to $\gamma$ in one field basis is equivalent to the set of
degenerate vacua with respect to $\omega=2\gamma$ in another
field basis.}.
This means that the MPP solution (\ref{2hdm26}) which corresponds to
$\sin\theta=\pm 1$ will not get, when one loop corrections are
included, a true continuum of degenerate vacua to accuracy $v^2 \Lambda^2$.
Thus if we interpret MPP to mean that we should
choose the solution with the largest number of degenerate
vacua, the solution (\ref{2hdm26}) is beaten by the solutions
with a Lie group symmetry such as the solution (\ref{2hdm18})
mainly studied in the present article. This is connected with
the fact that we do not have any custodial symmetry for solution
(\ref{2hdm26}), which also means that it does not exclude
FCNC and CP violation in the Higgs sector automatically
in contrast to the solution (\ref{2hdm18}).
So the solution (\ref{2hdm26}) for the set of vacua parameterized
by $\gamma$
is disfavoured by: 1) giving formally fewer vacua, 2) not
explaining the absence of FCNC and 3) generically having
CP violation in the Higgs sector.

Eq.~(\ref{2hdm28}) also illustrates the fact that the vacuum energy density
of the Higgs effective potential cannot be the same for different
values of $\theta$, because the masses of two gauge bosons depend
rather strongly on this parameter. Since the masses of the gauge
bosons change when $\theta$ varies, quantum corrections would
spoil the degeneracy of the MPP scale vacua with respect to
$\theta$.

Because at $\sin\theta=0$ the masses of the gauge bosons are
invariant under the variations of $\gamma$ and $\omega$, one can
try to find a vacuum configuration in which the energy density
goes to zero for arbitrary values of $\gamma$ and $\omega$ at the
MPP scale. Then the independence of the vacuum energy density on
the phase $\omega$ implies that the Yukawa and Higgs
self--couplings obey the MPP conditions
(\ref{2hdm21})--(\ref{2hdm22}). In this case the degeneracy of the
vacua with respect to $\gamma$ can be achieved only when
$\lambda_1(\Lambda)=\lambda_2(\Lambda)=0$. However, in our
previous publication \cite{Froggatt:2006zc}, we argued that either
$\lambda_1(\Phi)$ or $\tilde{\lambda}(\Phi)$ tends to be negative
just below the MPP scale if $\lambda_1(\Lambda)=\lambda_2(\Lambda)=0$.
As a result, near the scale $\Lambda$, there exists another minimum 
of the Higgs effective potential with a huge and negative vacuum 
energy density. This prohibits the self--consistent implementation 
of the MPP for arbitrary values of both $\gamma$ and $\omega$.

\section{Conclusions}

In this paper we have considered the application of the multiple
point principle (MPP) to the non-supersymmetric two Higgs doublet
extension of the SM. In general new couplings, which appear in this
model, give rise to potentially large flavour changing neutral
currents and CP--violation effects. We have argued that MPP can be
used as a mechanism for the suppression of FCNC and CP--violating
interactions. Indeed, MPP postulates the existence of a large
set of degenerate vacua, which are allowed by a given theory. These
vacua might not have exactly the same vacuum energy density. Here
we assumed that the vacua at the electroweak and at the high MPP scale
$\Lambda$ are degenerate with the accuracy $v^2\Lambda^2$.
Normally the presence of a large set of degenerate vacua is
associated with an enlarged global symmetry of the Lagrangian of
the considered model. This is also the case in the 2HDM. The most
favourable solution we found implies that
the quartic part of the Higgs potential and
the Lagrangian for the Higgs--fermion interactions are invariant
under the transformations of a set of global $U(1)$ symmetries
(\ref{2hdm229}), which forbid non--diagonal flavour transitions
and CP violating couplings. One example of such a custodial symmetry is a
Peccei--Quinn symmetry that contains $Z_2$ symmetry as a subgroup.
This $Z_2$ discrete symmetry ensures the suppression of FCNC
processes in the the 2HDM of type II. At the same time MPP allows
us to avoid problems that usually arise in the framework of the
two Higgs doublet models with exact global $U(1)$ or $Z_2$
symmetries. This is because we did not require the set of
vacua at the MPP scale to be exactly
degenerate. Therefore global custodial symmetries appearing in the
MPP inspired 2HDM can be approximate. As a consequence, in our
favourable MPP solution, the breakdown of electroweak symmetry
does not give rise to either an axion or domain walls. Meanwhile
the custodial symmetry violating couplings are expected to be
small $O(v^2/\Lambda^2)$. This leads to the suppression
of FCNC and CP--violating effects.

We explored the RG flow of the Yukawa and Higgs
couplings within the MPP inspired 2HDM with approximate custodial
symmetries and studied the phenomenology of the Higgs sector in
the framework of these models. In our analysis we concentrated on
the quasi--fixed point scenarios. The positions of the
quasi--fixed points at moderate and large values of $\tan\beta$
have been established. We argued that the quasi--fixed point
scenarios which correspond to large $\tan\beta$ lead to
unacceptably large values of the top quark running mass.
Nevertheless we found a self--consistent solution when only
$h_t(\Lambda)\gtrsim 1$, while all other Yukawa couplings are
small. In this case $\tan\beta$ can be chosen so that the
appropriate value of the top quark mass is reproduced. 
We also demonstrated that the RG solutions for the Higgs self--couplings
$\lambda_1(\mu)$, $\lambda_2(\mu)$ and $\lambda_3(\mu)$ are
focused in a narrow interval near the quasi--fixed points at low
energies, if the MPP scale is relatively high ($\Lambda\gtrsim
10^{13}\,\mbox{GeV}$). The solutions of the RG equations for
$\lambda_4(\mu)$ are attracted to the quasi--fixed point rather
weakly.

In the considered quasi--fixed point scenario, the spectrum and
couplings of the Higgs bosons depend on the MPP scale and
pseudoscalar Higgs mass $m_A$ predominantly. The masses of all
the Higgs states rise with increasing pseudoscalar mass. At
large values of $m_A\gtrsim 300\,\mbox{GeV}$ the heaviest
CP--even, CP--odd and charged Higgs bosons are almost degenerate
around $m_A$. When $m_A$ is large the lightest Higgs
boson mass approaches its theoretical upper bound. 
The results of our numerical studies show that $m_{h_1}$ does not exceed
$125\,\mbox{GeV}$ if $\Lambda\gtrsim 10^{10}\,\mbox{GeV}$. At the
same time, when $\Lambda\simeq 100-10\,\mbox{TeV}$, the lightest
Higgs boson mass can reach $180-220\,\mbox{GeV}$. With increasing
$m_A$ the heaviest CP--even, CP--odd and charged Higgs states
decouple and the couplings of the lightest Higgs boson approach
the SM ones. However our numerical analysis revealed that, for MPP
scales $\Lambda$ below $1000\,\mbox{TeV}$, there is a range of
$m_A$ where the couplings of the lightest Higgs scalar to the top
quark is considerably larger than in the SM. This leads to the
enhanced production of the lightest Higgs particle at the LHC,
which would allow us to distinguish the quasi--fixed point scenario
in the MPP inspired 2HDM from the SM and its supersymmetric
extensions.

We also discussed other possible scenarios, which appear in the
2HDM as a result of the implementation of the MPP.
In contrast to our favourable MPP solution, the other scenarios
do not result in either exact or approximate global symmetries.
Moreover, in most cases which we considered, it is extremely
difficult to reproduce the observed mass hierarchy in the quark
and lepton sector. Nevertheless we found one new scenario in which 
the corresponding MPP conditions can be fulfilled in the leading 
approximation. This new MPP solution leads to a set of vacua in 
which the energy density tends to zero for arbitrary values of 
the ratio of the Higgs vacuum expectation values, 
$\tan\gamma=\Phi_2/\Phi_1$, at the MPP scale 
($\Lambda^2=\Phi_1^2+\Phi_2^2$). In this scenario the SM--like Higgs 
boson attains its maximum value of $140\,\mbox{GeV}$ when 
$\Lambda\simeq M_{Pl}$ whereas low values of the MPP scale 
($\Lambda < 10^8\,\mbox{GeV}$) lead to a too light Higgs 
boson, which has already been ruled out by LEP.
However, because the considered MPP scenario is not
related with the invariance of the Lagrangian under global
symmetry transformations, the inclusion of the complete set of
one--loop corrections to the Higgs boson potential spoils the
degeneracy of vacua at the MPP scale. Therefore the degeneracy of
vacua with the accuracy $v^2\Lambda^2$ cannot be achieved.

\vspace{3mm}
\noindent
{\large \bf Acknowledgements}

\noindent The authors are grateful to L.~V.~Laperashvili and
M.~Sher for valuable comments and remarks. We would also like to
thank S.~F.~King, D.~J.~Miller, S.~Moretti, L.~B.~Okun,
M.~Shifman, D.~Sutherland and R.~St.Denis for fruitful
discussions. The authors acknowledge support from the SHEFC grant
HR03020 SUPA 36878.

\newpage
\section*{Appendix A: Renormalization of the Higgs self--couplings in general 2HDM}

\setcounter{equation}{0}
\def\theequation{A.\arabic{equation}}

The structure of the renormalization group equations for the Higgs
self--couplings in the 2HDM is fixed by the set of the
$\beta$--functions
\be
\frac{d\lambda_i}{dt}=\beta_{\lambda_i}\,.
\label{A1}
\ee
In Eq.(\ref{A1}) index $i$ runs from $1$ to $7$.
The variable $t$ is defined in the conventional way: $t=\ln\,
\mu$, where $\mu$ is the renormalization scale. When
$\lambda_6=\lambda_7=0$, we obtain:
\be
\ba{l}
\beta_{\lambda_1}=\ds\frac{1}{16\pi^2}\biggl[12\lambda_1^2+4\lambda_3^2+4\lambda_3\lambda_4+2\lambda_4^2+2|\lambda_5|^2+
\frac{9}{4}g_2^4+\frac{3}{2}g_2^2g_1^2+\frac{3}{4}g_1^4-\\[1mm]
\quad-\lambda_1\biggl(3(3g_2^2+g_1^2)-12|h_b|^2-4|h_{\tau}|^2\biggr)-12|h_b|^4-4|h_{\tau}|^4\biggr]\,,\\[1mm]
\beta_{\lambda_2}=\ds\frac{1}{16\pi^2}\biggl[12\lambda_2^2+4\lambda_3^2+4\lambda_3\lambda_4+2\lambda_4^2+2|\lambda_5|^2+
\frac{9}{4}g_2^4+\frac{3}{2}g_2^2g_1^2+\frac{3}{4}g_1^4-\\[1mm]
\quad-\lambda_2\biggl(3(3g_2^2+g_1^2)-12|h_t|^2-12|g_b|^2-
4|g_{\tau}|^2\biggr)-12|h_t|^4-12|g_b|^4-4|g_{\tau}|^4\biggr]\,,\\[1mm]
\beta_{\lambda_3}=\ds\frac{1}{16\pi^2}\biggl[2(\lambda_1+\lambda_2)(3\lambda_3+\lambda_4)+4\lambda_3^2+2\lambda_4^2+2|\lambda_5|^2+
\frac{9}{4}g_2^4-\frac{3}{2}g_2^2g_1^2+\frac{3}{4}g_1^4-\\[1mm]
\quad-\lambda_3\biggl(3(3g_2^2+g_1^2)-6|h_t|^2-6|g_b|^2-2|g_{\tau}|^2-6|h_b|^2-2|h_{\tau}|^2
\biggr)-12|h_t|^2|h_b|^2-\\[1mm]
\quad-12|h_b|^2|g_b|^2-4|h_{\tau}|^2|g_{\tau}|^2\biggr]\,,\\[1mm]
\beta_{\lambda_4}=\ds\frac{1}{16\pi^2}\biggl[2\lambda_4(\lambda_1+\lambda_2+4\lambda_3+2\lambda_4)+8|\lambda_5|^2+3g_2^2g_1^2-
\lambda_4\biggl(3(3g_2^2+g_1^2)-\\[1mm]
\quad-6|h_t|^2-6|g_b|^2-2|g_{\tau}|^2-6|h_b|^2-2|h_{\tau}|^2\biggr)+12|h_t|^2|h_b|^2-12|h_b|^2|g_b|^2-\\[0mm]
\quad-4|h_{\tau}|^2|g_{\tau}|^2\biggr]\,,\\[1mm]
\beta_{\lambda_5}=\ds\frac{1}{16\pi^2}\biggl[2\lambda_5(\lambda_1+\lambda_2+4\lambda_3+6\lambda_4)
-\lambda_5\biggl(3(3g_2^2+g_1^2)-6|h_t|^2-6|g_b|^2-\\[1mm]
\quad-2|g_{\tau}|^2-6|h_b|^2-2|h_{\tau}|^2\biggr)-12h_b^2g_b^{*2}-4h_{\tau}^2g_{\tau}^{*2}\biggr]\,,\\[1mm]
\beta_{\lambda_6}=\ds\frac{1}{16\pi^2}\biggl[(\lambda_1+\lambda_3+\lambda_4)(3g^{*}_bh_b+h_{\tau}g^{*}_{\tau})
+\lambda_5(3h_b^{*}g_b+h_{\tau}^{*}g_{\tau})-12|h_b|^2h_bg^{*}_b-\\[1mm]
\quad-4|h_{\tau}|^2h_{\tau}g^{*}_{\tau}\biggr]\,,\\[1mm]
\beta_{\lambda_7}=\ds\frac{1}{16\pi^2}\biggl[(\lambda_2+\lambda_3+\lambda_4)(3g^{*}_bh_b+h_{\tau}g^{*}_{\tau})
+\lambda_5(3h_b^{*}g_b+h_{\tau}^{*}g_{\tau})-12|g_b|^2h_bg^{*}_b-\\[1mm]
\quad-4|g_{\tau}|^2h_{\tau}g^{*}_{\tau}\biggr]\,. 
\ea 
\label{A2}
\ee
The Yukawa couplings $h_t$, $g_b$, $g_{\tau}$ and $h_b$,
$h_{\tau}$ appearing on the right--hand side of Eqs.(\ref{A2})
determine the strength of the interactions of the Higgs doublets
$H_2$ and $H_1$ with fermions (see (\ref{2hdm5})). Eqs.(\ref{A2})
are derived by assuming that only one Higgs doublet $H_2$ couples
to $t_R$. This can be easily achieved by the appropriate
redefinition of the Higgs doublets at the MPP scale. Although
$g_t(\Lambda)=0$, a non--zero value of this coupling can be generated
below the MPP scale, due to the renormalization group flow (see
(\ref{2hdm6})), in the absence of a custodial symmetry.

\section*{Appendix B: the degeneracy of vacua with respect to $\tan\gamma$ or $\theta$.}

\setcounter{equation}{0}
\def\theequation{B.\arabic{equation}}

In this section we consider possible sets of degenerate minima of
the Higgs effective potential with vanishing vacuum energy
density, which are not related with the presence of a Peccei Quinn
symmetry. In the case when there is a vacuum configuration in
which the energy density tends to zero for arbitrary values of the
ratio of the Higgs vacuum expectation values
$\tan\gamma=\Phi_2/\Phi_1$, the terms involving different powers
of $\Phi_2$ and $\Phi_1$ in the Higgs effective potential must go
to zero irrespective of each other. This leads to the conditions:
\be
\left\{ \ba{l}
\lambda_1(\Lambda)=\lambda_2(\Lambda)=0\\
\lambda_3(\Lambda)+\lambda_4(\Lambda)\cos^2\theta+\biggl(\ds\frac{\lambda_5(\Lambda)}{2}e^{2i\omega}\cos^2\theta+h.c.
\biggr)=0\\
\lambda_k(\Lambda)e^{i\omega}\cos\theta+h.c.=0\,, \ea \right.
\label{B1}
\ee
where $k=6,7$. The same should happen in the
conditions for the extrema of the Higgs effective potential, if it
attains its minimum at the considered vacuum expectation values of
the Higgs fields. Applying this requirement to the conditions
$\ds\frac{\partial V}{\partial \omega}=\frac{\partial V}{\partial
\theta}=0$, one obtains:
\be
\left\{ \ba{l}
\biggl(\lambda_5(\Lambda)e^{2i\omega}-h.c.\biggr)\cos^2\theta=0\\
\biggl(\lambda_k(\Lambda)e^{i\omega}-h.c.\biggr)\cos\theta=0\\
\biggl(\lambda_5(\Lambda)e^{2i\omega}+h.c.\biggr)\sin 2\theta=0\\
\biggl(\lambda_k(\Lambda)e^{i\omega}+h.c.\biggr)\sin\theta=0\,.
\ea \right. \label{B2}
\ee
In a similar way the minimization
conditions $\ds\frac{\partial V}{\partial \Phi_1}=\frac{\partial
V}{\partial \Phi_2}=0$ constrain the $\beta$--functions of the
Higgs self--couplings:
\be
\left\{ \ba{l}
\beta_{\lambda_1}(\Lambda)=\beta_{\lambda_2}(\Lambda)=0\\
\beta_{\lambda_3}(\Lambda)+\beta_{\lambda_4}(\Lambda)\cos^2\theta+
\biggl(\ds\frac{\beta_{\lambda_5}(\Lambda)}{2}e^{2i\omega}+h.c.\biggr)\cos^2\theta=0\\
\biggl(\beta_{\lambda_k}(\Lambda)e^{i\omega}+h.c.\biggr)\cos\theta=0\,.
\ea \right. \label{B3}
\ee
The relationships (\ref{B3}) for
$\beta_{\lambda_i}$ are deduced by assuming that the
conditions (\ref{B1}) are fulfilled.

Some of the MPP conditions (\ref{B1})--(\ref{B3}) are satisfied
when $\cos\theta$ goes to zero. For $\cos\theta=0$, the MPP
conditions (\ref{B1}) reduce to 
$\lambda_1(\Lambda)=\lambda_2(\Lambda)=\lambda_3(\Lambda)=0\,. $ In
this limit, the Higgs effective potential near the scale
$\Lambda$ can be written in the following form:
\be
\ba{c} V(H_1,
H_2)=\biggl(\lambda_4(\Lambda)+\left(\ds\frac{\lambda_5(\Lambda)}{2}e^{2i\omega}+h.c.
\right)\biggr)\Phi_1^2\Phi_2^2\cos^2\theta+\\[3mm]
+2\biggl(X^2+Y^2\biggr)^{1/2}\Phi_2\Phi_1\cos\theta\cos(\omega+\varphi)\,,
\ea
\label{B31}
\ee
where
$$
\ba{c}
\varphi=\tan^{-1}\left(\ds\frac{Y}{X}\right)\,,\qquad X=\mbox{Re}\,\biggl(\lambda_6(\Lambda)\Phi_1^2+\lambda_7(\Lambda)\Phi_2^2\biggr)\,,\\[3mm]
Y=\mbox{Im}\,\biggl(\lambda_6(\Lambda)\Phi_1^2+\lambda_7(\Lambda)\Phi_2^2\biggr)\,.
\ea
$$
The value of $\cos\theta=0$ corresponds to the minimum of the
scalar potential (\ref{B31}) only if
$\lambda_4(\Lambda)>|\lambda_5(\Lambda)|>0$ and $X=Y=0$.
Otherwise, $\cos\theta$ tends to get a non-zero value and the
vacuum energy becomes negative. The real and imaginary parts of
$\lambda_6(\Lambda)\Phi_1^2+\lambda_7(\Lambda)\Phi_2^2$ only get
zero values independently of $\tan\gamma$ in the case when both
$\lambda_6(\Lambda)$ and $\lambda_7(\Lambda)$ vanish identically.
For $\lambda_6(\Lambda)=\lambda_7(\Lambda)=0$, the Higgs scalar
potential at the MPP scale simplifies further, so that finally we
get:
\be
V(H_1,
H_2)=\biggl(\lambda_4(\Lambda)+\left(\ds\frac{\lambda_5(\Lambda)}{2}e^{2i\omega}+h.c.
\right)\biggr)\Phi_1^2\Phi_2^2\cos^2\theta\,. \label{B4}
\ee
The scalar potential (\ref{B4}) reaches a minimum at $\cos\theta=0$,
where $V(H_1, H_2)$ vanishes. Substituting $\cos\theta=0$ into
Eq.(\ref{B3}), we find the MPP conditions that provide a
degeneracy of vacua with different values of $\tan\gamma$:
\be
\left\{ \ba{l}
\lambda_4(\Lambda)>|\lambda_5(\Lambda)|\\
\lambda_1(\Lambda)=\lambda_2(\Lambda)=\lambda_3(\Lambda)=\lambda_6(\Lambda)=\lambda_7(\Lambda)=0\\
\beta_{\lambda_1}(\Lambda)=\beta_{\lambda_2}(\Lambda)=\beta_{\lambda_3}(\Lambda)=0\,.
\ea
\right.
\label{B5}
\ee

If $\cos\theta\ne 0$, then the degeneracy of vacua with respect to
$\tan\gamma$ can be achieved only when
$\lambda_6(\Lambda)=\lambda_7(\Lambda)=0$ (see
Eq.(\ref{B1})--(\ref{B2})). In this case the Higgs
effective potential at the MPP scale takes the form:
\be
V(H_1,
H_2)=\biggl[\lambda_3(\Lambda)+\lambda_4(\Lambda)\cos^2\theta+\left(\ds\frac{\lambda_5(\Lambda)}{2}e^{2i\omega}+h.c.
\right)\cos^2\theta\biggr]\Phi_1^2\Phi_2^2\,, \label{B6}
\ee
where
$\lambda_3(\Lambda), \lambda_4(\Lambda)$ and $\lambda_5(\Lambda)$
obey Eq.(\ref{B1})--(\ref{B2}). Minima of the scalar potential
(\ref{B6}) with a non--zero value of $\cos\theta$ arise when
$\lambda_4<|\lambda_5|$. In this part of parameter space, the
vacuum energy density decreases with increasing $\cos^2\theta$ and
reaches its minimum value at $\cos\theta=\pm 1$. If we redefine
the Higgs fields so that $\lambda_5(\Lambda)$ becomes real and
negative, the minimum of the Higgs effective potential (\ref{B6})
corresponds to $\omega=0$. Then the MPP conditions
(\ref{B1})--(\ref{B3}) reduce to:
\be
\left\{ \ba{l}
\lambda_4(\Lambda)<|\lambda_5(\Lambda)|\\
\lambda_1(\Lambda)=\lambda_2(\Lambda)=\lambda_6(\Lambda)=\lambda_7(\Lambda)=0\\
\lambda_3(\Lambda)+\lambda_4(\Lambda)+\lambda_5(\Lambda)=0\\
\beta_{\lambda_1}(\Lambda)=\beta_{\lambda_2}(\Lambda)=\mbox{Re}\,\beta_{\lambda_6}(\Lambda)=\mbox{Re}\,\beta_{\lambda_7}(\Lambda)=0\\
\beta_{\lambda_3}(\Lambda)+\beta_{\lambda_4}(\Lambda)+\mbox{Re}\,\beta_{\lambda_5}(\Lambda)=0\,.
\ea
\right.
\label{B7}
\ee

Let us now consider vacuum configurations in which the energy density
vanishes for arbitrary values of $\theta$. The Higgs effective
potential will not depend on $\cos\theta$ near the MPP scale, only
if the following conditions are satisfied:
\be
\ba{l}
\lambda_4(\Lambda)+\biggl(\ds\frac{\lambda_5(\Lambda)}{2}e^{2i\omega}+h.c.\biggr)=0\\
\biggl(\lambda_6(\Lambda)\Phi_1^2+\lambda_7(\Lambda)\Phi_2^2\biggr)e^{i\omega}+h.c.=0\,.
\ea \label{B8}
\ee
If the relationships (\ref{B8}) between the
Higgs self--couplings are fulfilled, the scalar potential can be
written as:
\be
\ba{rcl} V(H_1,H_2)\simeq
\ds\frac{1}{2}\biggl(\sqrt{\lambda_1(\Lambda)}\Phi_1^2-
\sqrt{\lambda_2(\Lambda)}\Phi_2^2\biggr)^2+\left(\sqrt{\lambda_1(\Lambda)\lambda_2(\Lambda)}+\lambda_3(\Lambda)\right)\Phi_1^2\Phi_2^2\,.
\ea \label{B9}
\ee
Its minimum value goes to zero when
\be
\bar{\lambda}(\Lambda)=\sqrt{\lambda_1(\Lambda)\lambda_2(\Lambda)}+\lambda_3(\Lambda)=0\,.
\label{B10}
\ee
If the Higgs effective potential has a local
minimum at the MPP scale then
$\ds\frac{d\bar{\lambda}}{d\Phi}\biggl|_{\Phi=\Lambda}$ vanishes
as well.

The presence of the set of degenerate vacua with respect to
$\theta$ implies that the terms which are proportional to
$\cos\theta$ and $\cos^2\theta$ in the conditions for the extrema
$\ds\frac{\partial V}{\partial \omega}=\frac{\partial V}{\partial
\Phi_1}=\frac{\partial V}{\partial \Phi_2}=0$ should vanish
separately. It imposes extra constraints on the Yukawa and Higgs
self--couplings, which are given by:
\be
\left\{ \ba{l}
\ds\frac{\lambda_5(\Lambda)}{2}e^{2i\omega}-h.c.=0\\[3mm]
\biggl(\lambda_6(\Lambda)\Phi_1^2+\lambda_7(\Lambda)\Phi_2^2\biggr)e^{i\omega}-h.c.=0\\[3mm]
\beta_{\lambda_4}(\Lambda)+\biggl[\ds\frac{\beta_{\lambda_5}(\Lambda)}{2}e^{2i\omega}+h.c.\biggr]=0\\[3mm]
\biggl[(3\lambda_6(\Lambda)\Phi_1^2+\lambda_7(\Lambda)\Phi_2^2)\Phi_2+\biggl(\beta_{\lambda_6}(\Lambda)\Phi_1^2+
\beta_{\lambda_7}(\Lambda)\Phi_2^2\biggr)\ds\frac{\Phi_1^2\Phi_2}{\Phi^2}\biggr]e^{i\omega}+h.c.=0\\[3mm]
\biggl[(\lambda_6(\Lambda)\Phi_1^2+3\lambda_7(\Lambda)\Phi_2^2)\Phi_1+\biggl(\beta_{\lambda_6}(\Lambda)\Phi_1^2+
\beta_{\lambda_7}(\Lambda)\Phi_2^2\biggr)\ds\frac{\Phi_1\Phi_2^2}{\Phi^2}\biggr]e^{i\omega}+h.c.=0
\ea \right. \label{B11}
\ee
The first two relationships between the
$\lambda_i(\Lambda)$ in Eq.(\ref{B11}) come from
$\ds\frac{\partial V}{\partial \omega}=0$, whereas the other three
conditions follow from the assumption that $\ds\frac{\partial
V}{\partial \Phi_1}$ and $\ds\frac{\partial V}{\partial \Phi_2}$
are independent of $\cos\theta$ near the minima of the Higgs
scalar potential at the MPP scale. The derivative of $V(H_1,
\,H_2)$ with respect to $\theta$ vanishes automatically, if the
MPP conditions (\ref{B8}) are satisfied.

To simplify the analysis, we restrict our consideration to real
and negative values of $\lambda_5(\Lambda)$, which can be arranged
by the appropriate redefinition of the Higgs fields. Then the
first MPP condition in Eq.(\ref{B11}) enforces $\omega$ to take a
discrete set of values $\ds\frac{\pi}{2} n$, where $n$ is an integer
number. However only even values of $n$ correspond to a minimum of
the Higgs scalar potential. Substituting $\omega=\pi m$ into
Eq.(\ref{B8}) and Eq.(\ref{B10})--(\ref{B11}), we find:
\be
\left\{ \ba{l}
\bar{\lambda}(\Lambda)=\beta_{\bar{\lambda}}(\Lambda)=0\,,\\
\lambda_4(\Lambda)+\lambda_5(\Lambda)=0\,,\\
\mbox{Re}\,\lambda_6(\Lambda)=\mbox{Re}\,\lambda_7(\Lambda)=0\,,\\
\mbox{Im}\,\biggl[\lambda_6(\Lambda)\Phi_1^2+\lambda_7(\Lambda)\Phi_2^2\biggr]=0\,,\\
\beta_{\lambda_4}(\Lambda)+\mbox{Re}\,\beta_{\lambda_5}(\Lambda)=0\,,\\
\mbox{Re}\,\biggl[\beta_{\lambda_6}(\Lambda)\Phi_1^2+\beta_{\lambda_7}(\Lambda)\Phi_2^2\biggr]=0\,.
\ea \right. \label{B12}
\ee
Relying on the MPP conditions
(\ref{B12}) and taking into account that $V(H_1,\,H_2)$ vanishes
near the MPP scale when
$\Phi_2^2=\ds\sqrt{\lambda_1(\Lambda)/\lambda_2(\Lambda)}\Phi_1^2$,
one can easily deduce the complete expression for the vacuum
energy density at the scale $\Lambda$:
\be
\ba{rcl}
V(H_1,H_2)&\simeq
&\ds\frac{1}{2}\biggl(\sqrt{\lambda_1(\Lambda)}\Phi_1^2-\sqrt{\lambda_2(\Lambda)}\Phi_2^2+
\varkappa \Phi_1\Phi_2\cos\theta\sin\omega\biggr)^2+\\[2mm]
&+&2\left(\lambda_4(\Lambda)-\ds\frac{\varkappa^2}{4}\right)\Phi_1^2\Phi_2^2\cos^2\theta
\sin^2\omega\,,
\label{B13}
\ea
\ee
where
$$
\varkappa=\frac{-2\,\mbox{Im}\,\lambda_6(\Lambda)}{\sqrt{\lambda_1(\Lambda)}}=
\frac{2\,\mbox{Im}\,\lambda_7(\Lambda)}{\sqrt{\lambda_2(\Lambda)}},
$$
which is valid for arbitrary values of $\gamma$, $\theta$ and
$\omega$. From Eq.(\ref{B13}) it is obvious that the stable minima
at $\omega=\pi m$, which lead to degeneracy of the vacuua with
respect to $\theta$, are attained only for positive values of
$\lambda_4(\Lambda)$ when $\lambda_4(\Lambda)>\varkappa^2/4$.

\newpage
\noindent
{\Large \bf Figure captions}
\vspace{5mm}

\noindent {\bf Fig.1.}\,\,(a) The top quark Yukawa coupling at
$\mu=M_t$ versus the MPP scale $\Lambda$. (b) The dependence of
$\tan\beta$ on the scale $\Lambda$. The solid and dashed curves
correspond to $h_t^2(\Lambda)=10$ and $h_t^2(\Lambda)=2.25$. 
The dash--dotted lines represent the quasi--fixed point solution 
(\ref{2hdm32}). Here we set $m_t(M_t)=161.6\,\mbox{GeV}$ and
$\alpha_3(M_Z)=0.117$. The MPP scale $\Lambda$ is given in GeV.\\
\\
{\bf Fig.2.}\,\,The allowed range of $R_1(\Lambda)$ and
$R_2(\Lambda)$ for (a) $\Lambda=M_{Pl}$ and (b)
$\Lambda=10\,\mbox{TeV}$. The solid line corresponds to the upper
bound on $R_2(\Lambda)$, which comes from the requirement that
$\lambda_4^2(\Lambda)\geq 0$. The dotted curve represents the
lower bound on $R_2(\Lambda)$ caused by the vacuum stability
condition. The open circle indicates the position of the fixed
point in the gaugeless limit. Other parameters are fixed as
follows: $h_t(\Lambda)=3$, $m_t(M_t)=161.6\,\mbox{GeV}$
and $\alpha_3(M_Z)=0.117$.\\
\\
{\bf Fig.3.}\,\,The renormalisation group flow of $R_i(\mu)$ from
the Planck scale to the electroweak scale in (a) the $(R_1,\,R_2)$
plane, (b) the $(R_1,\,R_3)$ plane and (c) the $(R_1,\,R_4)$
plane, for different initial values of $R_1(\Lambda)$ and
$R_2(\Lambda)$ from the allowed part of parameter space (see
Fig.2). The open circle indicates the position of the quasi--fixed 
point (\ref{2hdm48}). Here we set $h_t(\Lambda)=3$. The initial 
values of $\lambda_3(\Lambda)$ and $\lambda_4(\Lambda)$ satisfy 
the MPP conditions (\ref{2hdm49}) and (\ref{2hdm50}).\\
\\
{\bf Fig.4.}\,\, The dependence of (a) $R_1(M_t)$, (b) $R_2(M_t)$,
(c) $R_3(M_t)$ and (d) $R_4(M_t)$ on the MPP scale near the
quasi--fixed point. Solid and dashed lines correspond to
$h_t^2(\Lambda)=10$ and $h_t^2(\Lambda)=2.25$ respectively. The
Higgs self--couplings $\lambda_1(\Lambda)$ and
$\lambda_2(\Lambda)$ are fixed so that $R_1(\Lambda)=0.75$ and
$R_2(\Lambda)=0.883$, whereas $\lambda_3(\Lambda)$ and
$\lambda_4(\Lambda)$ obey the MPP conditions
(\ref{2hdm49}) and (\ref{2hdm50}).\\
\\
{\bf Fig.5.}\,\, Higgs masses and couplings for $\Lambda=M_{Pl}$,
$h^2_t(M_{Pl})=10$, $R_1(M_{Pl})=0.75$ and $R_2(M_{Pl})=0.883$.
(a) The dependence of the spectrum of Higgs bosons on the
pseudoscalar Higgs mass $m_A$. The dash--dotted and dashed lines
correspond to the CP--even Higgs boson masses, while the solid
line represents the mass of the charged Higgs states. All masses
are given in $\mbox{GeV}$. (b) Absolute values of the relative
couplings $R_{ZZh_i}$ of the Higgs scalars to $Z$ pairs. The solid
and dashed--dotted curves represent the dependence of the
couplings of the lightest and heaviest CP--even Higgs states to Z
pairs on $m_A$. (c) Absolute values of the relative couplings
$R_{t\bar{t}h_i}$ of the lightest (solid curve) and heaviest
(dashed--dotted curve) CP--even Higgs bosons to the top quark as a
function of $m_A$. Here the Higgs self--couplings
$\lambda_3(\Lambda)$ and $\lambda_4(\Lambda)$
satisfy the MPP conditions (\ref{2hdm49}) and (\ref{2hdm50}).\\
\\
{\bf Fig.6.}\,\,Upper bound on the mass of the SM--like Higgs
boson versus the MPP scale $\Lambda$ in the quasi--fixed point
scenario. The solid and dashed curves correspond to
$h_t^2(\Lambda)=10$ and $h_t^2(\Lambda)=2.25$. The Higgs
self--couplings $\lambda_1(\Lambda)$ and $\lambda_2(\Lambda)$ are
fixed so that $R_1(\Lambda)=0.75$ and $R_2(\Lambda)=0.883$,
whereas $\lambda_3(\Lambda)$ and $\lambda_4(\Lambda)$ obey the MPP
conditions (\ref{2hdm49}) and (\ref{2hdm50}). The value of
$\tan\beta$ is chosen so that
$m_t(M_t)=161.6\,\mbox{GeV}$. The MPP scale $\Lambda$ is given in GeV.\\
\\
{\bf Fig.7.}\,\, Higgs masses and couplings for
$\Lambda=100\,\mbox{TeV}$, $h^2_t(\Lambda)=10$, $R_1(\Lambda)=0.75$
and $R_2(\Lambda)=0.883$. (a) Spectrum of Higgs bosons versus
$m_A$. (b) Absolute values of the relative couplings $R_{ZZh_i}$
of the Higgs scalars to $Z$ pairs. (c) Absolute values of the
relative couplings $R_{t\bar{t}h_i}$ of the CP--even Higgs bosons
to the top quark as a function of $m_A$. Here $\lambda_3(\Lambda)$
and $\lambda_4(\Lambda)$ obey MPP conditions.
The notations are the same as in Fig.~5.\\
\\
{\bf Fig.8.}\,\,The running of $\lambda_1(\mu)$, $\lambda_2(\mu)$
and $\widehat{\lambda}(\mu)$ in the MPP scenario which implies the
existence of a set of vacua degenerate with respect to
$\tan\gamma$ at the scale $\Lambda$. (a) The renormalisation group
flow of these Higgs self--couplings below $\Lambda=M_{Pl}$. (b)
The evolution of $\lambda_1(\mu)$, $\lambda_2(\mu)$ and
$\widehat{\lambda}(\mu)$ below $\Lambda=10\,\mbox{TeV}$. The solid,
dashed and dash--dotted lines correspond to $\lambda_1(\mu)$,
$\lambda_2(\mu)$ and $\widehat{\lambda}(\mu)$ respectively. Here we
fix $g_b(\Lambda)=g_{\tau}(\Lambda)=\lambda_5(\Lambda)=0$,
$m_t(M_t)=165\,\mbox{GeV}$ and $\alpha_3(M_Z)=0.117$. 
Other parameters are specified in Table 3.\\

\newpage
{\large \hspace{-1.5cm}$h_t(M_t)$}
\begin{center}
{\hspace*{-20mm}\includegraphics[height=100mm,keepaspectratio=true]{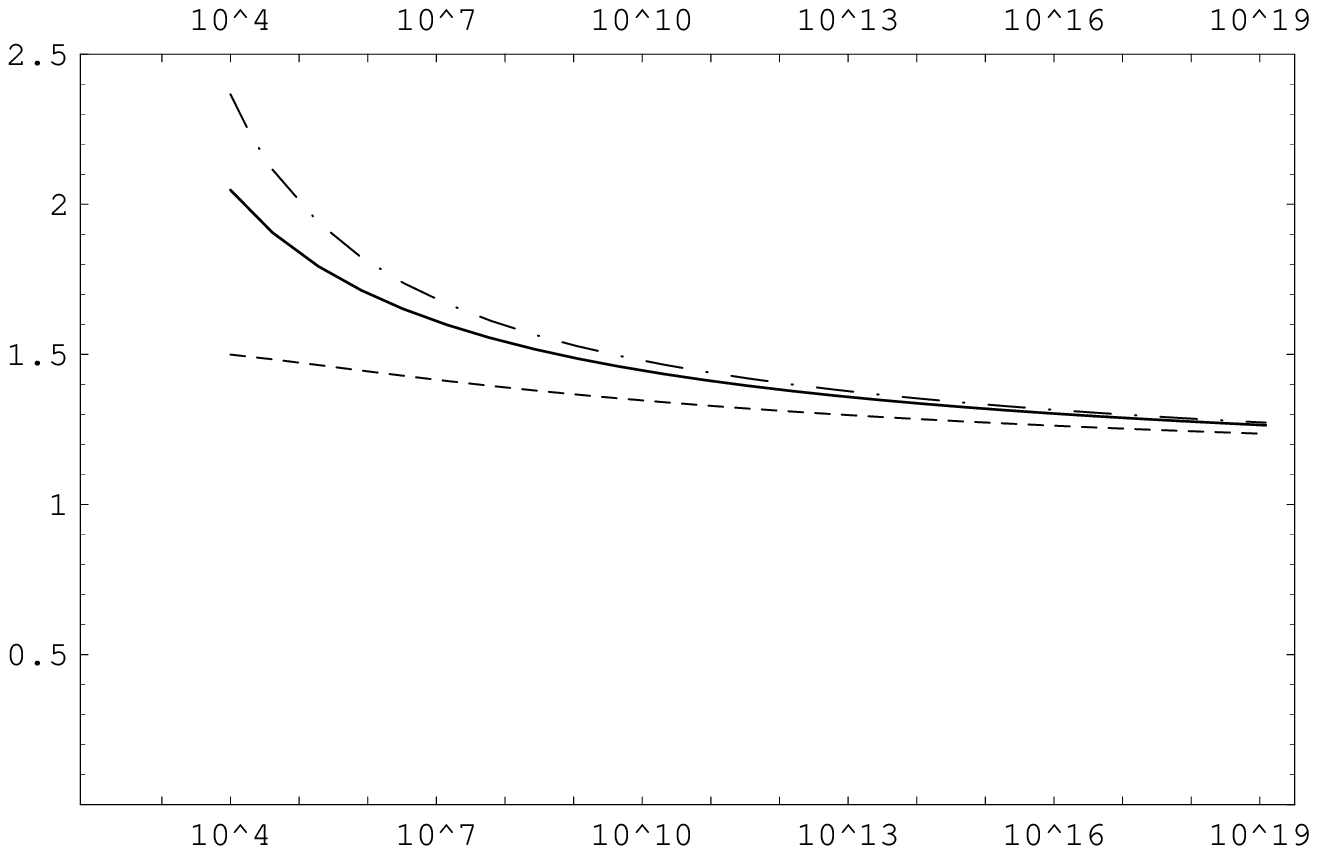}}\\
{\large $ \Lambda $ }\\[0mm]
{\large\bfseries Fig.1a}
\end{center}
{\large \hspace{-1cm}$\tan\beta$}
\begin{center}
{\hspace*{-20mm}\includegraphics[height=100mm,keepaspectratio=true]{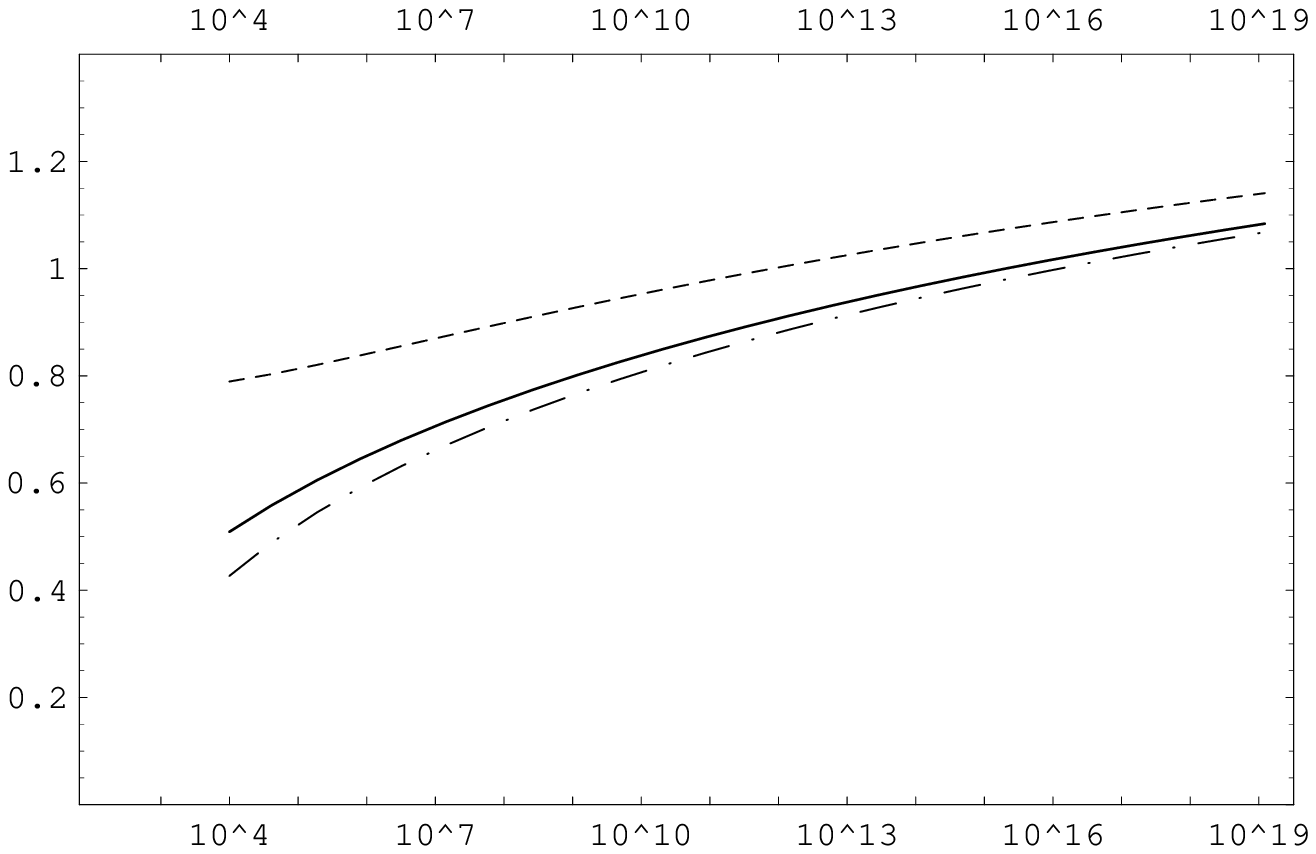}}\\
{\large $ \Lambda $}\\[0mm] {\large\bfseries Fig.1b}
\end{center}

\newpage
{\large \hspace{-1.5cm}$R_2(\Lambda)$}
\begin{center}
{\hspace*{-20mm}\includegraphics[height=100mm,keepaspectratio=true]{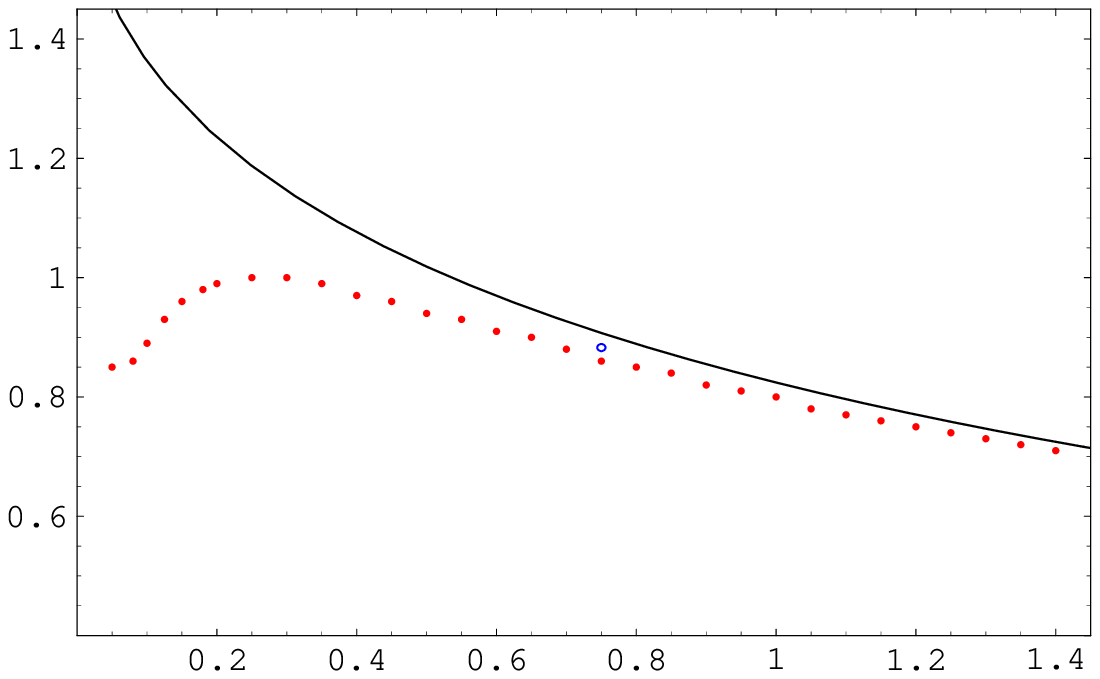}}\\
{\large $R_1(\Lambda)$ }\\[0mm]
{\large\bfseries Fig.2a}
\end{center}
{\large \hspace{-1cm}$R_2(\Lambda)$}
\begin{center}
{\hspace*{-20mm}\includegraphics[height=100mm,keepaspectratio=true]{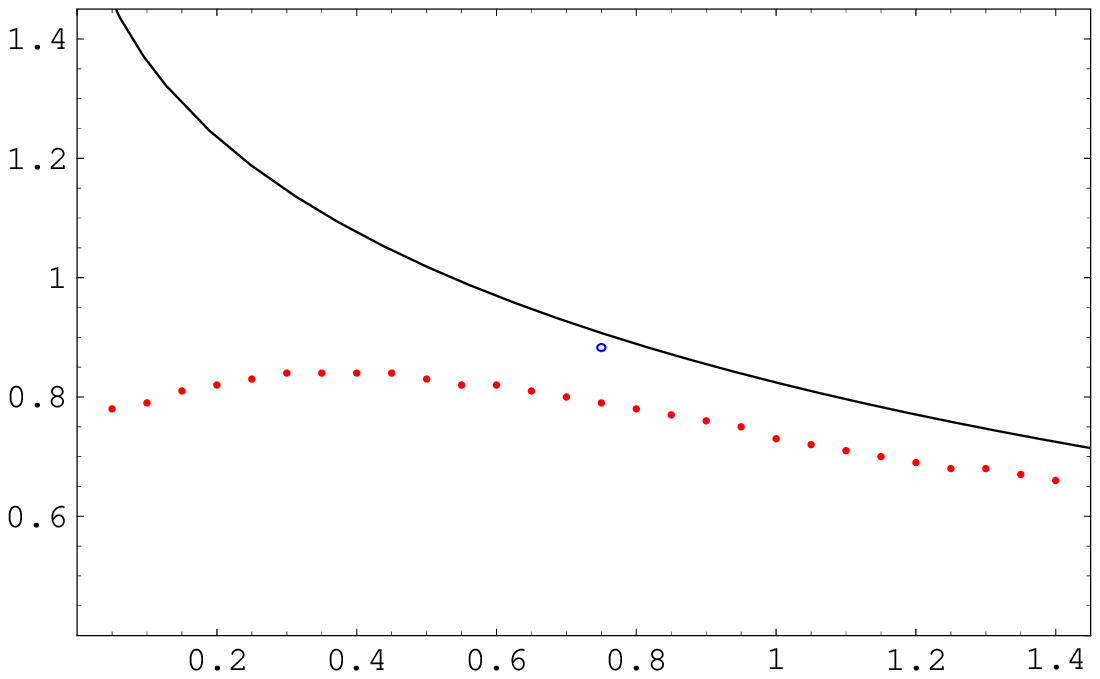}}\\
{\large $R_1(\Lambda)$}\\[0mm] {\large\bfseries Fig.2b}
\end{center}

\newpage
{\large \hspace{-1.5cm}$R_2(\mu)$}
\begin{center}
{\hspace*{-20mm}\includegraphics[height=100mm,keepaspectratio=true]{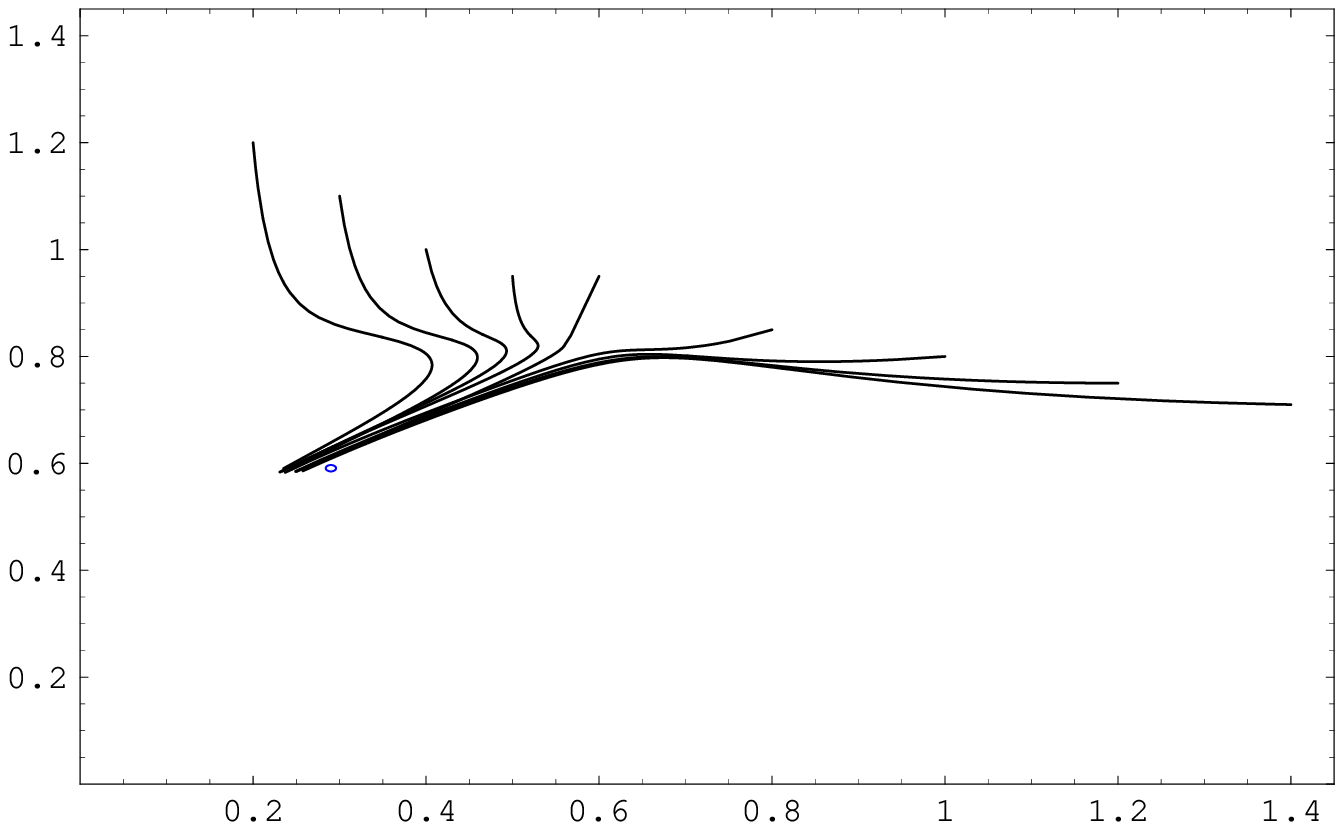}}\\
{\large $ R_1(\mu) $ }\\[0mm]
{\large\bfseries Fig.3a}
\end{center}
{\large \hspace{-1cm}$R_3(\mu)$}
\begin{center}
{\hspace*{-20mm}\includegraphics[height=100mm,keepaspectratio=true]{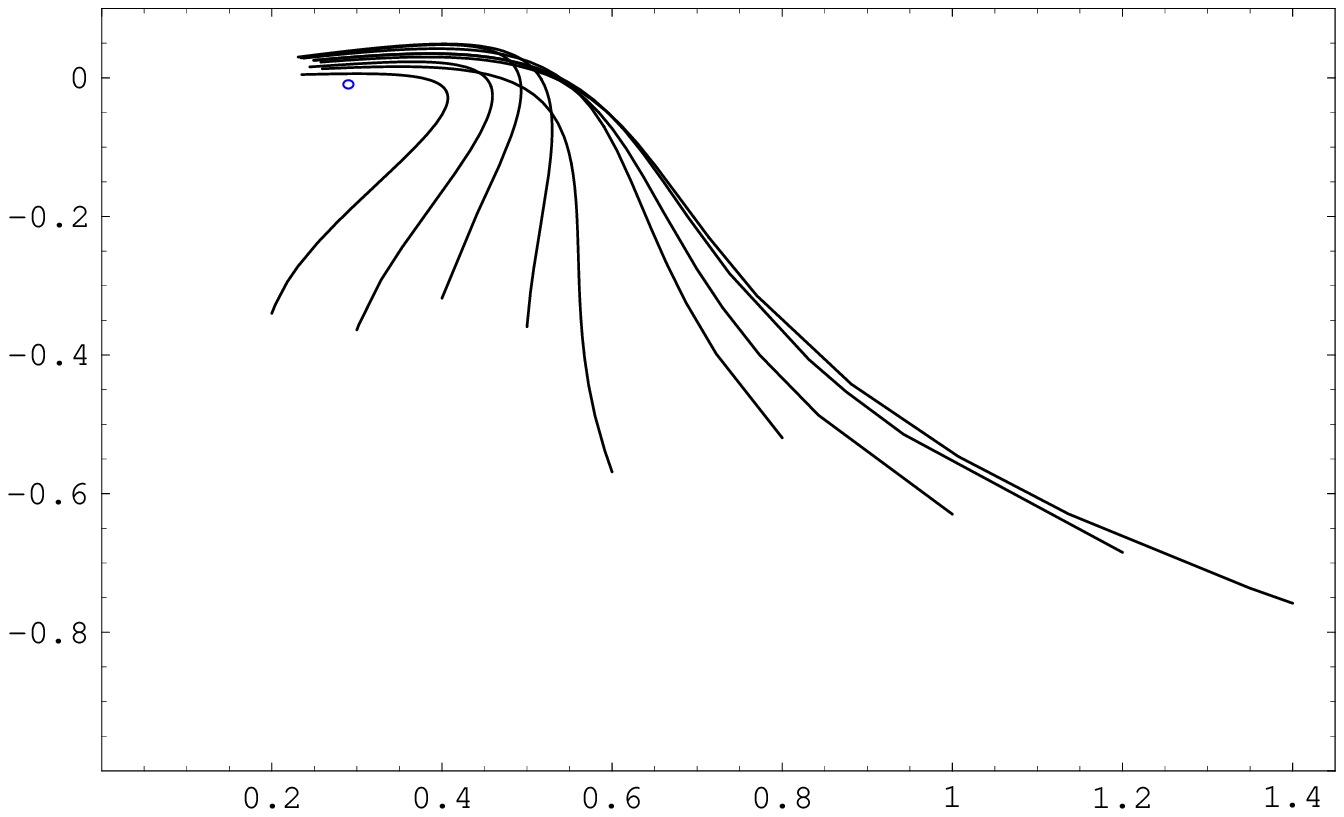}}\\
{\large $ R_1(\mu) $}\\[0mm]
{\large\bfseries Fig.3b}
\end{center}

\newpage
{\large \hspace{-1.5cm}$R_4(\mu)$}
\begin{center}
{\hspace*{-20mm}\includegraphics[height=100mm,keepaspectratio=true]{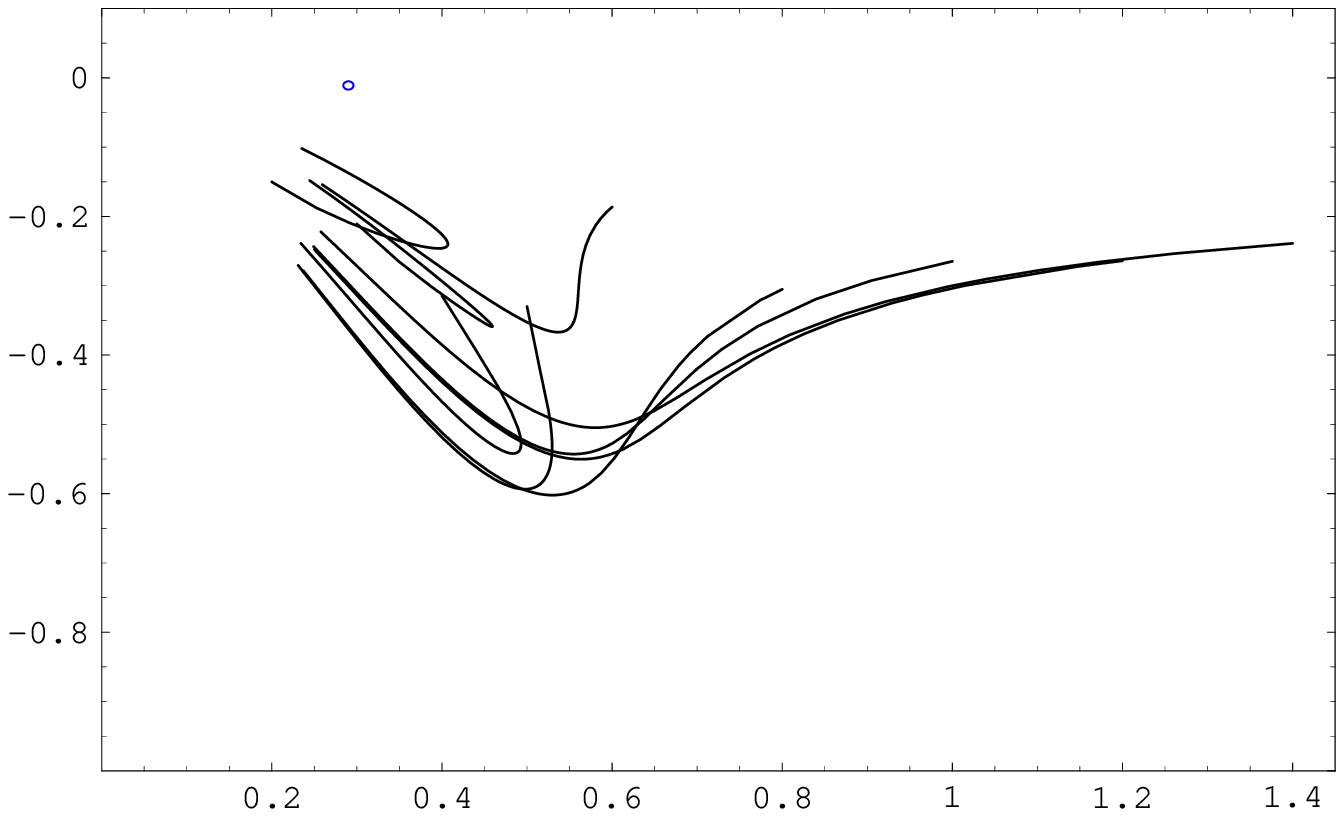}}\\
{\large $R_1(\mu)$}\\[0mm]
{\large\bfseries Fig.3c}
\end{center}

\newpage
{\large \hspace{-1.5cm}$R_1(M_t)$}
\begin{center}
{\hspace*{-20mm}\includegraphics[height=100mm,keepaspectratio=true]{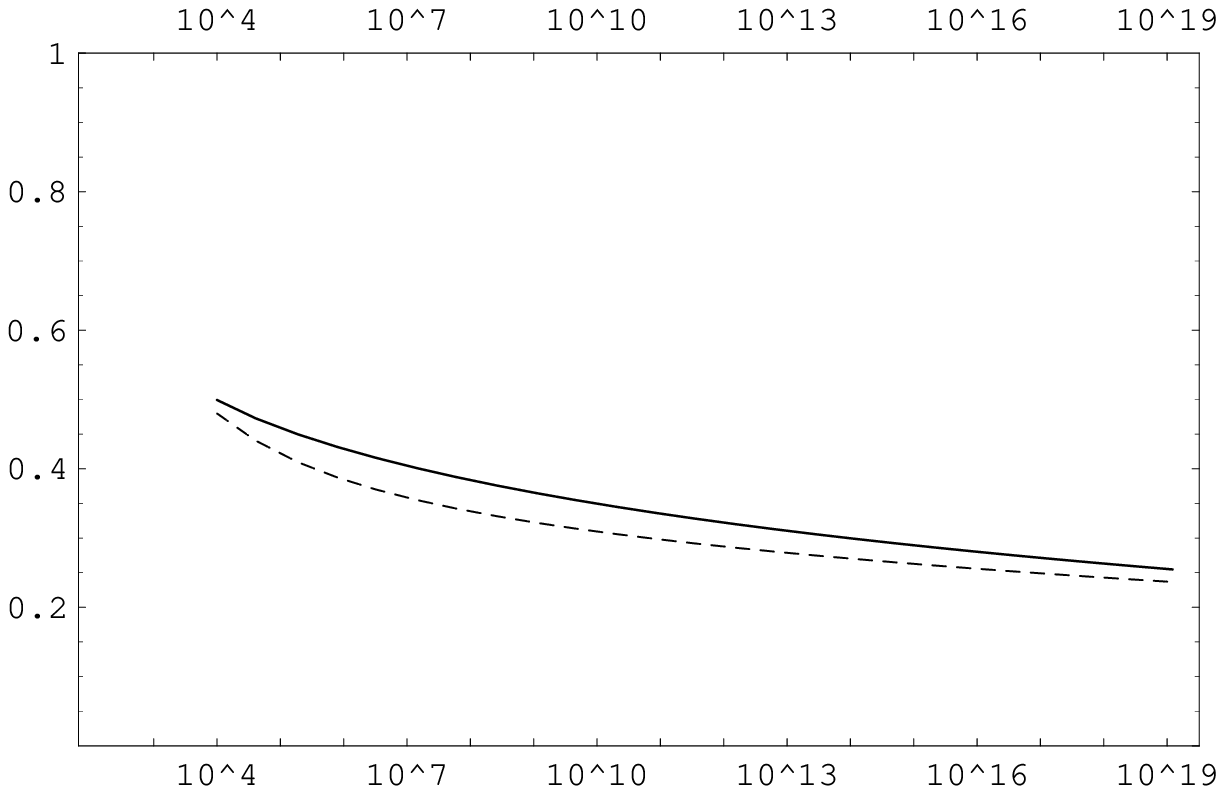}}\\
{\large $ \Lambda $ }\\[0mm]
{\large\bfseries Fig.4a}
\end{center}
{\large \hspace{-1cm}$R_2(M_t)$}
\begin{center}
{\hspace*{-20mm}\includegraphics[height=100mm,keepaspectratio=true]{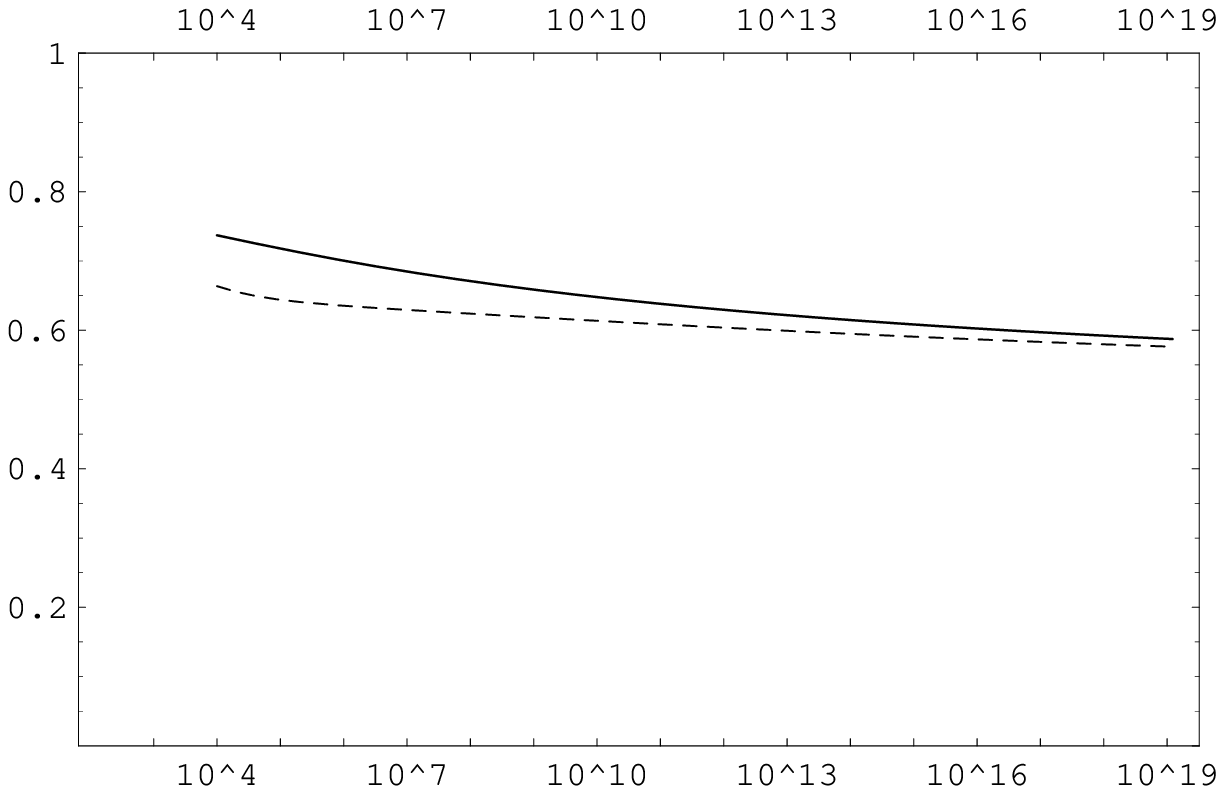}}\\
{\large $ \Lambda $}\\[0mm] {\large\bfseries Fig.4b}
\end{center}

\newpage
{\large \hspace{-1.5cm}$R_3(M_t)$}
\begin{center}
{\hspace*{-20mm}\includegraphics[height=100mm,keepaspectratio=true]{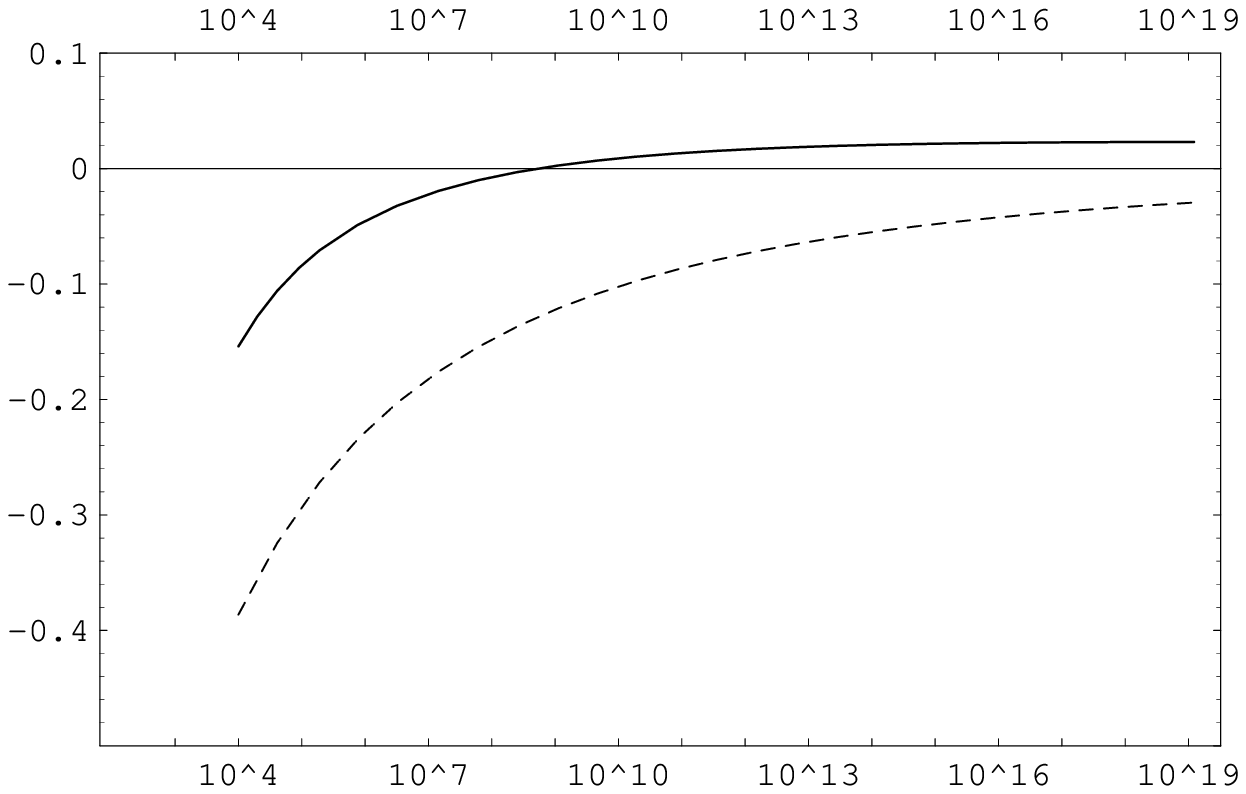}}\\
{\large $ \Lambda $ }\\[0mm]
{\large\bfseries Fig.4c}
\end{center}
{\large \hspace{-1cm}$R_4(M_t)$}
\begin{center}
{\hspace*{-20mm}\includegraphics[height=100mm,keepaspectratio=true]{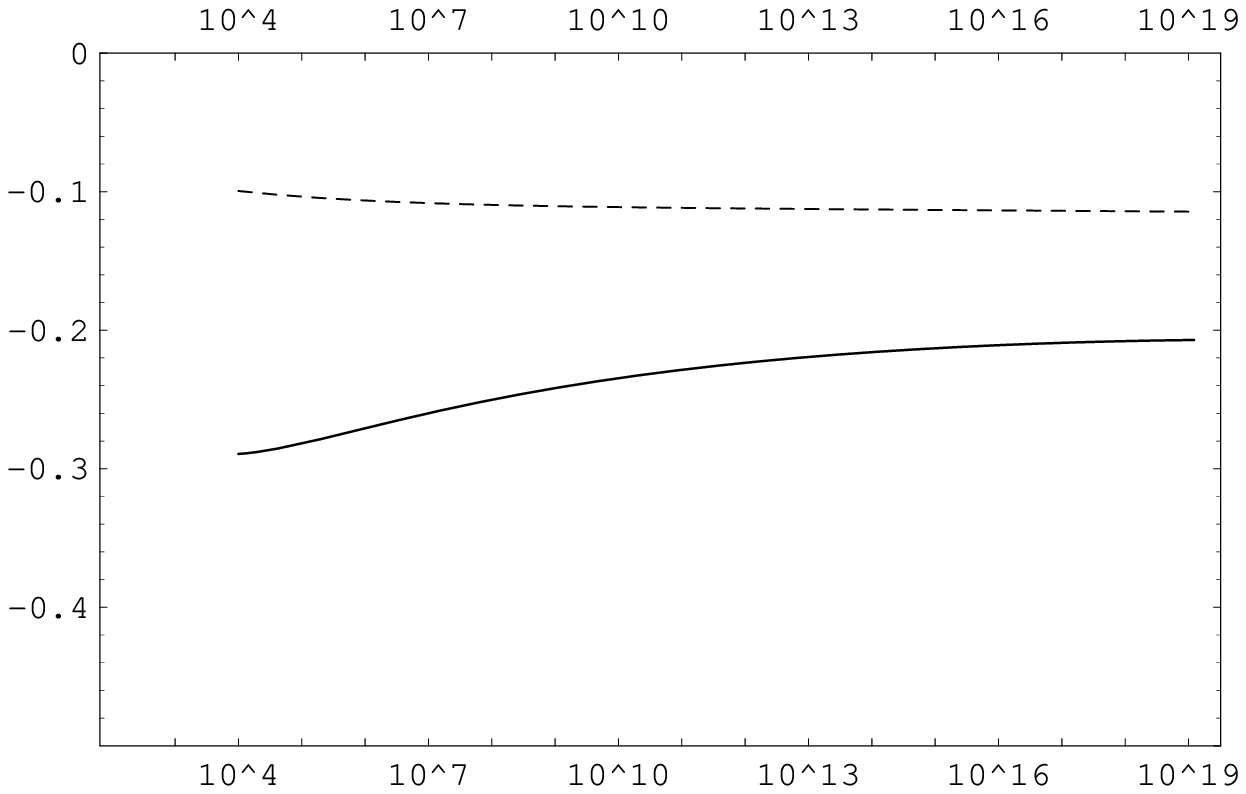}}\\
{\large $ \Lambda $}\\[0mm] {\large\bfseries Fig.4d}
\end{center}

\newpage
{\large \hspace{-1.5cm}$m_{h_i}$}
\begin{center}
{\hspace*{-20mm}\includegraphics[height=100mm,keepaspectratio=true]{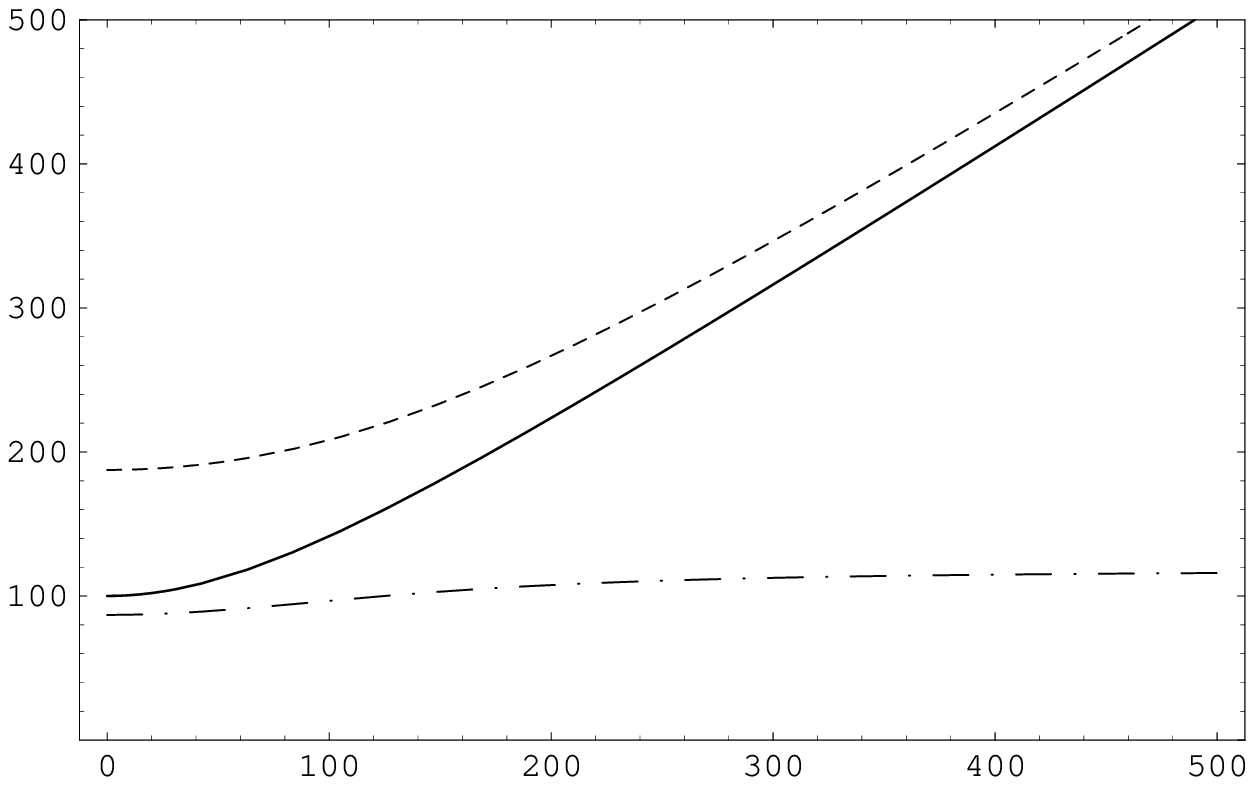}}\\
{\large $m_A$}\\[0mm]
{\large\bfseries Fig.5a}
\end{center}
\vspace{-5mm}
{\large \hspace{-1cm}$|R_{ZZh_i}|$}
\begin{center}
{\hspace*{-20mm}\includegraphics[height=100mm,keepaspectratio=true]{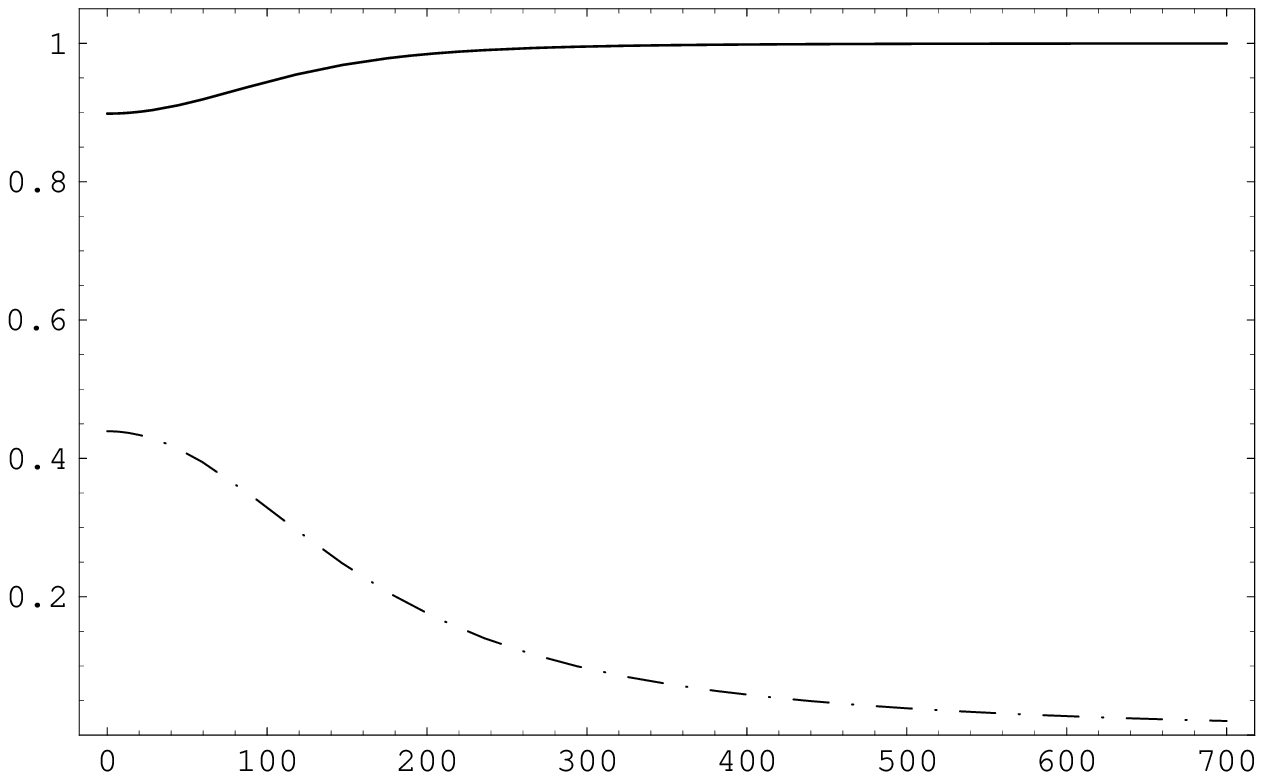}}\\
{\large $m_A$}\\[0mm]
{\large\bfseries Fig.5b}
\end{center}

\newpage
{\large \hspace{-1.5cm}$|R_{t\bar{t}h_i}|$}
\begin{center}
{\hspace*{-20mm}\includegraphics[height=100mm,keepaspectratio=true]{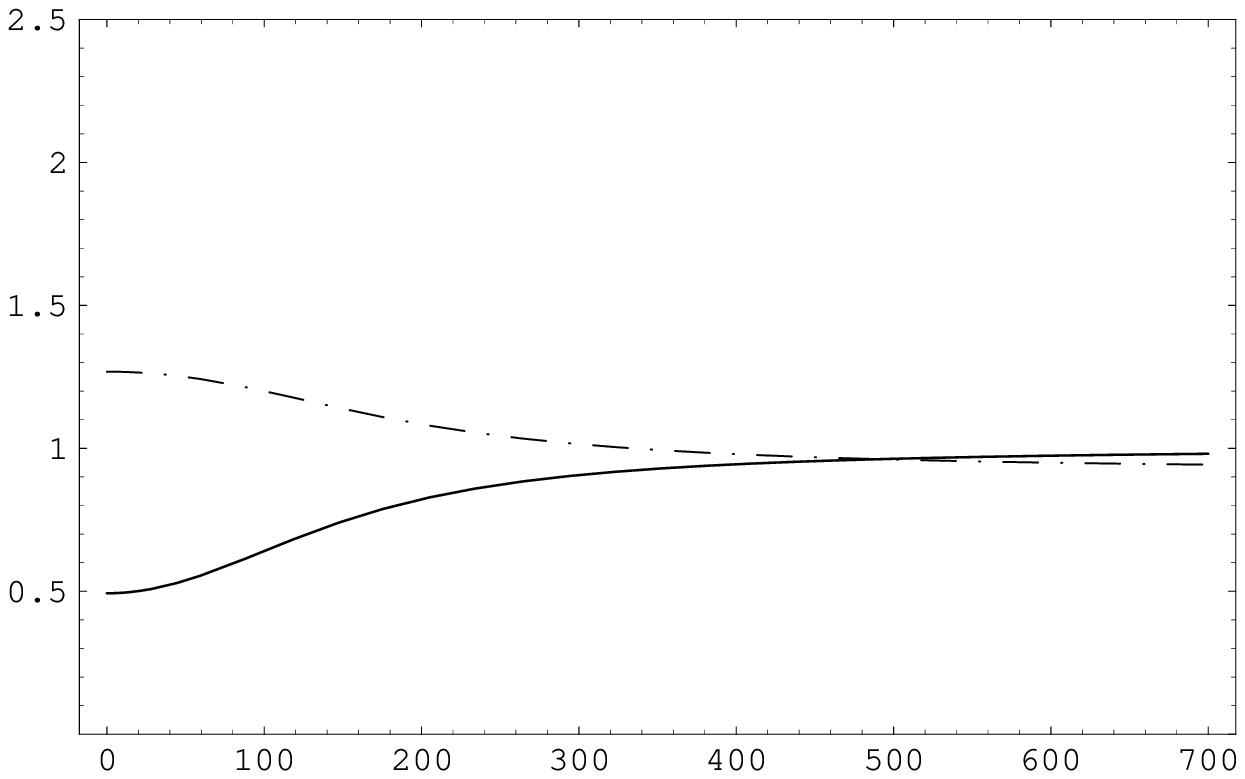}}\\
{\large $m_A$}\\[0mm]
{\large\bfseries Fig.5c}
\end{center}

\newpage
{\large \hspace{-1cm}$m_h$}
\begin{center}
{\hspace*{-20mm}\includegraphics[height=100mm,keepaspectratio=true]{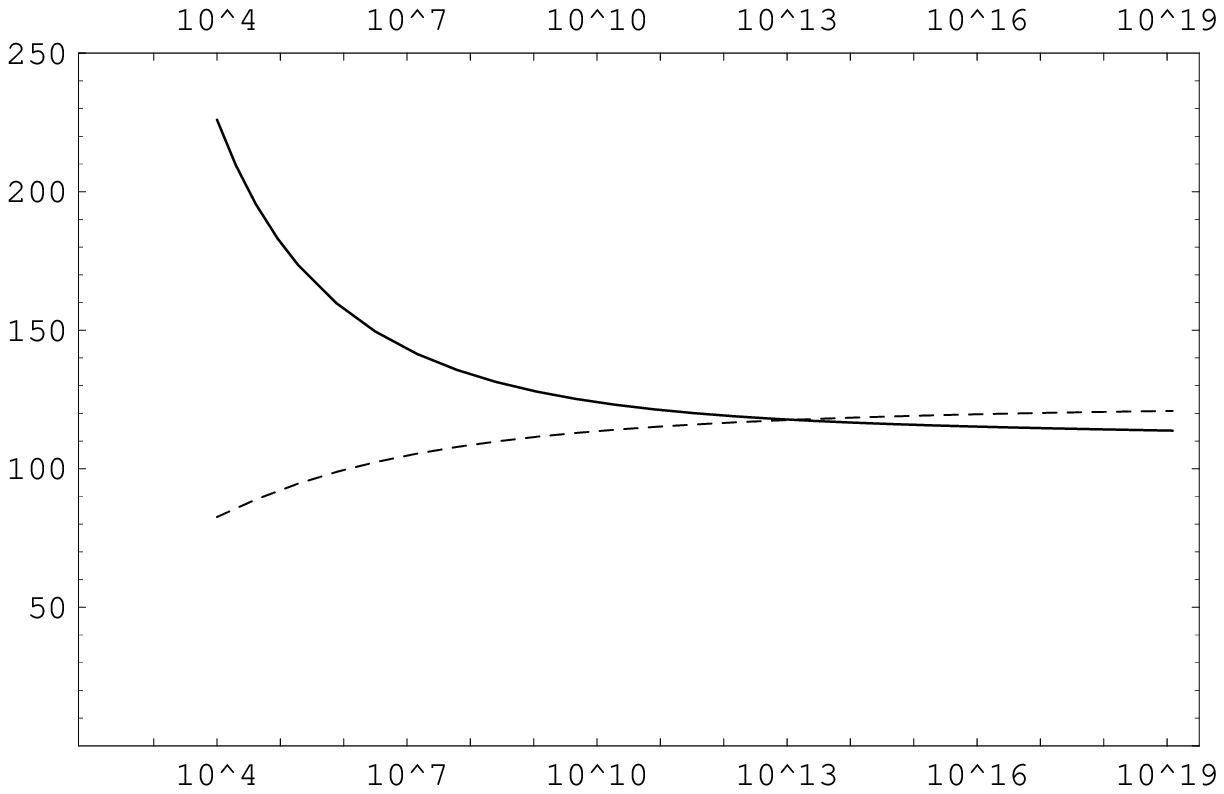}}\\
{\large $\Lambda$ }\\[0mm]
{\large\bfseries Fig.6}
\end{center}

\newpage
{\large \hspace{-1.5cm}$m_{h_i}$}
\begin{center}
{\hspace*{-20mm}\includegraphics[height=100mm,keepaspectratio=true]{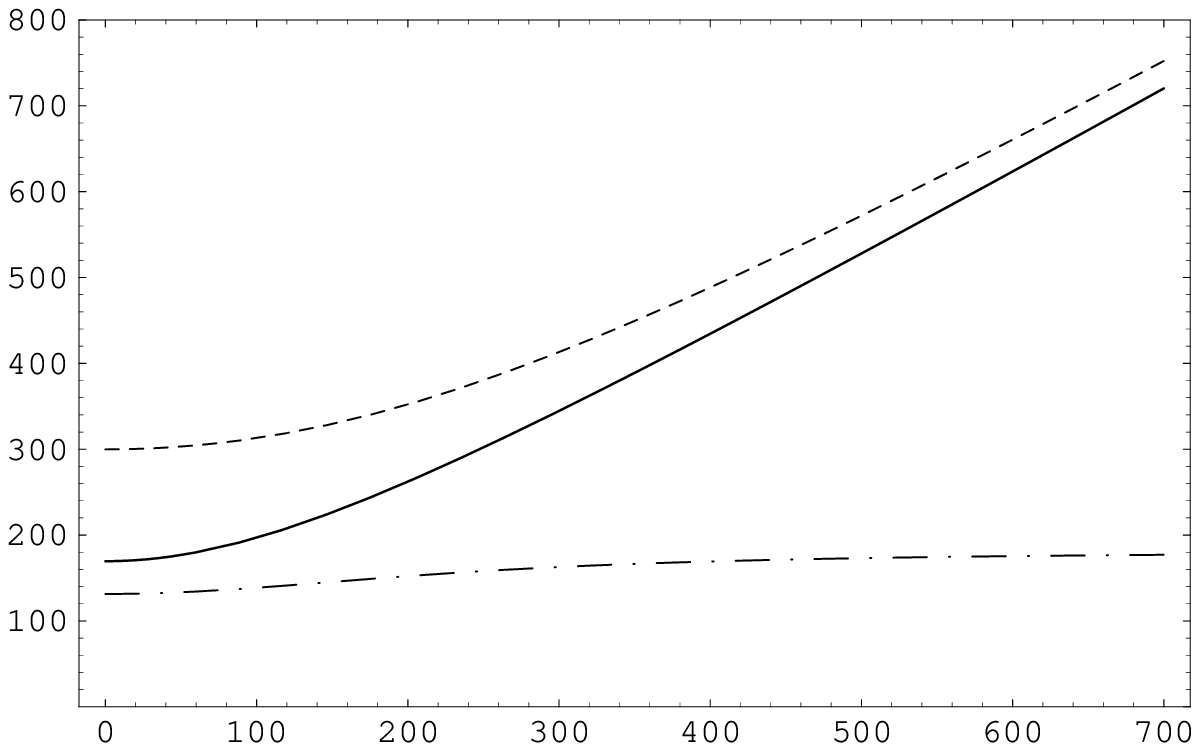}}\\
{\large $m_A$}\\[0mm]
{\large\bfseries Fig.7a}
\end{center}
\vspace{-5mm}
{\large \hspace{-1cm}$|R_{ZZh_i}|$}
\begin{center}
{\hspace*{-20mm}\includegraphics[height=100mm,keepaspectratio=true]{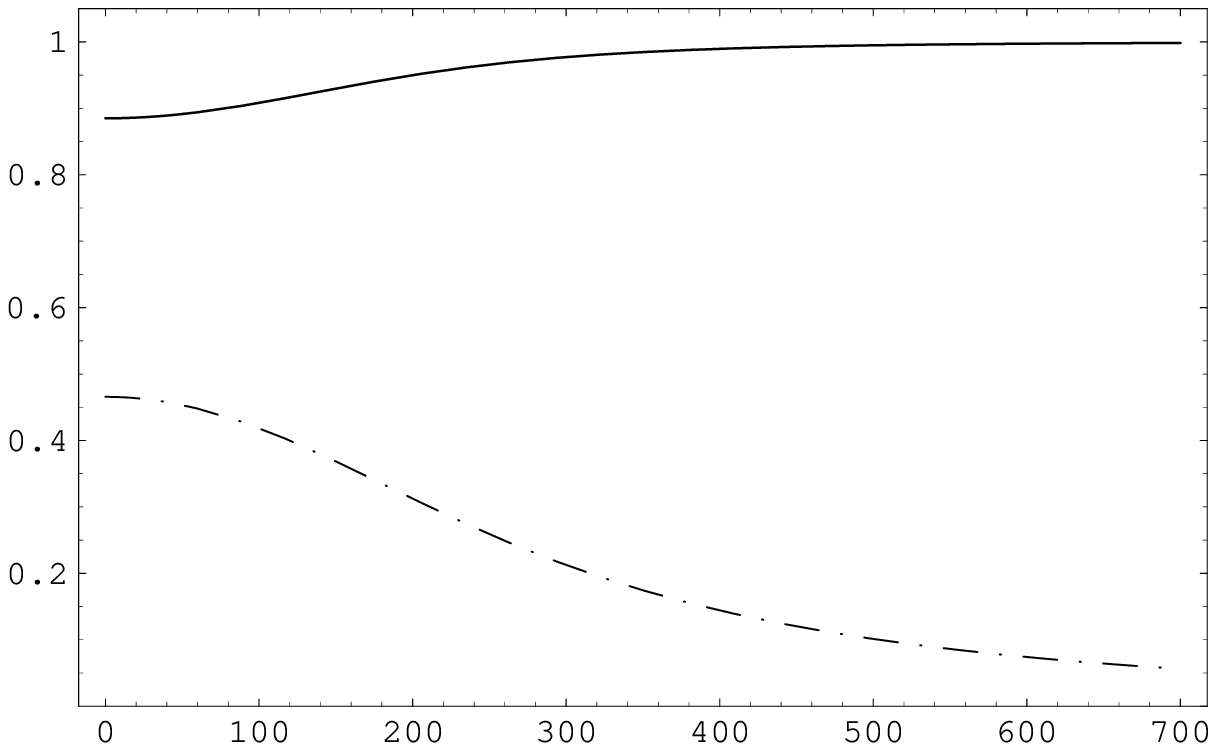}}\\
{\large $m_A$}\\[0mm]
{\large\bfseries Fig.7b}
\end{center}

\newpage
{\large \hspace{-1.5cm}$|R_{t\bar{t}h_i}|$}
\begin{center}
{\hspace*{-20mm}\includegraphics[height=100mm,keepaspectratio=true]{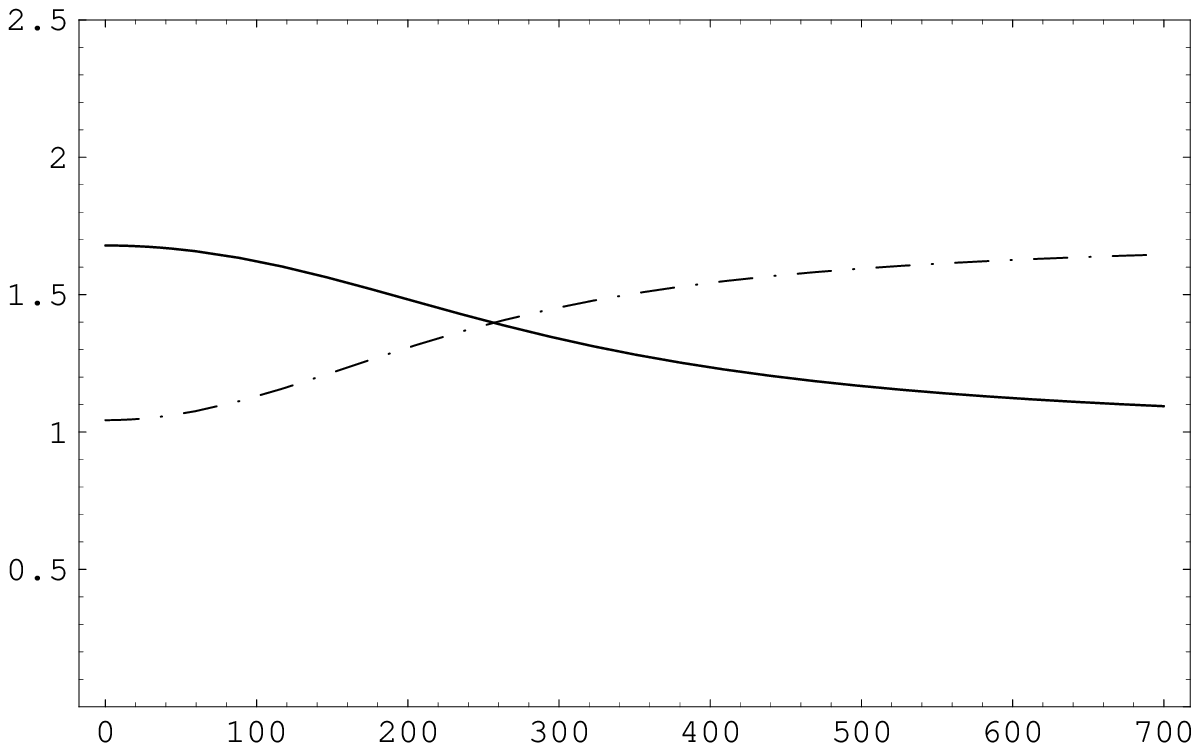}}\\
{\large $m_A$}\\[0mm]
{\large\bfseries Fig.7c}
\end{center}

\newpage
{\large \hspace{-1.5cm}$\lambda_i(\mu)$}
\begin{center}
{\hspace*{-20mm}\includegraphics[height=100mm,keepaspectratio=true]{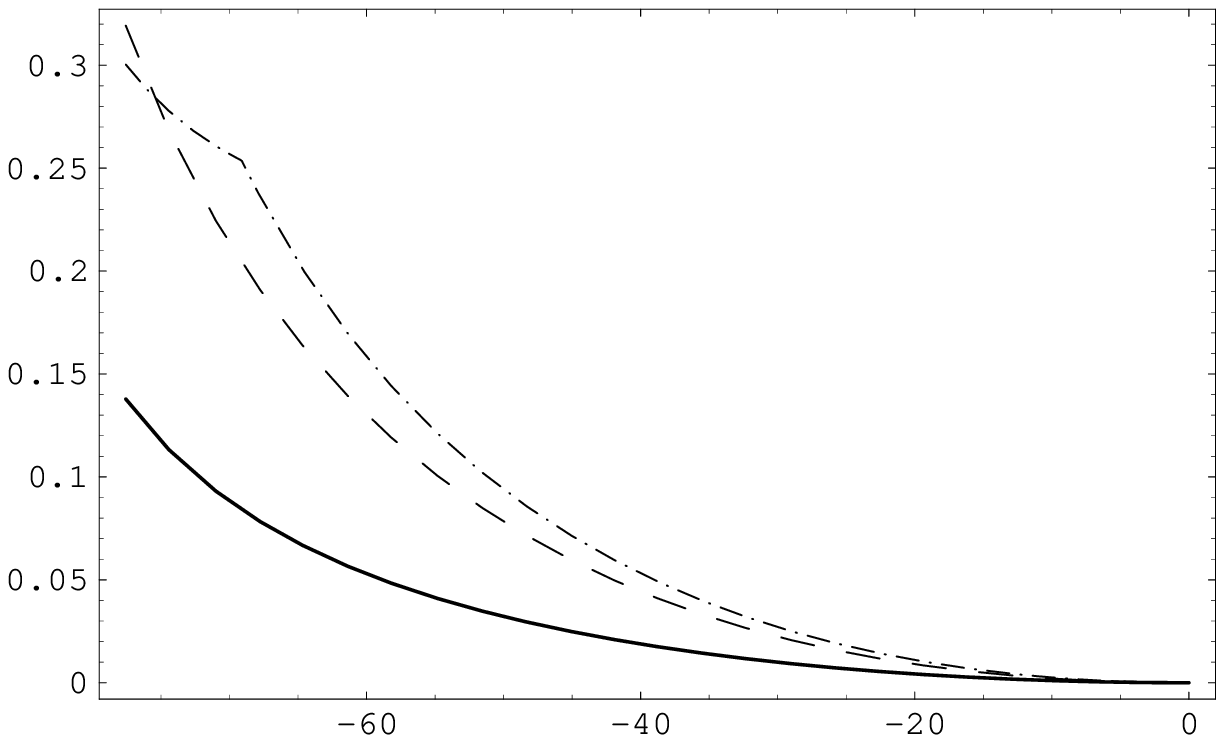}}\\
{\large $\log[\mu^2/M_{Pl}^2]$}\\[0mm]
{\large\bfseries Fig.8a}
\end{center}
{\large \hspace{-0.5cm}$\lambda_i(\mu)$}
\begin{center}
{\hspace*{-20mm}\includegraphics[height=100mm,keepaspectratio=true]{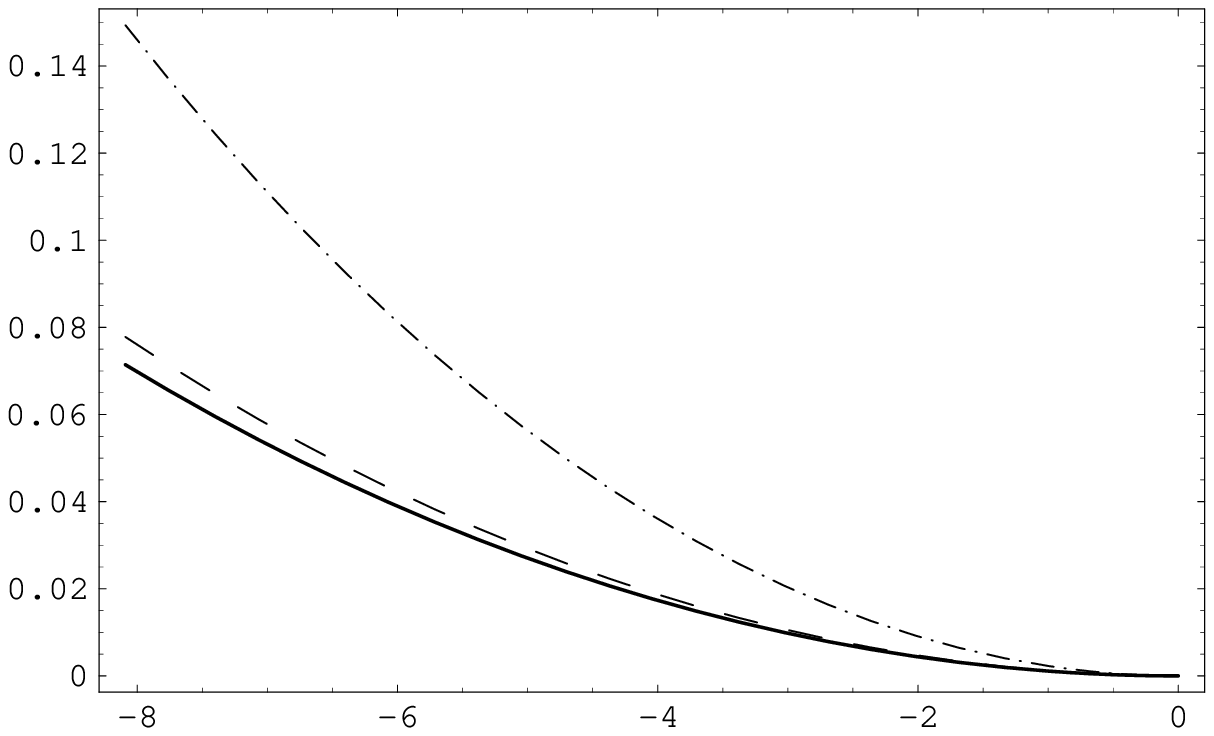}}\\
{\large $\log[\mu^2/\Lambda^2]$}\\[0mm] {\large\bfseries Fig.8b}
\end{center}

\begin{table}[qfp]
  \centering
  \begin{tabular}{|c|c|c|c|c|c|c|}
\hline
            & $\Lambda$                & $\rho_t(M_t)$ & $\rho_b(M_t)$ & $\rho_{\tau}(M_t)$ & $m_t(M_t)$ & $m_b(M_t)$ \\
\hline
            & $M_{Pl}$                 & $1.012$       & $0.707$       &  $0.391$           & $213-203$  & $2.39-2.09$ \\
\cline{2-7}
model $II$  & $10^{13}\,\mbox{GeV}$    & $1.173$       & $0.782$       &  $0.586$           & $229-207$  & $2.05-1.79$ \\
\cline{2-7}
            & $10^{7}\,\mbox{GeV}$     & $1.664$       & $0.986$       &  $1.287$           & $273-208$  & $1.56-1.34$ \\
\hline
            & $M_{Pl}$                 & $1.081$       &  ---          &  $0.732$           & $220-208$  &  --- \\
\cline{2-7}
model $III$ & $10^{13}\,\mbox{GeV}$    & $1.248$       &  ---          &  $1.002$           & $237-212$  &  --- \\
\cline{2-7}
            & $10^{7}\,\mbox{GeV}$     & $1.754$       &  ---          &  $1.868$           & $281-211$  &  --- \\
\hline
            & $M_{Pl}$                 & $0.976$       &  $0.949$      &   ---              & $209-199$  &  --- \\
\cline{2-7}
model $IV$  & $10^{13}\,\mbox{GeV}$    & $1.128$       &  $1.107$      &   ---              & $225-203$  &  --- \\
\cline{2-7}
            & $10^{7}\,\mbox{GeV}$     & $1.593$       &  $1.578$      &   ---              & $267-205$  &  --- \\
\hline
\end{tabular}
\caption{Predictions for $\rho_t(M_t) = (h_t(M_t)/g_3(M_t))^2$,
$\rho_b(M_t) = (h_b(M_t)/g_3(M_t))^2$, $\rho_{\tau}(M_t) =
(h_{\tau}(M_t)/g_3(M_t))^2$ and running quark masses obtained near
the quasi--fixed points of the RG equations at large values of
$\tan\beta$, in different MPP scenarios with approximate
Peccei--Quinn and $Z_2$--symmetries. All masses are given in GeV.
The specified ranges of quark masses correspond to the variation
of $h_t^2(\Lambda)$ from $10$ to 1.}
  \label{qfp}
\end{table}

\begin{table}[qfp1]
  \centering
  \begin{tabular}{|c|c|c|c|}
\hline
 $R_1^0$      &     $R_2^0$     &   $R_3^0$  &  $R_4^0$\\
\hline
  $0$       &     $-1.133$  &   $0$    &  $0$  \\
\hline
  $0$       &     $0.883$   &   $0$    &  $0$  \\
\hline
  $0.750$   &     $-1.133$  &   $0$    &  $0$  \\
\hline
  $0.750$   &     $0.883$   &   $0$    &  $0$  \\
\hline
  $0.742$   &     $0.880$   & $-0.160$ &  $0.259$ \\
\hline
  $0.742$   &     $0.880$   & $0.099$  &  $-0.259$ \\
\hline
\end{tabular}
\caption{ The six fixed points of the RG equations for
$g_1(\mu)=g_2(\mu)=g_3(\mu)=0$ and
$h_b(\mu)=h_{\tau}(\mu)=g_b(\mu)=g_{\tau}(\mu)=g_t(\mu)=0$.}
\label{qfp1}
\end{table}

\begin{table}[qfp2]
  \centering
  \begin{tabular}{|c|c|c|c|c|}
\hline
 $\Lambda$      &  $h_t(\Lambda)$  & $\tan\beta$  &  $m_b(M_t)$  & $m_h$\\
\hline
  $10^4$\, GeV  &   0.811          &   $49.84$    &  $3.24$ & $69.0$ \\
\hline
  $10^8$\, GeV  &   0.645          &   $47.64$    &  $3.28$ & $115.7$ \\
\hline
$10^{12}$\, GeV &   0.549          &   $47.41$    &  $3.18$ & $130.1$ \\
\hline
$10^{16}$\, GeV &   0.480          &   $48.53$    &  $2.94$ & $136.3$ \\
\hline
  $M_{Pl}$      &   0.435          &   $50.43$    &  $2.61$ & $138.9$ \\
\hline
\end{tabular}
\caption{The dependence of the upper bound $m_h$ for the lightest Higgs
mass, $\tan\beta$ and $m_b(M_t)$ on the MPP scale, in the MPP
scenario that implies the existence of a set of vacua degenerate
with respect to $\tan\gamma$ at the scale $\Lambda$. The
Peccei--Quinn symmetry violating Yukawa couplings $g_b(\Lambda)$
and $g_{\tau}(\Lambda)$, as well $\lambda_5(\Lambda)$, are set to
zero. All masses are given in GeV.} \label{qfp2}
\end{table}

\begin{table}[qfp3]
  \centering
  \begin{tabular}{|c|c|c|c|}
\hline
 $\Lambda$      &  $h_t(\Lambda)$   & $\ds\frac{\lambda_5(\Lambda)}{\lambda_4(\Lambda)}$  & $m_h$\\
\hline
  $10^4$\, GeV  &   0.811           & $0$                                                 & $69.0$ \\
\hline
  $10^4$\, GeV  &   0.811           & $0.5$                                               & $72.4$ \\
\hline
  $10^4$\, GeV  &   0.811           & $0.8$                                               & $74.3$ \\
\hline
  $10^4$\, GeV  &   0.811           & $-0.5$                                              & $72.1$ \\
\hline
  $10^4$\, GeV  &   0.811           & $-0.8$                                              & $73.9$ \\
\hline
  $10^8$\, GeV  &   0.645           & $0$                                                 & $115.7$ \\
\hline
  $10^8$\, GeV  &   0.645           & $0.5$                                               & $117.9$ \\
\hline
  $10^8$\, GeV  &   0.645           & $0.8$                                               & $119.1$ \\
\hline
  $10^8$\, GeV  &   0.645           & $-0.5$                                              & $117.9$ \\
\hline
  $10^8$\, GeV  &   0.645           & $-0.8$                                              & $119.0$ \\
\hline
  $M_{Pl}$      &   0.435           & $0$                                                 & $138.9$ \\
\hline
  $M_{Pl}$      &   0.435           & $0.5$                                               & $139.2$ \\
\hline
  $M_{Pl}$      &   0.435           & $0.8$                                               & $139.2$ \\
\hline
  $M_{Pl}$      &   0.435           & $-0.5$                                              & $139.2$ \\
\hline
  $M_{Pl}$      &   0.435           & $-0.8$                                              & $139.2$ \\
\hline
\end{tabular}
\caption{The dependence of the upper bound $m_h$ for the lightest Higgs
mass on the Higgs self--couplings $\lambda_5(\Lambda)$ and
$\lambda_4(\Lambda)$ for different values of the scale $\Lambda$,
in the MPP scenario that implies the existence of a set of vacua
degenerate with respect to $\tan\gamma$ at the MPP scale. The
Peccei--Quinn symmetry violating Yukawa couplings $g_b(\Lambda)$
and $g_{\tau}(\Lambda)$ are set to zero. The theoretical
restriction on the SM--like Higgs mass is given in GeV.}
\label{qfp3}
\end{table}

\end{document}